\documentclass{aastex}

\usepackage{graphicx}
\usepackage{latexsym}
\usepackage{amsfonts}
\usepackage{amsmath}
\usepackage{amssymb}
\usepackage{revsymb}
\usepackage{bm}
\usepackage{color}
\usepackage{cancel}
\usepackage{epstopdf}
\usepackage{epsfig}

\newcommand{\be}{\begin{equation}}
\newcommand{\ee}{\end{equation}}
\newcommand{\bea}{\begin{eqnarray}}
\newcommand{\eea}{\end{eqnarray}}

\def\hi{H\,{\sc i}}

\begin{document}

\date{\center{Submitted \today}}

\title{Testing MONDian Dark Matter with Galactic Rotation Curves}

\author{%
Doug Edmonds \altaffilmark{1,2},
Duncan Farrah \altaffilmark{2},
Chiu Man Ho \altaffilmark{3,4},
\\
Djordje Minic \altaffilmark{2},
Y. Jack Ng \altaffilmark{5},
and
Tatsu Takeuchi \altaffilmark{2,6}
}
\email{%
dedmonds@ehc.edu,
farrah@vt.edu,
chiuman.ho@vanderbilt.edu,
\\
dminic@vt.edu,
yjng@physics.unc.edu,
takeuchi@vt.edu
}

\altaffiltext{1}{Department of Physics, Emory \& Henry College, Emory, VA 24327, USA}
\altaffiltext{2}{Department of Physics, Virginia Tech, Blacksburg, VA 24061, USA}
\altaffiltext{3}{Department of Physics and Astronomy, Vanderbilt University, Nashville, TN 37235, USA}
\altaffiltext{4}{Department of Physics and Astronomy, Michigan State University, East Lansing, MI 48824, USA}
\altaffiltext{5}{Institute of Field Physics, Department of Physics and Astronomy, University of North Carolina, Chapel Hill, NC 27599, USA}
\altaffiltext{6}{Kavli Institute for the Physics and Mathematics of the Universe (WPI), University of Tokyo, Kashiwa-shi, Chiba-ken 277-8583, Japan}

\begin{abstract}
MONDian dark matter (MDM) is a new form of dark matter quantum that naturally accounts for Milgrom's scaling, usually associated with modified Newtonian dynamics (MOND), and theoretically behaves like cold dark matter (CDM) at cluster and cosmic scales. In this paper, we provide the first observational test of MDM by fitting rotation curves to a sample of 30 local spiral galaxies ($z \approx 0.003$). For comparison, we also fit the galactic rotation curves using MOND, and CDM. We find that all three models fit the data well. The rotation curves predicted by MDM and MOND are virtually indistinguishable over the range of observed radii ($\sim$1 to 30 kpc). The best-fit MDM and CDM density profiles are compared. We also compare with MDM the dark matter density profiles arising from MOND if Milgrom's formula is interpreted as Newtonian gravity with an extra source term instead of as a modification of inertia. We find that discrepancies between MDM and MOND will occur near the center of a typical spiral galaxy. In these regions, instead of continuing to rise sharply, the MDM mass density turns over and drops as we approach the center of the galaxy. Our results show that MDM, which restricts the nature of the dark matter quantum by accounting for Milgrom's scaling, accurately reproduces observed rotation curves.
\end{abstract}

\maketitle


\section{Introduction}

The cold dark matter (CDM) model \citep[e.g.,][]{dark} successfully explains several astrophysical phenomena. These include flat galactic rotation curves\footnote{In reality, rotation curves are not all flat, they display a variety of properties\citep[e.g.,][]{Persic1,Persic2,Catinella}.} \citep[e.g.,][]{Rubin80},
gravitational lensing \citep[e.g.,][]{bullet},
elemental abundances from big bang nucleosynthesis (BBN, e.g., \citealt{Cyburt04}),
and the power spectrum of cosmic microwave background anisotropies \citep[e.g.,][]{Lineweaver97,planck}.
This consistency has led to the widespread acceptance of the $\Lambda$CDM paradigm,
in which the Universe also exhibits a cosmological constant $\Lambda$.
CDM does, however, have remaining tensions with observations, especially on $\lesssim$Mpc scales.
These include inconsistency with the observed asymptotic velocity-mass ($v^4 \propto M$) scaling in the Tully-Fisher relation \citep{TF,Gentile,McGaugh,McGaugh2}, and the over-prediction of the number of satellite galaxies \citep{cen}.

Efforts have been made to construct theories that better match observations on galactic length scales than CDM. The most prominent of these is modified Newtonian dynamics \citep[MOND:][]{mond,teves,fmrev}.
In MOND, Newton's equation of motion $F=ma$ is modified to
\begin{equation}
F \;=\;
\begin{cases}
ma       & \mbox{for $a\gg a_c$} \\
ma^2/a_c & \mbox{for $a\ll a_c$}
\end{cases}
\end{equation}
where $a_c$ is the critical acceleration\footnote{In the literature on MOND, this is usually denoted $a_0$.}, which separates the two regions of different behavior.
For a given source mass $M$, we have $F=m(GM/r^2)\equiv ma_N$, where $a_N$ is the usual Newtonian acceleration without dark matter. Thus $a=a_N=GM/r^2$ when $a\gg a_c$, while $a=\sqrt{a_c a_N}$ when $a\ll a_c$.
On the outer edges of a galaxy of mass $M$ where gravity is weak, we can therefore expect
\begin{equation}
v^2 \;=\; r a \;\xrightarrow{r\rightarrow\infty}\; r\sqrt{a_c a_N} \;=\; \sqrt{a_c GM} \;\equiv\; v^2_\infty
\;,
\end{equation}
leading to asymptotically flat rotation curves, and $v^4\propto M$, the Tully-Fisher relation \citep{dsmond}.

The two regions of acceleration, $a\gg a_c$ and $a\ll a_c$,
are connected by a smooth interpolating function $\mu(x \equiv a/a_c)$ such that
\begin{equation}
\mu(x) \;=\;
\begin{cases}
1    & \mbox{for $x\gg 1$} \\
x    & \mbox{for $x\ll 1$}
\end{cases}
\end{equation}
and the MOND equation of motion is $F=ma\mu(x)$.
The choice for $\mu(x)$ is not unique. For example, some possible choices are
\begin{equation}
\mu(x) \;=\; \dfrac{x}{(1+x^n)^{1/n}}\;,\qquad
\dfrac{1}{x}\left(\sqrt{\dfrac{1}{4}+x^2}-\dfrac{1}{2}\right)\;,\qquad
.
\label{MONDinterps}
\end{equation}
Milgrom's original choice was the first expression with $n=2$.
The formula with $n=1$ was adopted by \citep{Famaey05} while the second expression in Equation (\ref{MONDinterps}) was studied by
\citep{interpol}. Fitting galactic rotation curves with MOND
by tuning the critical acceleration yields consistent values for $a_c$ \citep{Begeman91},
found to be numerically related to the speed of light $c$ and the Hubble parameter $H$ as
$a_c \approx c \,H/(2\, \pi) \sim 10^{-8} \mathrm{cm/s^2}$.

As mentioned above,
MOND naturally explains both
the observed non-keplarian galactic rotation curves and the
Tully-Fisher relation \citep{dsmond}.
MOND however struggles to reproduce observations at cluster and cosmological length scales \citep{bullet,Angus07,planck}.
Hence, CDM is usually preferred over MOND, with efforts ongoing to reconcile CDM with observations on $\lesssim$Mpc scales (e.g., \citealt{Swaters03,Frenk,Strigari}).


While CDM and MOND are often viewed as competing theories, it is notable that they actually complement each other, one being phenomenologically successful on length scales where the other is less so.
This complementarity motivated three of us (Ho, Minic, and Ng) to combine the salient successful features of both CDM and MOND into a unified scheme, MONDian dark matter \citep[MDM;][]{HMN, HMN2, HMN3},
which is a dark matter model that behaves like CDM at cluster and cosmological scales
while reproducing MOND at galactic scales.
This also provides an astrophysically motivated restriction on the nature of the dark matter quantum.


There has been an attempt \citep{turner}
to theoretically explain the Milgrom (MOND) scaling from the standard CDM scenario.
However, such an explanation has turned out to be quite complicated \citep{turner}. A more 
astrophysical attempt, which required some fine-tuning, was conducted in \citep{vandenBosch}.
Other approaches that try to unite MOND’s successes at galaxy scales to cosmology include \citet{Blanchet} and \citet{Klinkhamer}. 
In \citet{Blanchet}, the phenomenology of MOND is explained by
the concept of gravitational polarization. In \citet{Klinkhamer}, the MOND-type acceleration is explained by introducing a fundamental minimum temperature. Distinct from the existing approaches, the concept of MDM realizes a deep duality \footnote{See the last paragraph in Appendix A, right after Equation (A14).} between MOND and dark matter. For MDM, the appearance of Milgrom’s scaling is simple, but it has profound consequences on the properties of dark matter. Our approach differs from these papers in the crucial fact that it assumes
an actual source of dark matter, which, however, is not of a traditional particle type
(i.e. it is not described by a local effective quantum field theory).
These other approaches to which our approach could be compared do not have this central feature, and they operate with the assumption which amounts to an effective change of either the gravitational equations of motion, or the inertial properties.


In this paper, we provide the first observational test of MDM by comparing predicted rotation curves to those observed for a sample of spiral galaxies. The remainder of the paper is organized as follows. In $\S$2, we discuss the theoretical background for MDM. In $\S$3, we fit the above MONDian dark matter profile to galactic data,
tuning the mass to luminosity ratio $\alpha$.
Comparison of MDM and CDM density profiles is also presented.
In $\S$4, we summarize and discuss the results of the data fitting.
A physical motivation for the mass profile of MDM is given in the appendix.

\section{Theoretical Constructs}

MDM is a new form of dark matter, and while the equations governing MDM (Equation~\ref{modG}) can be mathematically rewritten in a form that is essentially MOND with a particular interpolating function, it is not physically sensible to do so. The form of MDM is rooted in entropic gravity (see the appendix). To rewrite MDM in a form resembling MOND would belie the entropic origin of MDM. However, as shown below, we can rewrite MOND as having dark matter and no modification of gravity. This is akin to MDM in a universe where $\Lambda$ vanishes, and allows us to directly compare MOND and MDM. We would also like to compare the CDM mass profiles with the MDM mass profiles on galactic scales.

In this section we will discuss the MDM mass profile and show how it can accommodate Milgrom's scaling found in a context apparently completely different than that of CDM, viz, the approach of MOND.
We assume the dark matter is spherically distributed about the ordinary matter of mass $M$. Galactic disks would presumably form in a halo of MDM similar to the way they form in simulations using CDM. Investigation of the actual shape of MDM halos is a topic for future works. In the following, the radial dependence of the baryonic mass is implicit unless otherwise stated.
In the next section, we will review the CDM mass profiles and then compare the CDM results to the MDM mass profile on the galactic scales.

A dark matter model that naturally reproduces MOND can be constructed in the following way. We take the equation of motion of MOND
without dark matter and associate the interpolating function $\mu(a/a_c)$ with the force term:
\begin{equation}
\dfrac{1}{\mu(a/a_c)}\,F\;=\;
\dfrac{1}{\mu(a/a_c)}\left(G\dfrac{mM}{r^2}\right)\;=\; ma\;.
\label{milgrom}
\end{equation}
The usual equation of motion with ordinary matter $M$ and dark matter $M'(r)$,
enclosed within a sphere of radius $r$ around $M$, is
\begin{equation}
G\dfrac{m(M+M'(r))}{r^2} \;=\; ma\;,
\end{equation}
so both equations will lead to the same prediction for $a(r)=v^2(r)/r$ if we have an
integrated dark matter profile such that
\begin{equation}
\dfrac{M'(r)}{M} \;=\; \dfrac{1}{\mu(a(r)/a_c)}-1\;,
\end{equation}
from which we can infer the dark matter density profile assuming a spherically symmetric distribution:
\begin{equation}
\rho'(r) \;=\; \dfrac{1}{4\pi r^2}\dfrac{d}{dr}M'(r)\;.
\end{equation}
The required dark matter profile $\rho'(r)$ can thus be reverse engineered from the required
acceleration profile $a(r)$.

Thus, we can interpret Equation~(\ref{milgrom}) as a modification of inertia or as Newtonian gravity with an additional source term, the dark matter mass $M'$. This duality was noted in \citet{HMN} for MDM, and we show here it extends to (non-relativistic) MOND.
Whether such a dark matter profile can be realized dynamically will depend on the detailed
nature of the MDM quanta and their interactions with ordinary matter.
We have presented several ideas on how this may be arranged in previous publications \citep{HMN, HMN2, HMN3},
but such an endeavor would be pointless if the predicted dark matter profile did not match
observations.
Thus, in this paper we fit our MDM model to actual galactic data,
in the hopes that the model's predictions can be checked against data from experiments such as Fermi-LAT \citep{fermi}. If Fermi-LAT is able to map out the dark matter profile from gamma ray observations, we will be able to compare the observed profile to that predicted by the MDM model.\footnote{We note that the signals from Fermi-LAT could also have a regular astrophysical origin such as pulsars \citep{Profumo}.}

\begin{figure}[t]
\includegraphics[height=5.5cm]{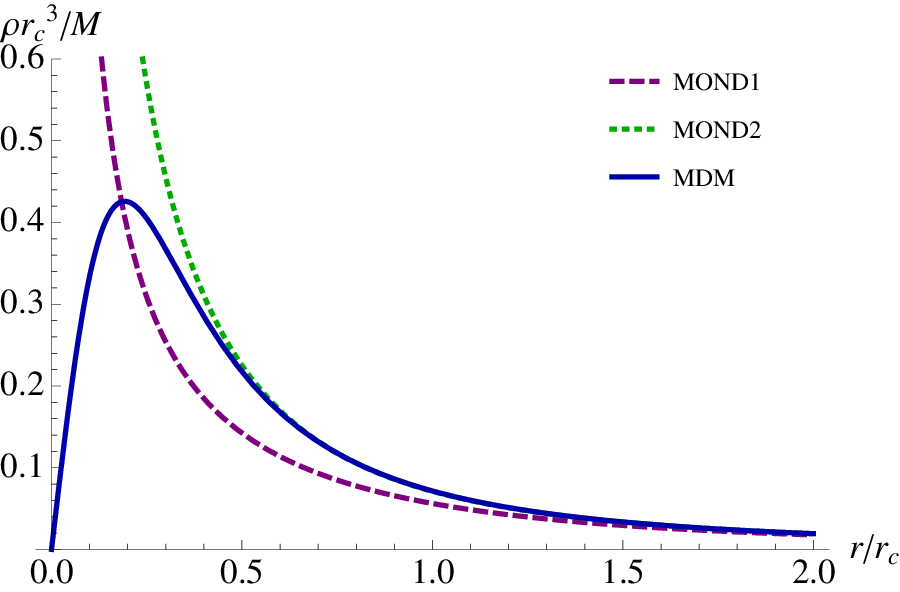}
\includegraphics[height=5.5cm]{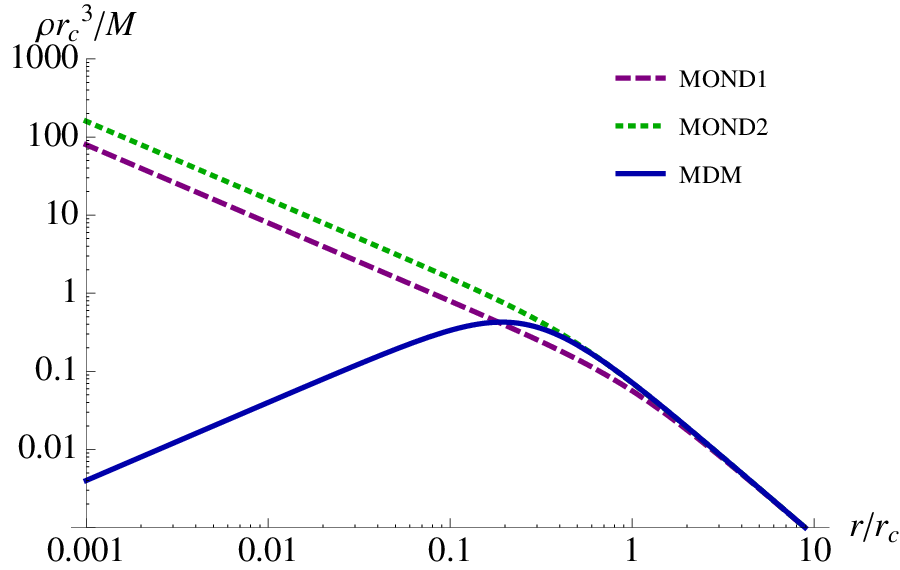}
\caption{Dark matter density distributions around a point baryonic mass $M$ at the origin
which reproduce MOND shown in linear (left) and log-log (right) plots.
Blue (solid) line: the MONDian Dark Matter distribution,
Purple (dashed) line: Equation~(\ref{rho1}),
Green (dotted) line: Equation~(\ref{rho2}).
Note the different behaviors in the region $r/r_c \ll 1$.
}
\label{MONDvsMDM}
\end{figure}

The form of the dark matter profile $\rho'(r)$ that reproduces MOND will of course depend on the
choice of interpolating function $\mu(a/a_c)$.
For instance, if we choose
\begin{equation}
\mu(x)\;=\; \dfrac{1}{x}\left(\sqrt{\dfrac{1}{4}+x^2}-\dfrac{1}{2}\right)\;,
\label{MOND1}
\end{equation}
for a point like mass $M$ at the origin
MOND will predict
\begin{equation}
v^2(r) \;=\;
r\,a(r)
\;=\; r\,a_N\sqrt{1+\dfrac{a_c}{a_N}}
\;=\; v_\infty^2 \sqrt{1+\left(\dfrac{r_c}{r}\right)^2}
\;,
\end{equation}
%
where $v^2_\infty=\sqrt{a_c GM}$, $r_c=\sqrt{GM/a_c}$,
and the corresponding dual dark matter profile will be
\begin{equation}
\dfrac{M'(r)}{M}
\;=\; \dfrac{a(r)}{a_N}-1
\;=\; \sqrt{1+\dfrac{a_c}{a_N}}-1
\;=\; \sqrt{1+\left(\dfrac{r}{r_c}\right)^2}-1\;,
\end{equation}
or equivalently
\begin{equation}
\rho'(r) \;=\; \dfrac{M}{4\pi r_c^3}\;\dfrac{1}{(r/r_c)\sqrt{1+(r/r_c)^2}} \;.
\label{rho1}
\end{equation}
%
%
%
Alternatively, if we choose the so-called ``simple'' interpolating function \citep{fmrev}
\begin{equation}
\mu(x) \;=\; \dfrac{x}{1+x}\;,
\label{standardMU}
\end{equation}
MOND predicts
\begin{equation}
v^2(r)
\;=\; r\,a(r)
\;=\; r\,a_N\;\dfrac{1+\sqrt{1+4a_c/a_N}}{2}
\;=\; v^2_\infty \;\dfrac{1+\sqrt{1+4\,(r/r_c)^2}}{2\,(r/r_c)}
\;,
\end{equation}
and the corresponding dark matter profile will be given by
\begin{equation}
\dfrac{M'(r)}{M}
\;=\; \dfrac{a_c}{a(r)}
\;=\; \dfrac{2\,(r/r_c)^2}{1+\sqrt{1+4\,(r/r_c)^2}}
\;,
\end{equation}
or equivalently
\begin{equation}
\rho'(r) \;=\; \dfrac{M}{4\pi r_c^3}\;\dfrac{2}{(r/r_c)\sqrt{1+4(r/r_c)^2}} \;.
\label{rho2}
\end{equation}
Note that both interpolating functions yield
$\rho'(r) \approx (M/4 \pi r_c)/r^2$ for $r \gg r_c$, and
$\rho'(r) \sim (M/r_c^2)/r$ for $r\ll r_c$ (See Figure~\ref{MONDvsMDM}).

The MDM mass profile can be written as
\begin{equation}
\dfrac{M'(r)}{M}
\;=\; f(r)
\;,
\end{equation}
with $f(r)$ given as the solution to
\begin{equation}
\sqrt{\dfrac{1}{\pi f} + 1}-1 \;=\; \dfrac{1}{A}
\left( 1 + f \right)\;,
\label{feq}
\end{equation}
with
\begin{equation}
A
\;=\; \dfrac{2\pi a_c}{a_N}
\;=\; \dfrac{2\pi a_c}{GM}\,r^2
\;=\; \dfrac{2\pi r^2}{r_c^2}
\;.
\end{equation}
The motivation for this equation is explained in the appendix.
Equation~\eqref{feq} is a cubic equation in $f$, and thus has three solutions for each value
of $A\ge 0$.
Two of these solutions are always negative and converge to $-1$ in the limit $A\rightarrow 0$.
The third solution is always positive and vanishes as $A^2/\pi$ as $A\rightarrow 0$.
This third solution is the one we need.
Thus for $r \ll r_c$, the MDM model yields
$\rho'(r) \sim (M/r_c^4)r$, quite different from what the
above two interpolating function choices give.
For $A \gg 1$ (i.e.,
$r \gg r_c$), however, the cubic equation in $f$ becomes a quadratic equation,
yielding
$f \rightarrow \sqrt{A/2\pi} = r/r_c$ and hence $M' \rightarrow M r/r_c $
(Ho, Minic \& Ng 2011), i.e.,
$\rho'(r)\approx (M/4 \pi r_c)/r^2$, the same
as what the above two interpolating functions gave.
This difference in behavior for the three different cases is shown in Figure~\ref{MONDvsMDM}
corresponding to a point baryonic mass $M$ at the origin.

As an example, we consider the Milky Way. We estimate the mass interior to the solar orbit as $10^{11}~M_\odot$ \citep[e.g.,][and references therein]{Sommer06}. The dark matter density profiles diverge around $r/r_c = 0.1$. With $r_c = \sqrt{GM/a_c}$, these values imply that differences in dark matter distributions between MDM and Equations~(\ref{rho1}) and (\ref{rho2}) will occur at radii $r < 1$~kpc for a Milky Way sized disk.

The MDM prediction for the rotation curves is then
\begin{equation}
\dfrac{v^2(r)}{r}
\;=\; \dfrac{G(M+M'(r))}{r^2}
\;=\; \dfrac{GM}{r^2}\left(1+\dfrac{M'(r)}{M}\right)
\;=\; a_N \left(\,1+f(r)\,\right)\;,
\label{MONDianV2}
\end{equation}
where $a_N=GM/r^2$ is the Newtonian acceleration due to ordinary matter only.
%
%
The baryonic mass $M$ consists of contributions from
the stellar disk and the interstellar gas:
\begin{equation}
M \;=\; M_\mathrm{disk} + M_\mathrm{gas}\;.
\end{equation}
The mass of the interstellar gas can be inferred from 21~cm observations of \hi.
The mass of the stellar disk must be inferred indirectly from its luminosity $L_\mathrm{disk}$,
\begin{equation}
M_\mathrm{disk} \;=\; \alpha L_\mathrm{disk}\;,
\end{equation}
which introduces an extra free parameter $\alpha$, the stellar mass-to-light ratio ($M/L$), into the fit.


\section{Data Fitting}
\label{sec:data}

\subsection{Sample Selection}

In order to test MDM with galactic rotation curves, we fit computed rotation curves to the sample of Ursa Major galaxies given in \citet{Sanders98}. Of the 79 members of the Ursa major cluster identified by \citet{Tully96}, 62 are brighter than $M_B \approx -16.5$. These form a complete optically selected sample and have been imaged in the $B$, $R$, $I$, and $K'$ bands \citep{Tully96}. \citet{Sanders98} identify 30 galaxies from this sample that have appropriate LOS inclination angles for kinematic studies while being neither poor in neutral hydrogen nor strongly interacting. These galaxies are listed in Table~\ref{tab:galaxies}. The sample contains both high surface brightness (HSB) and low surface brightness (LSB) galaxies. LSBs are particularly useful for testing MOND theories of gravity since the acceleration is below $a_c$ at all radii (e.g., \citealt{Swaters10}). In Table~\ref{tab:galaxies}, LSB galaxies are denoted with an ``L'' beside the galaxy name. Of the 30 galaxies in our sample, 7 have velocity fields with significant deviation from circular motion \citep{Verheijen97} and are marked with an asterisk.

Using the $K'$--band surface photometry of \citet{Tully96} and the 21~cm line data from \citet{Verheijen97}, \citet{Sanders98} compared rotation curves predicted by MOND to those observed in the sample listed in Table~\ref{tab:galaxies}. They used a tilted rings fitting procedure \citep{Begeman87} to estimate rotation curves from velocity measurements assuming a thin, axisymmetric disk and an adopted distance of 15.5~Mpc for each galaxy in the sample. The dispersion of up to a few Mpc in actual galactic distances introduces some error in the data fitting. However, we expect the errors in velocity to be less than about 15~km~s$^{-1}$ (e.g., see Figure~2 of \citealt{Bottema}), which is within error for most of the data. Assuming $M/L$ is constant with radius for a given galaxy but allowed to vary between galaxies, they find MOND is in general agreement with the observed rotation curves. After removing the 7 galaxies with disturbed velocity fields and one outlier (NGC~3992), they determine a mean $M/L$ ratio in the near-infrared of $0.92 \pm 0.25$.

\subsection{Model Comparisons}
In this section, we discuss fits to galactic rotation curves for our sample using CDM, MOND, and MDM. The data fits are shown in Figures~2--6, and dark matter density profiles for CDM and MDM are shown in Figures~7--11. The physical mechanisms that give rise to the galactic rotation curves for each model differ. In the CDM paradigm, accelerations are determined by General Relativity and the addition of massive particles that interact only gravitationally, dark matter. In (non-relativistic) MOND, inertia is modified for accelerations near or below the critical acceleration ($a_c$), and no non-baryonic particles are introduced. In MDM, the critical acceleration is introduced through the cosmological constant $\Lambda$, and the dark matter quanta give rise to the observed galactic rotation curves. In essence, CDM and MDM introduce dark matter quanta to explain (nearly) flat galactic rotation curves, while MOND modifies inertia. 

\begin{figure}[!ht]
  \includegraphics[angle=90,width=0.45\textwidth]{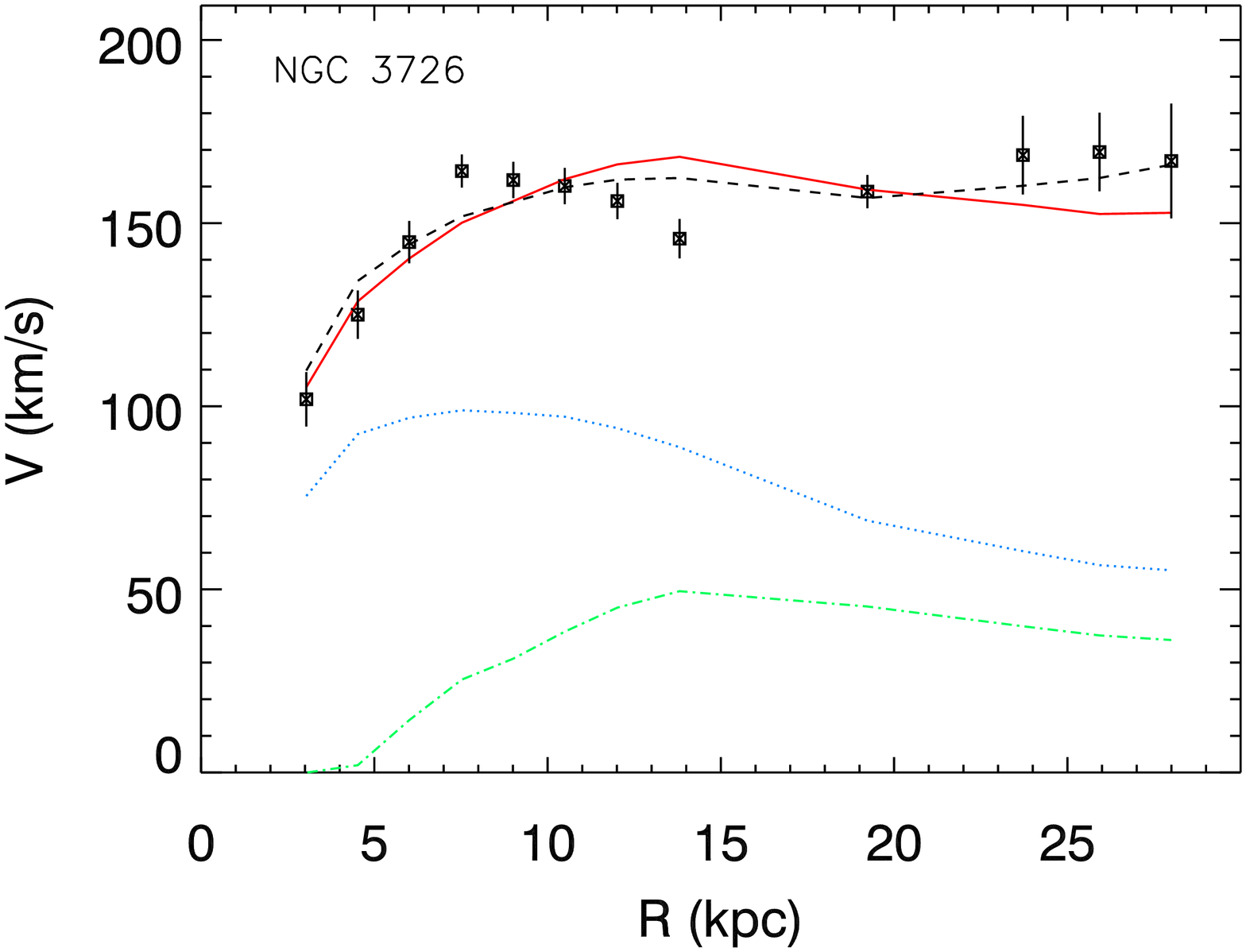}
  \includegraphics[angle=90,width=0.45\textwidth]{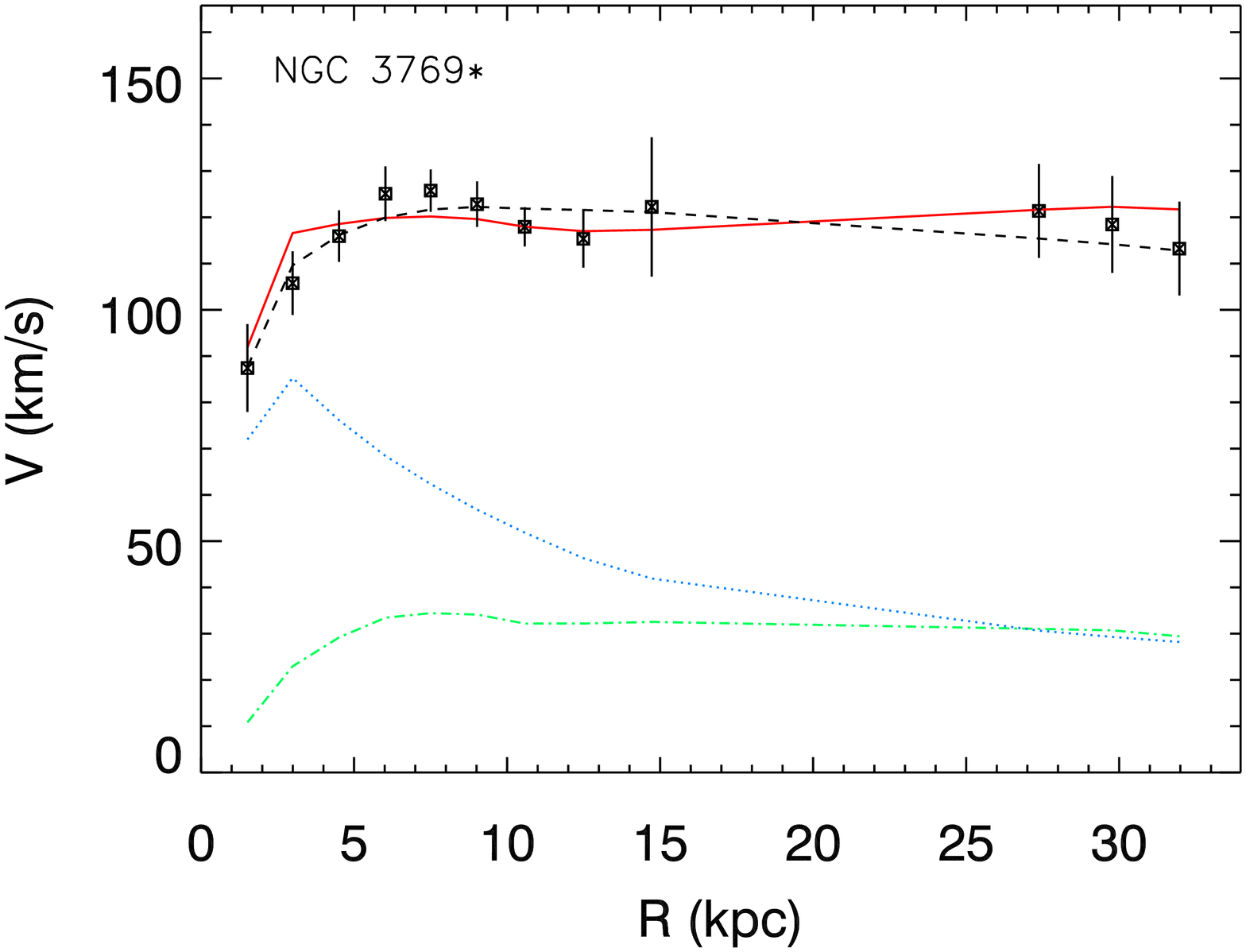}\vspace{3mm}\\
  \includegraphics[angle=90,width=0.45\textwidth]{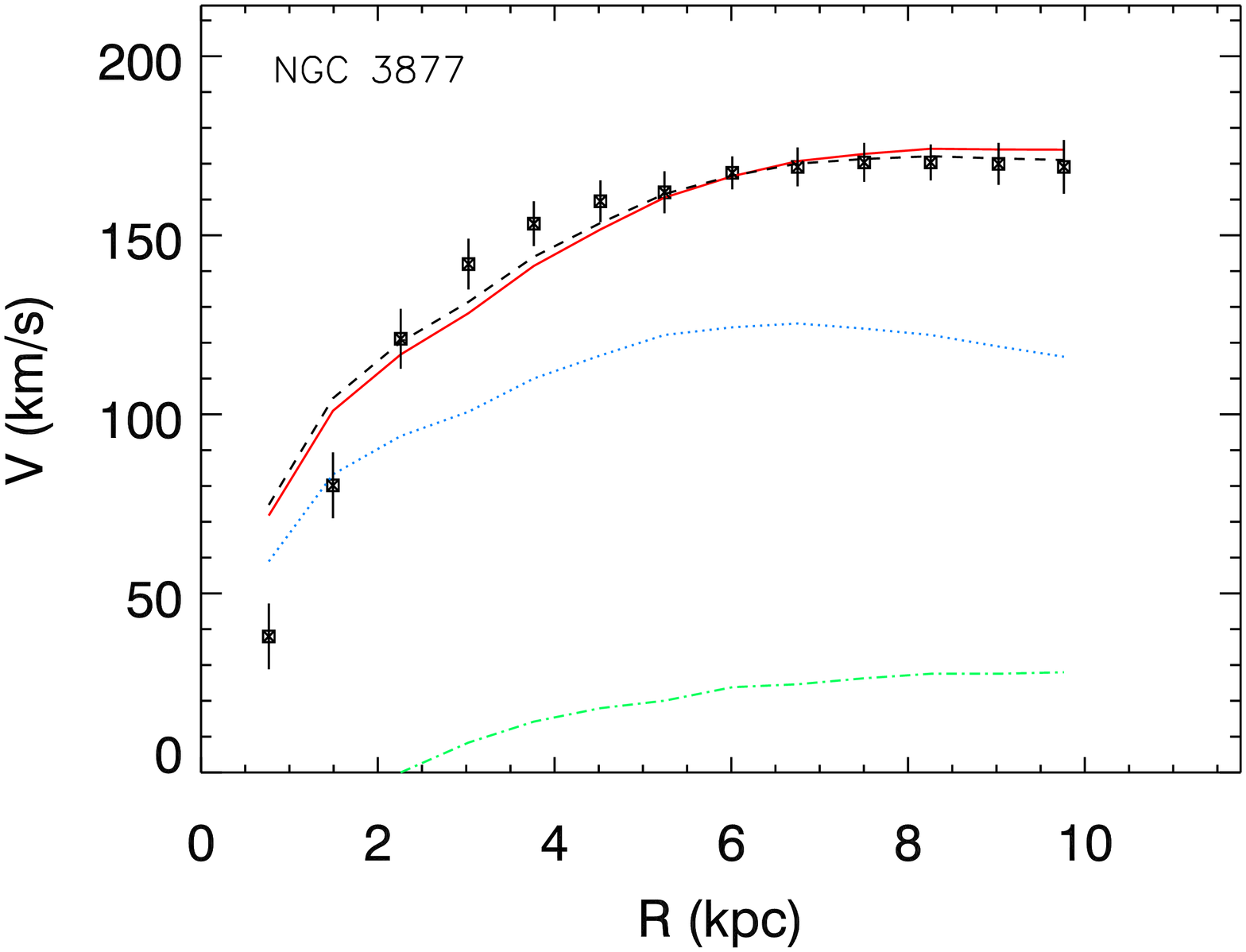}
  \includegraphics[angle=90,width=0.45\textwidth]{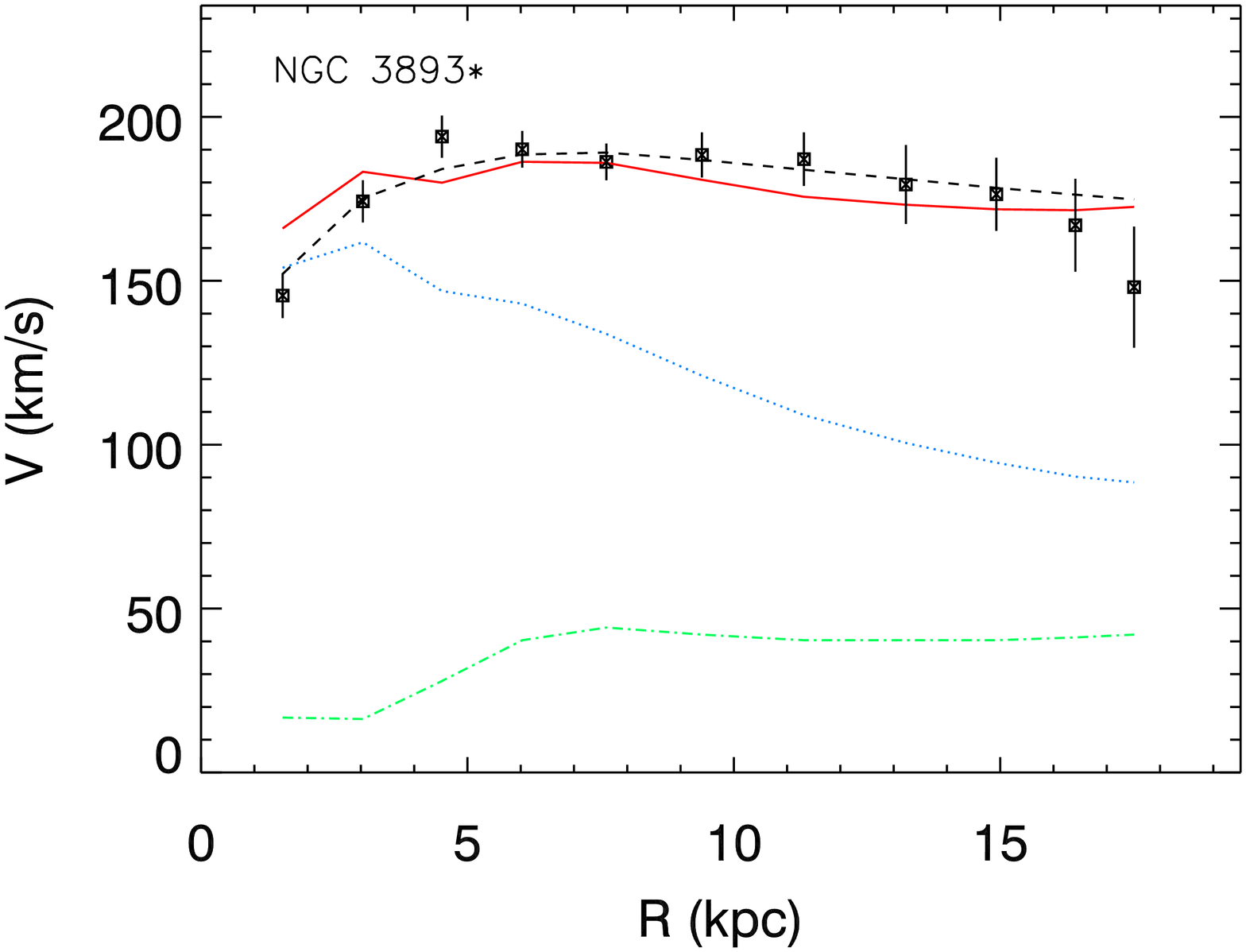}\vspace{3mm}\\
  \includegraphics[angle=90,width=0.45\textwidth]{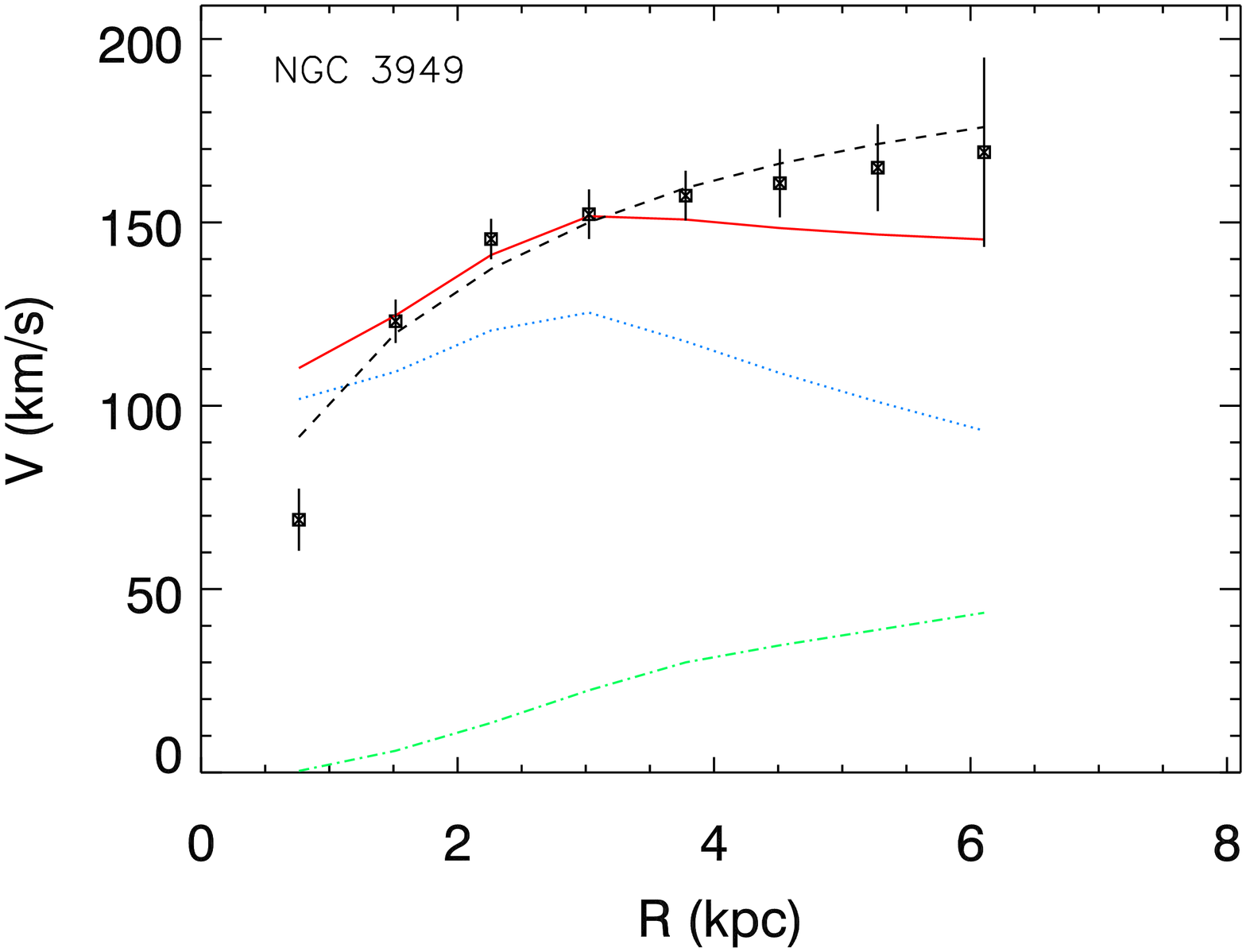}
  \includegraphics[angle=90,width=0.45\textwidth]{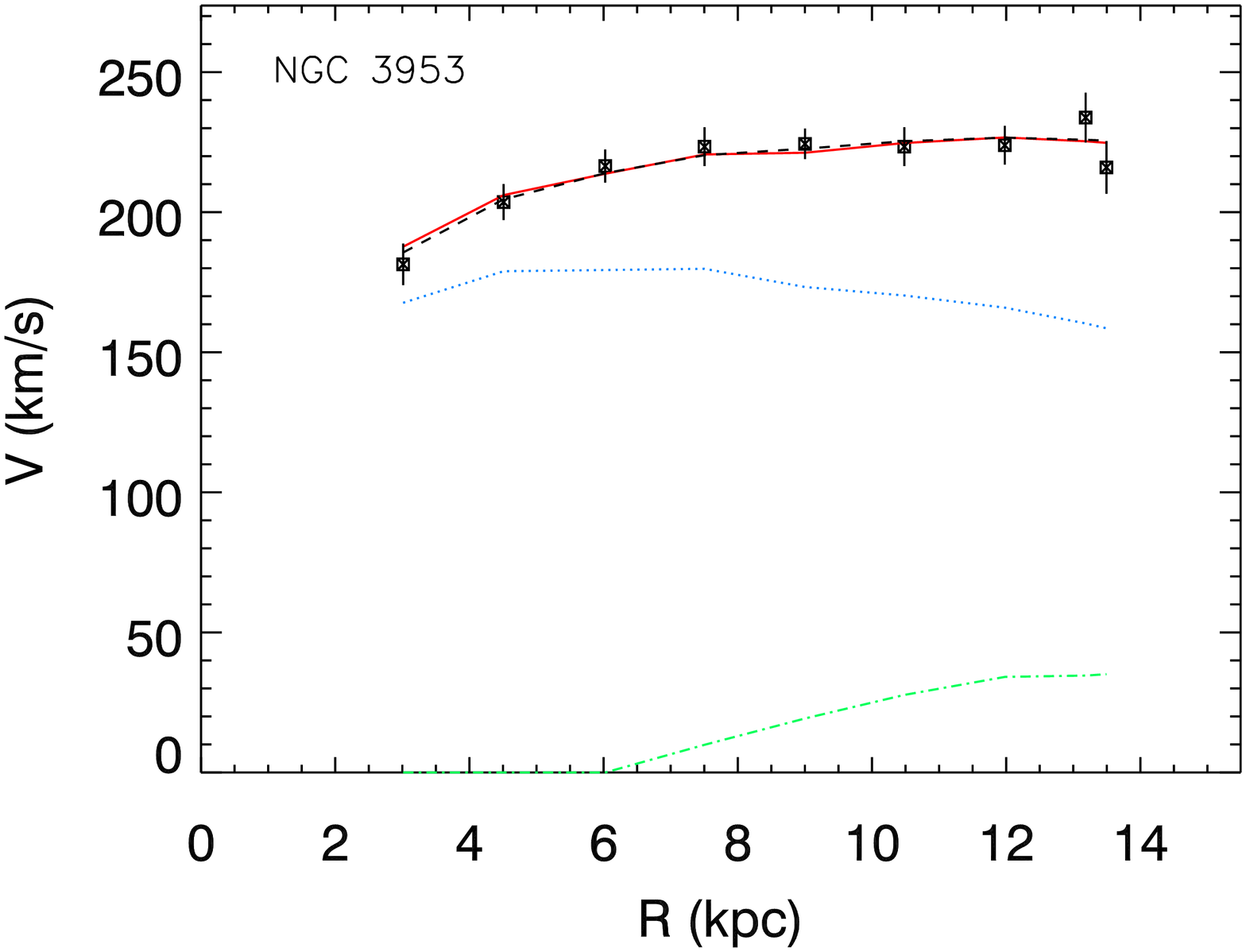}
  \caption{HSB galactic rotation curves. The observed rotation curve is depicted by points with error bars. The solid and dashed lines are the MDM and CDM rotation curves, respectively. MOND fits are nearly identical to the MDM fits and are therefore not shown. Newtonian curves for the stellar and gas components of the baryonic matter are depicted by dotted and dot-dashed lines, respectively. The plotted stellar component is derived from the $M/L$ ratio determined from MDM fits to the rotation curve.}
  \label{fig:hsb}
\end{figure}

The rotation curves predicted by MDM are given by Equation~(\ref{MONDianV2}). We fit these to the observed rotation curves as determined in \citet{Sanders98} using a least-squares fitting routine. As in \citet{Sanders98}, $\alpha = M/L$, which is our only fitting parameter for MDM and MOND, is assumed constant for a given galaxy but allowed to vary between galaxies. Newtonian rotation curves for the stellar disk and the interstellar gas given in \citet{Sanders98} are used to determine the baryonic mass $M(\alpha,r)$. We use the value of the critical acceleration determined in \citet{Begeman91}; $a_c = 1.2 \times 10^{-8}$~cm~s$^{-1}$. Rotation curves predicted by MDM for each galaxy are shown in Figures~\ref{fig:hsb}---\ref{fig:hsblast} and \ref{fig:lsb}---\ref{fig:lsblast} for HSB galaxies and LSB galaxies, respectively. In these figures, observed rotation curves are depicted as filled circles with error bars, and the dotted and dash-dotted lines show the stellar and interstellar gas rotation curves, respectively. The dashed lines are rotation curves predicted by the standard cold dark matter (CDM) paradigm discussed below. The $M/L$ ratios for MDM, MOND, and CDM predicted by our data fits are given in Table~\ref{tab:galaxies}. Uncertainties in fitting parameters for each galaxy are determined by randomizing the data within the errors, obtaining fits for one hundred randomized data sets, and determining the standard deviation for each parameter. For MOND, we use the interpolation function in Equation~(\ref{standardMU}). The rotation curves for MOND are not shown in the figures since they are virtually indistinguishable from the curves predicted by MDM.
We note that different interpolating functions for MOND yield different $M/L$ ratios, but the fits to data are very similar. For comparison, \citet{Sanders98} present rotation curve fits and $M/L$ ratios using the first interpolating function in Equation~(\ref{MONDinterps}) with $n=2$.

\begin{figure}[!ht]
  \includegraphics[angle=90,width=0.45\textwidth]{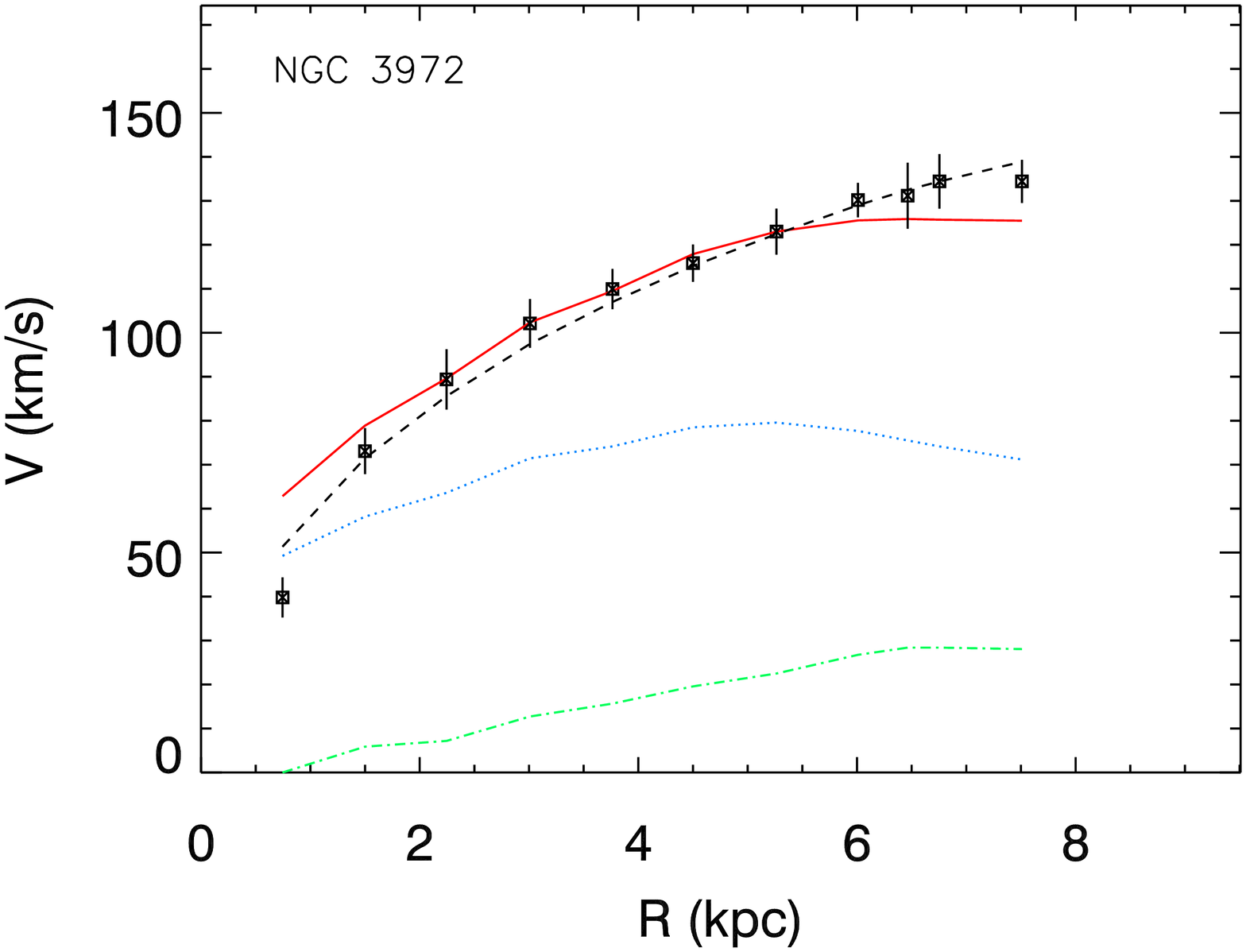}
  \includegraphics[angle=90,width=0.45\textwidth]{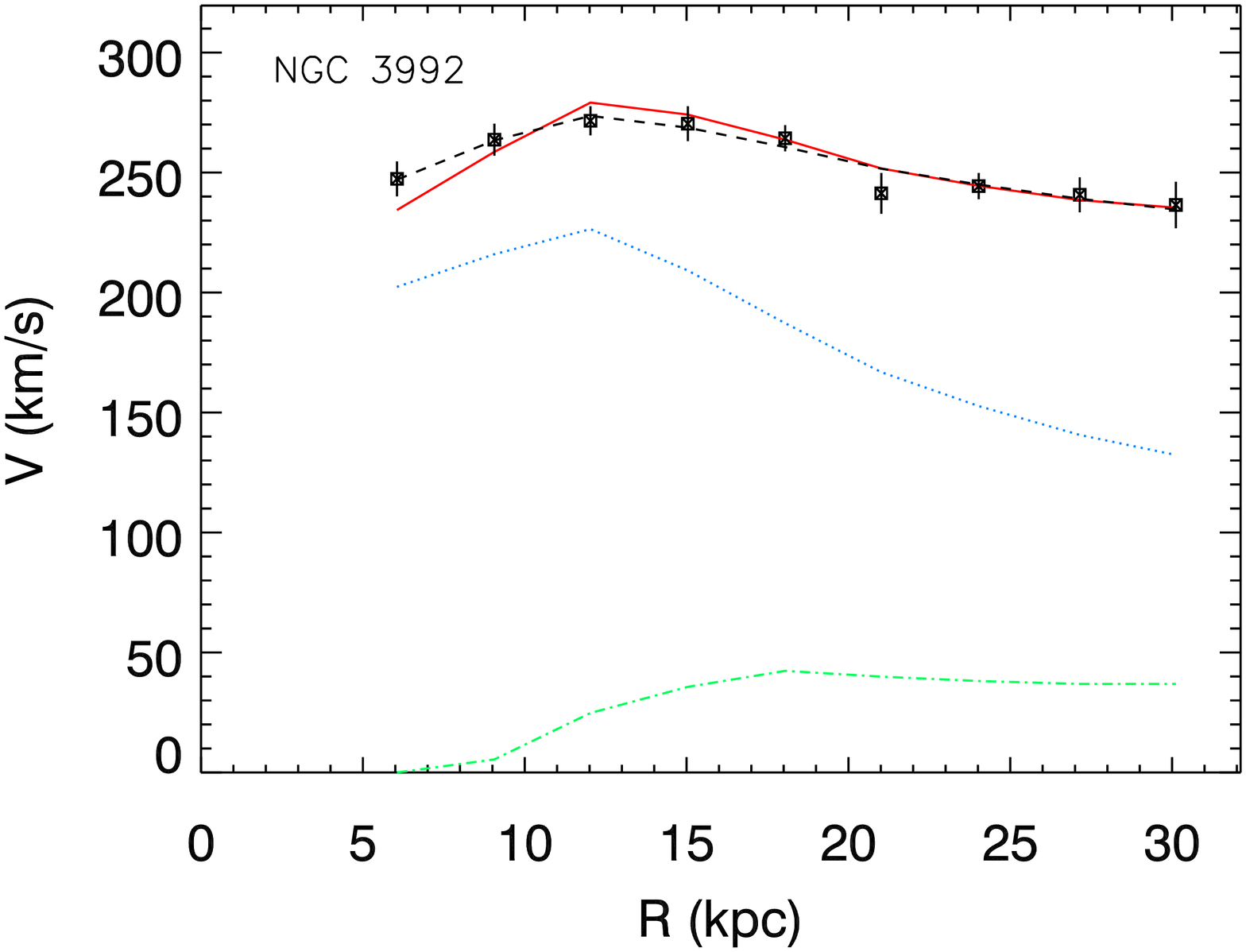}\vspace{3mm}\\
  \includegraphics[angle=90,width=0.45\textwidth]{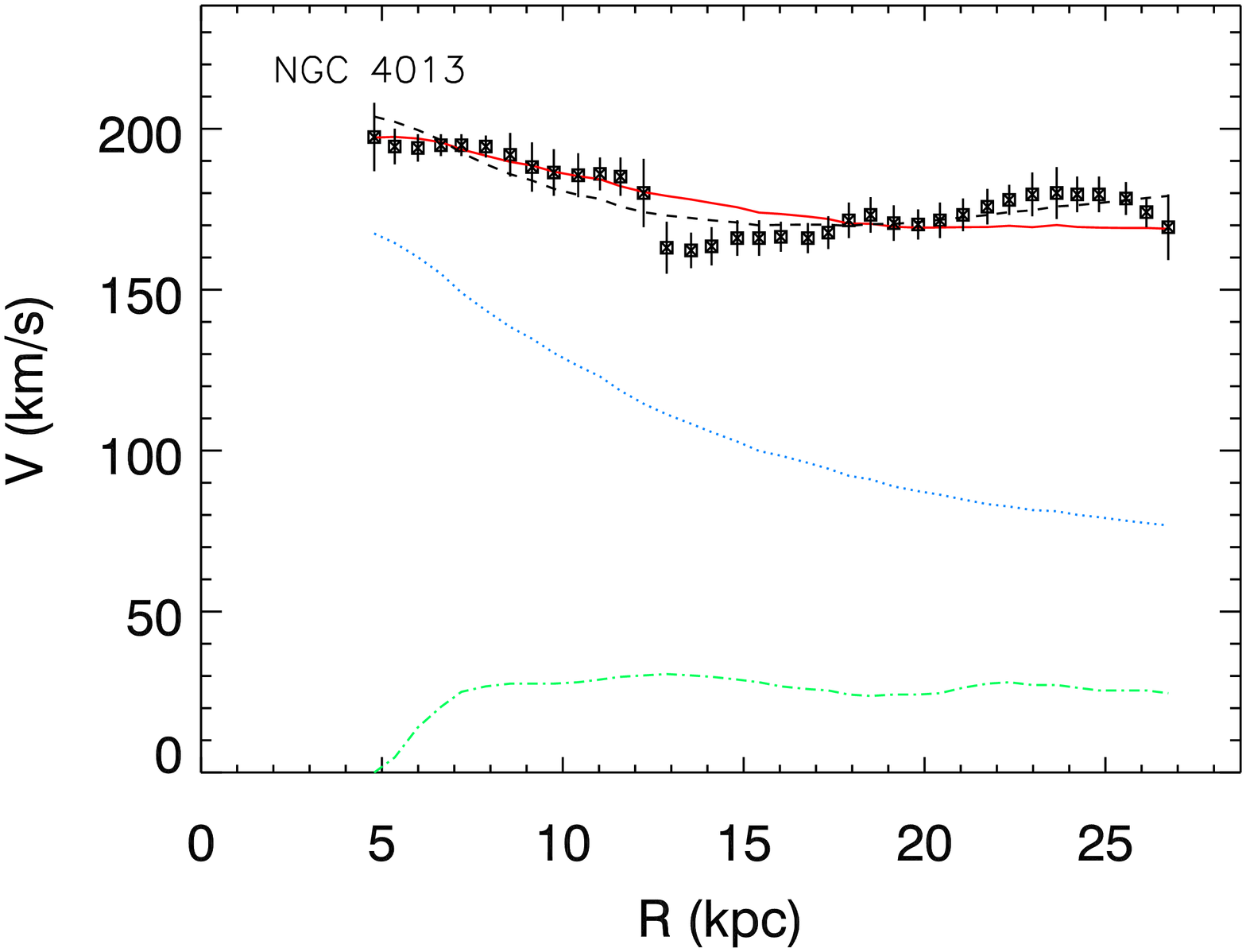}
  \includegraphics[angle=90,width=0.45\textwidth]{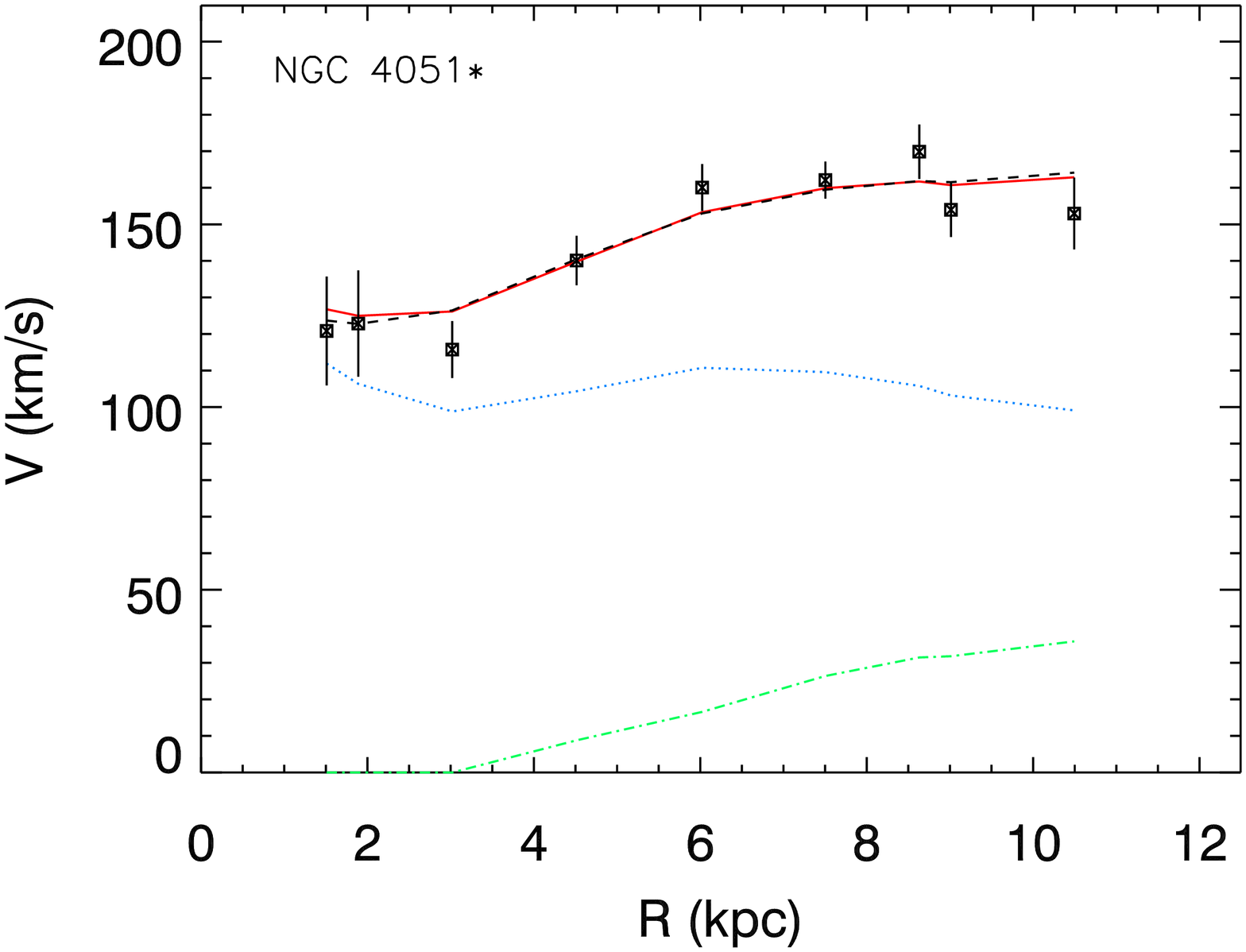}\vspace{3mm}\\
  \includegraphics[angle=90,width=0.45\textwidth]{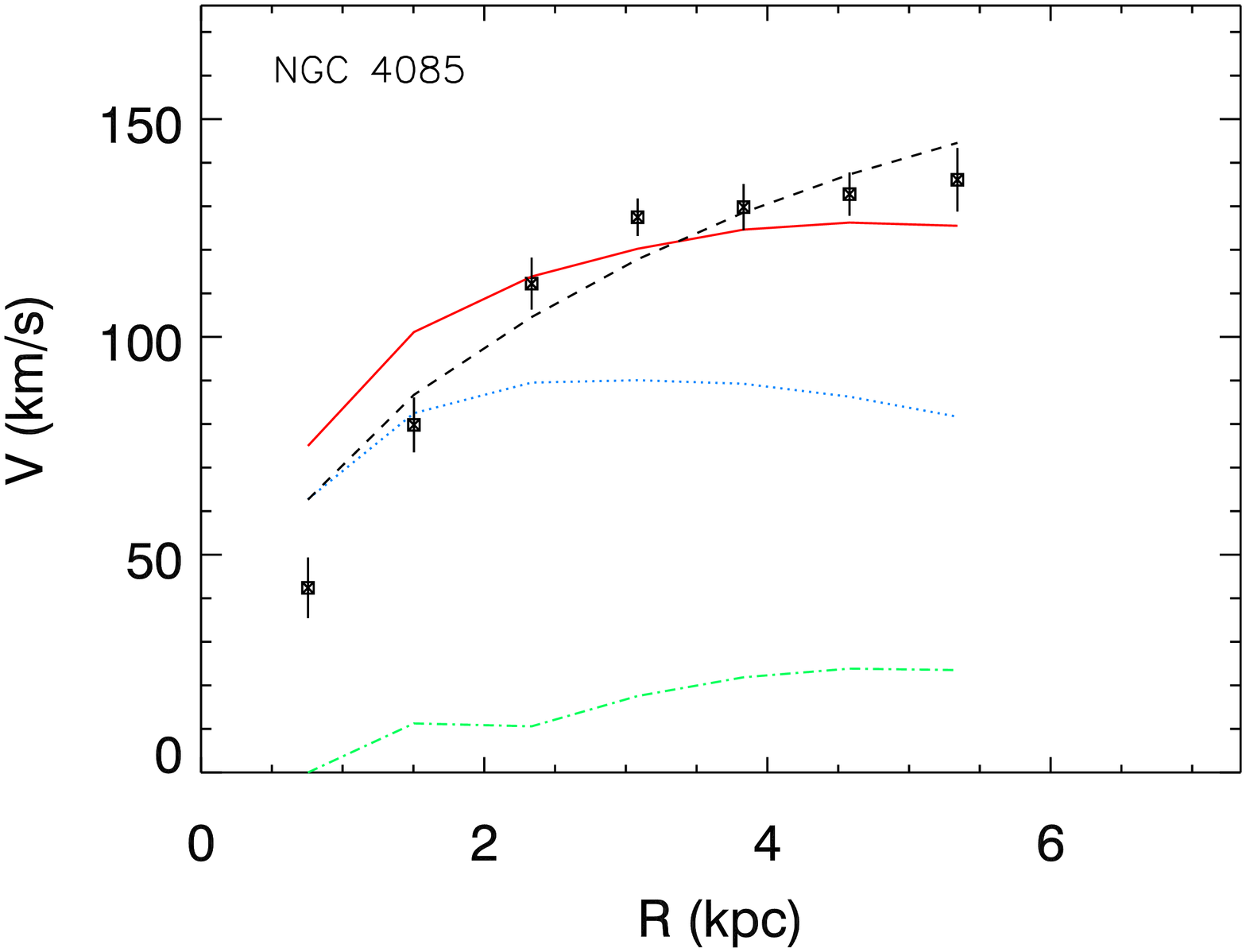}
  \includegraphics[angle=90,width=0.45\textwidth]{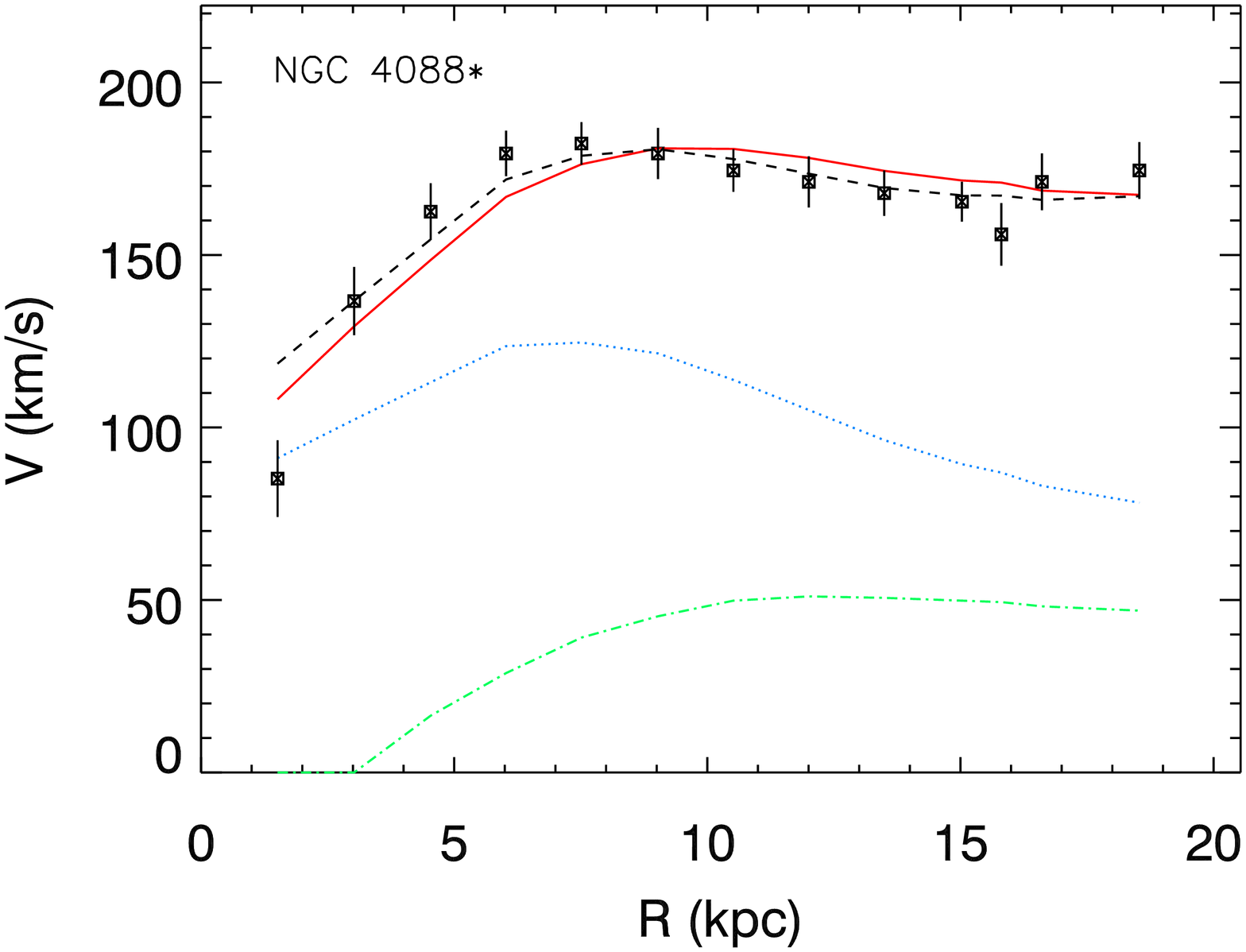}
  \caption{Figure~\ref{fig:hsb} continued.}
\end{figure}

\begin{figure}[!ht]
  \includegraphics[angle=90,width=0.45\textwidth]{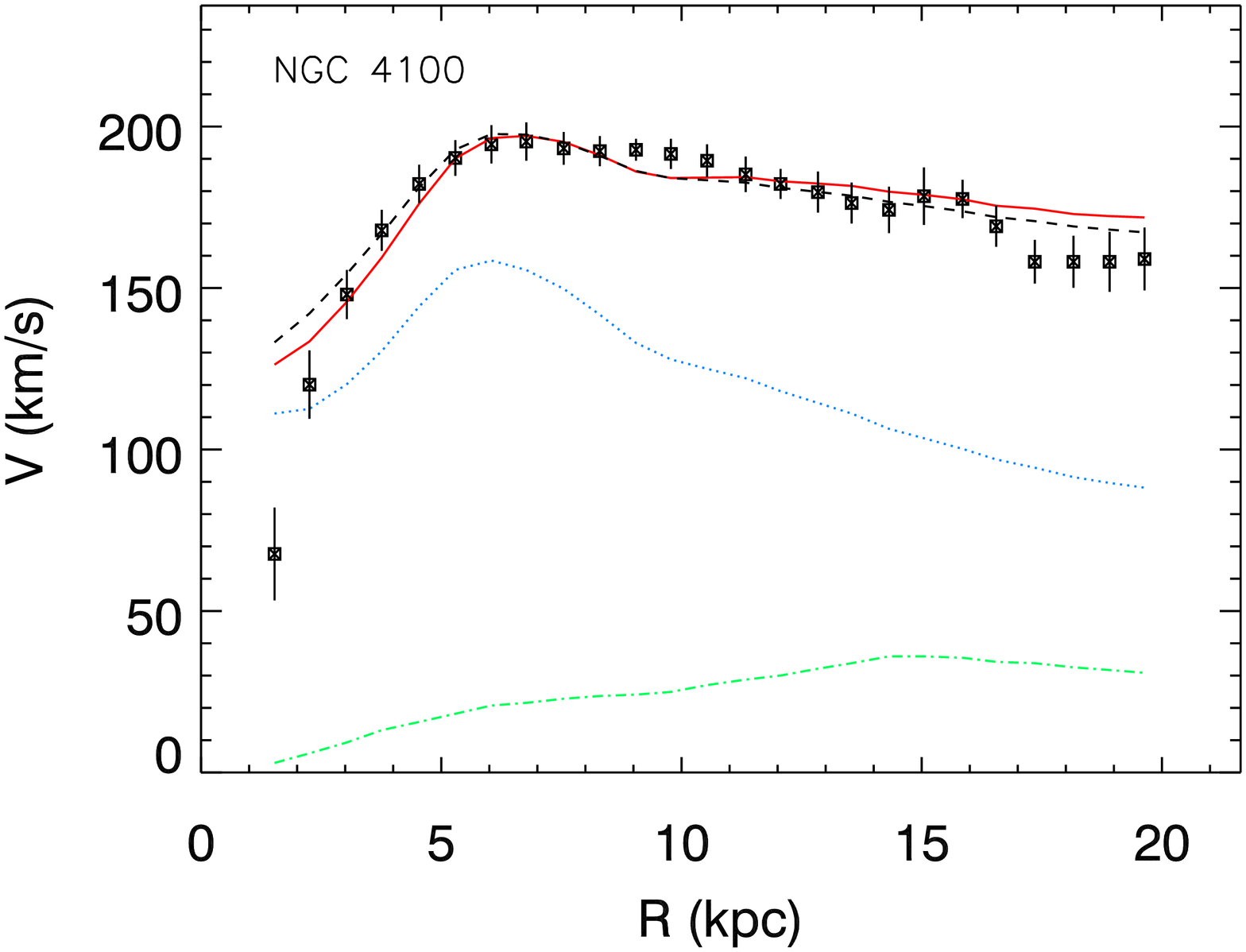}
  \includegraphics[angle=90,width=0.45\textwidth]{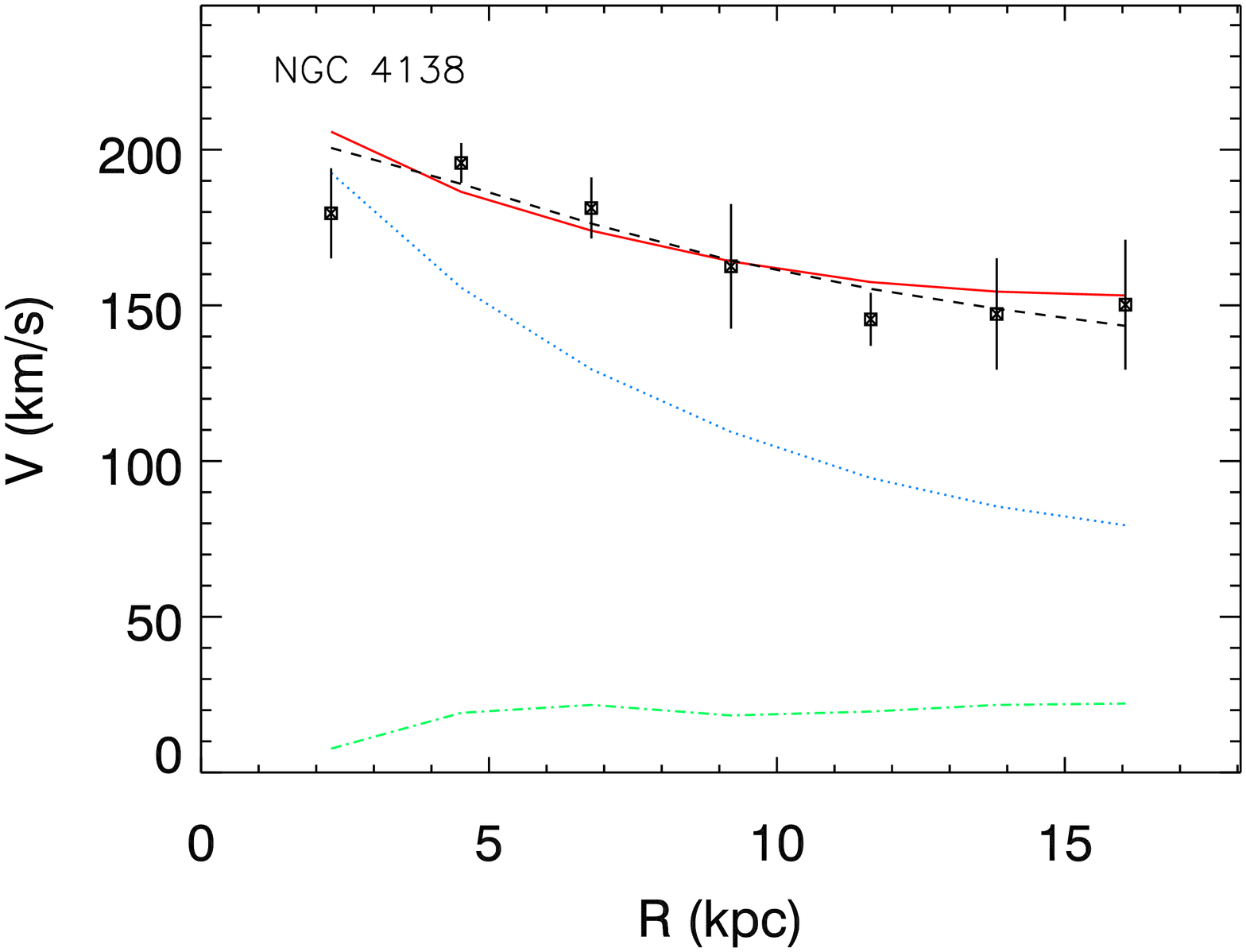}\vspace{3mm}\\
  \includegraphics[angle=90,width=0.45\textwidth]{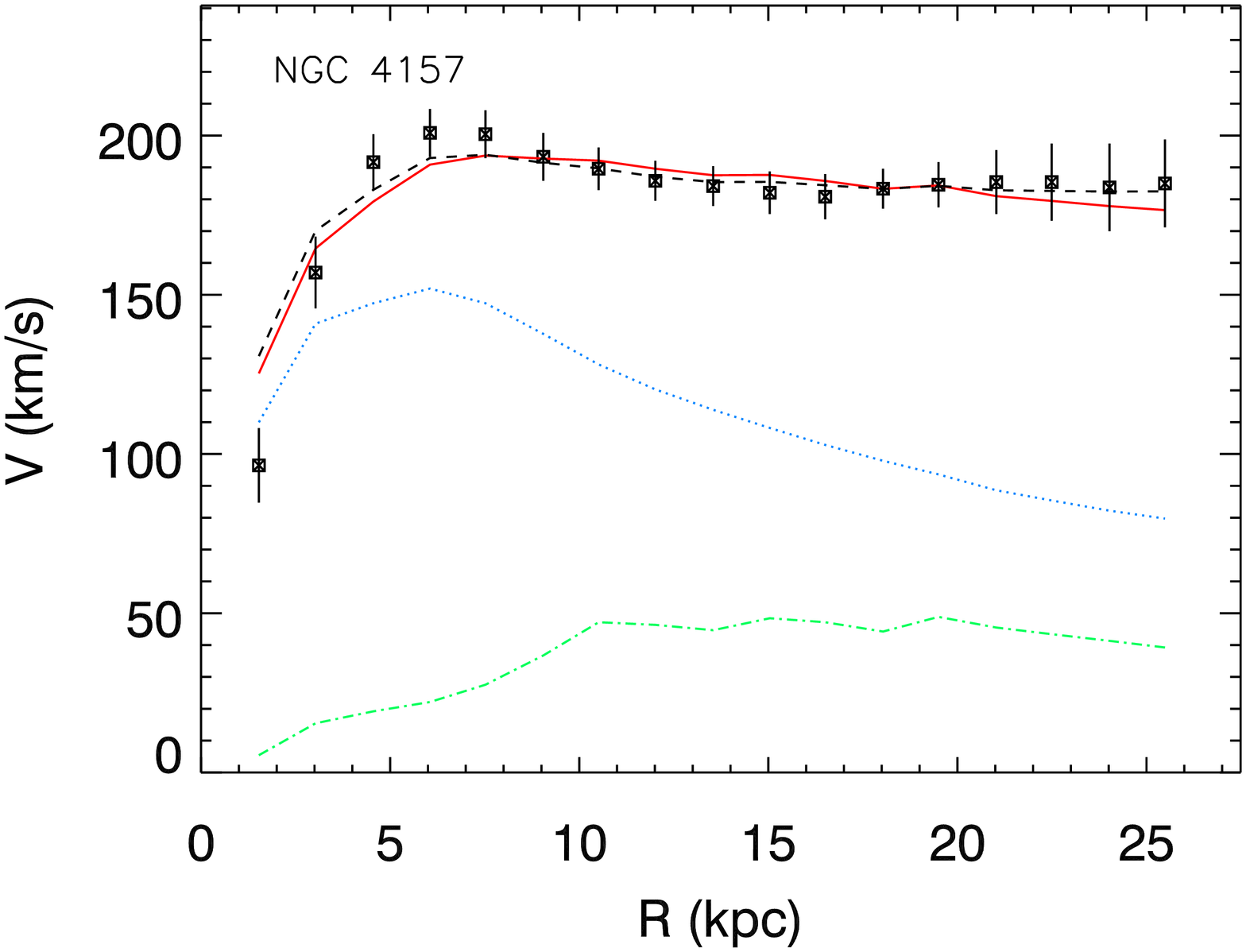}
  \includegraphics[angle=90,width=0.45\textwidth]{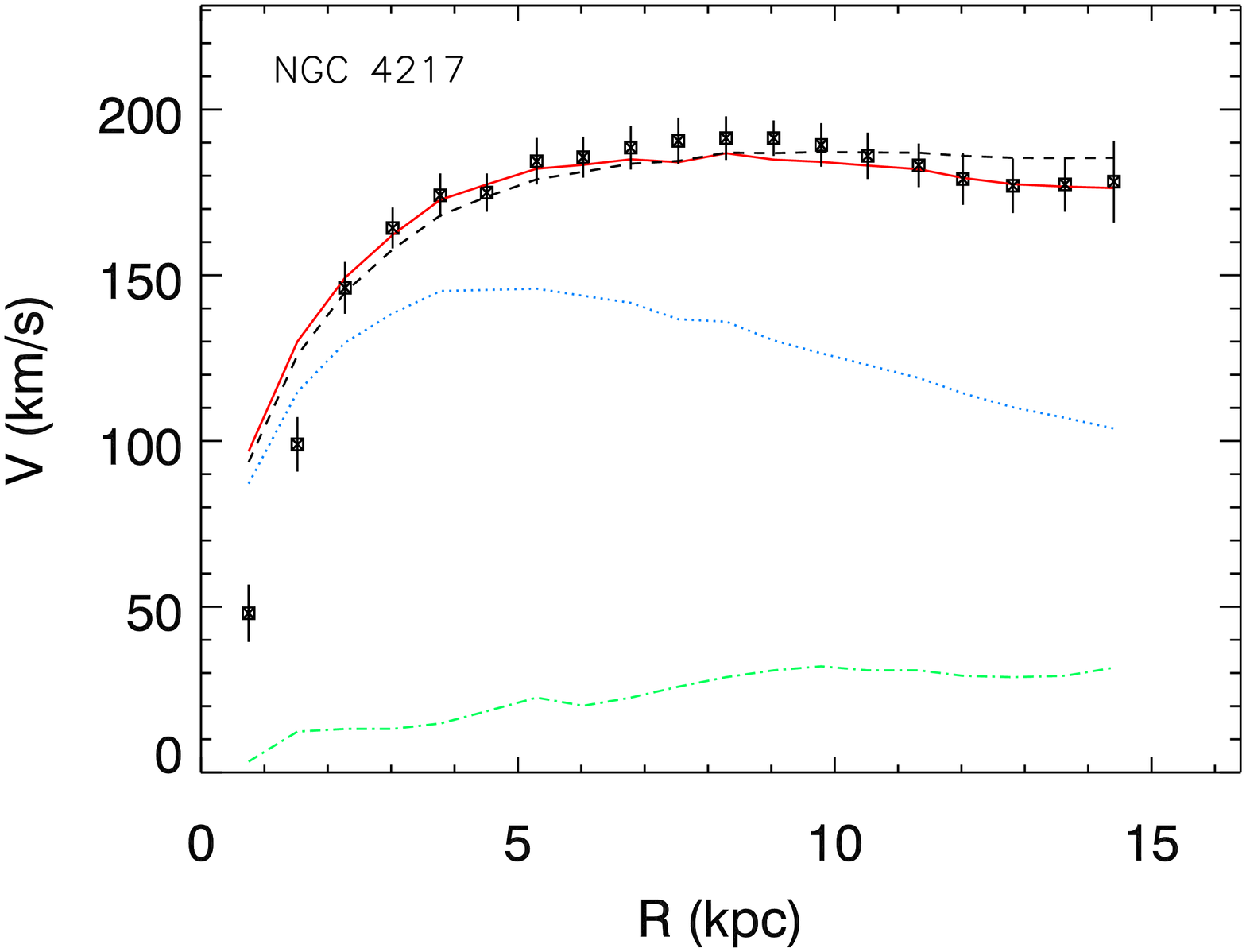}\vspace{3mm}\\
  \includegraphics[angle=90,width=0.45\textwidth]{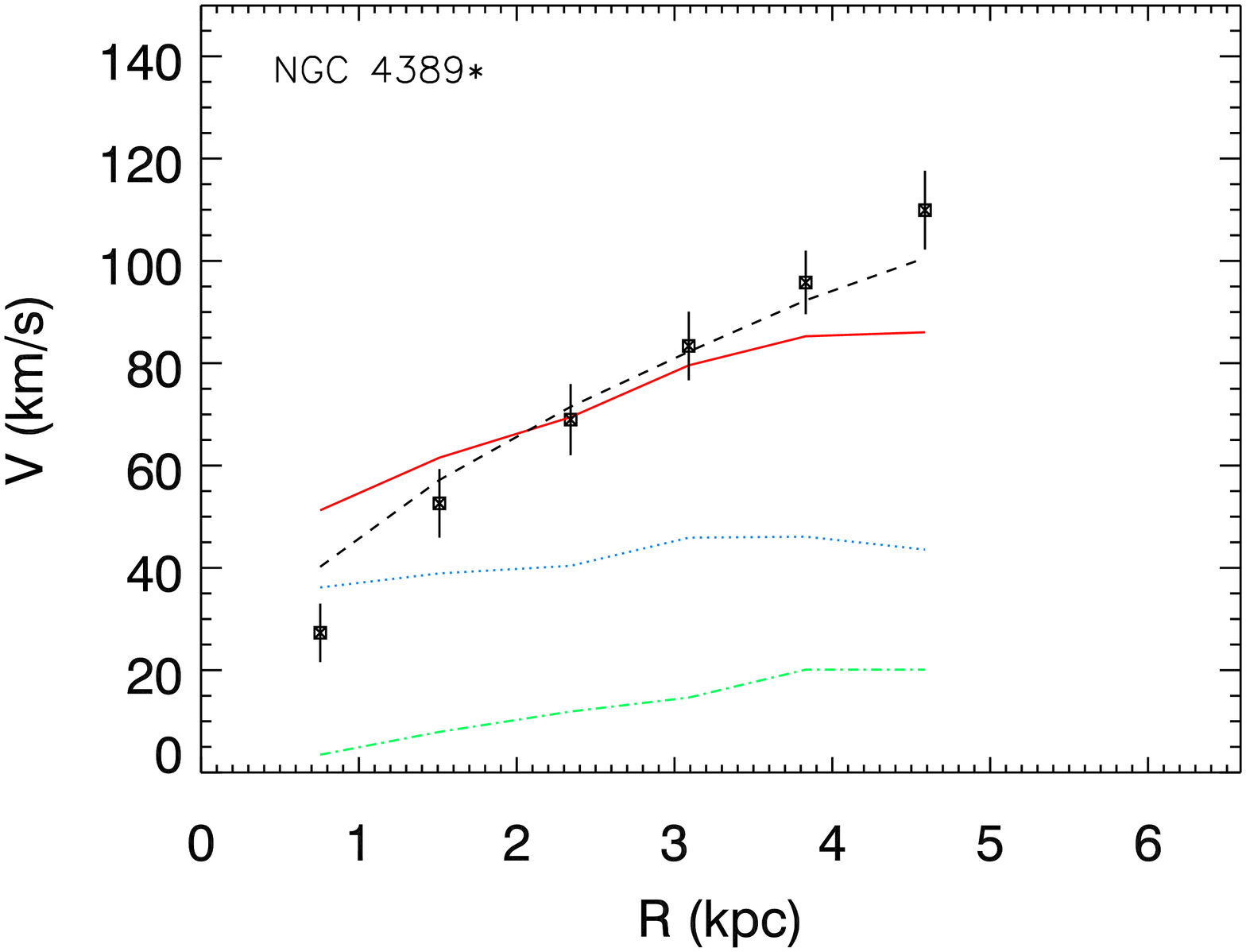}
  \includegraphics[angle=90,width=0.45\textwidth]{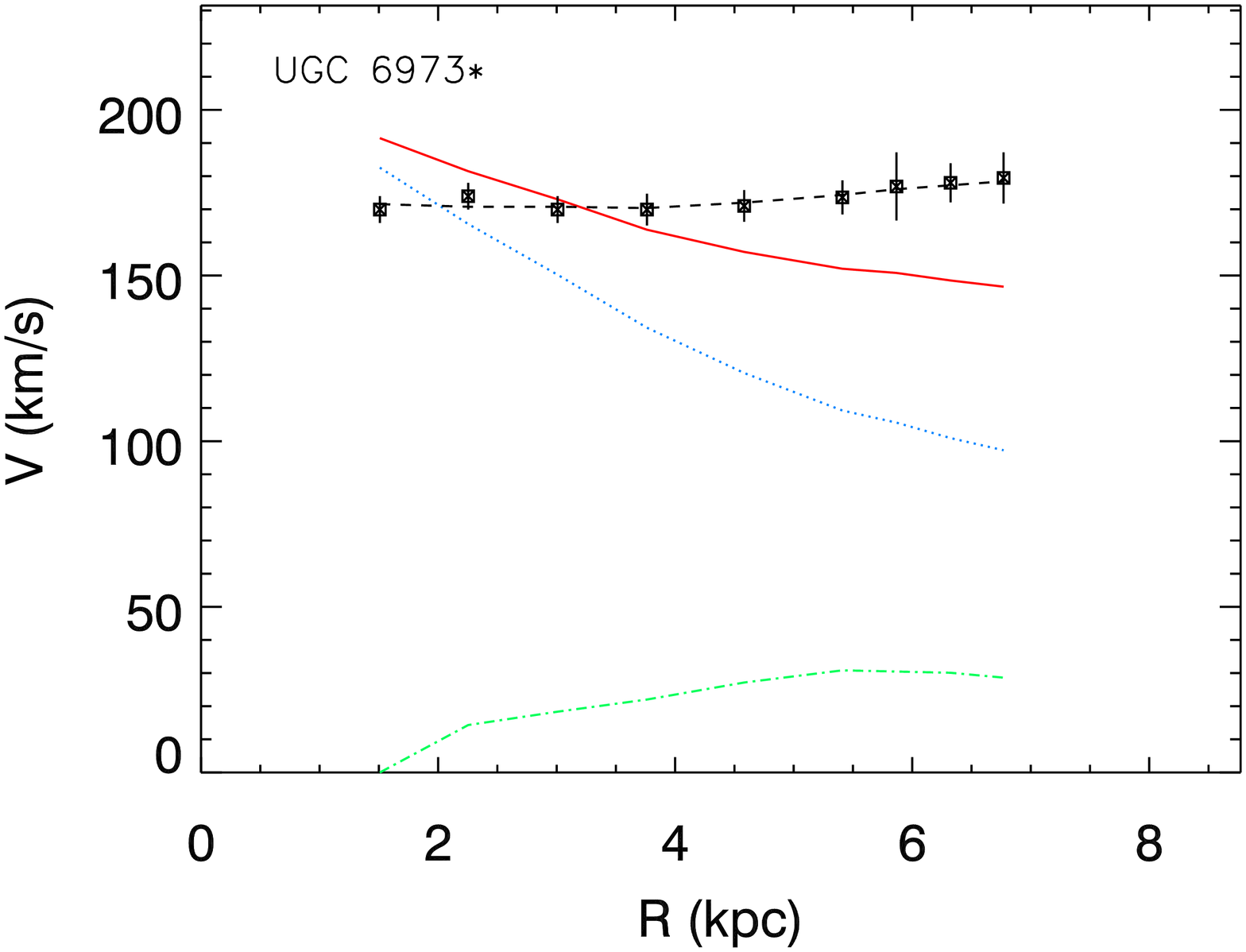}
  \caption{Figure~\ref{fig:hsb} continued.}
\label{fig:hsblast}
\end{figure}

\begin{figure}[!ht]
  \includegraphics[angle=90,width=0.45\textwidth]{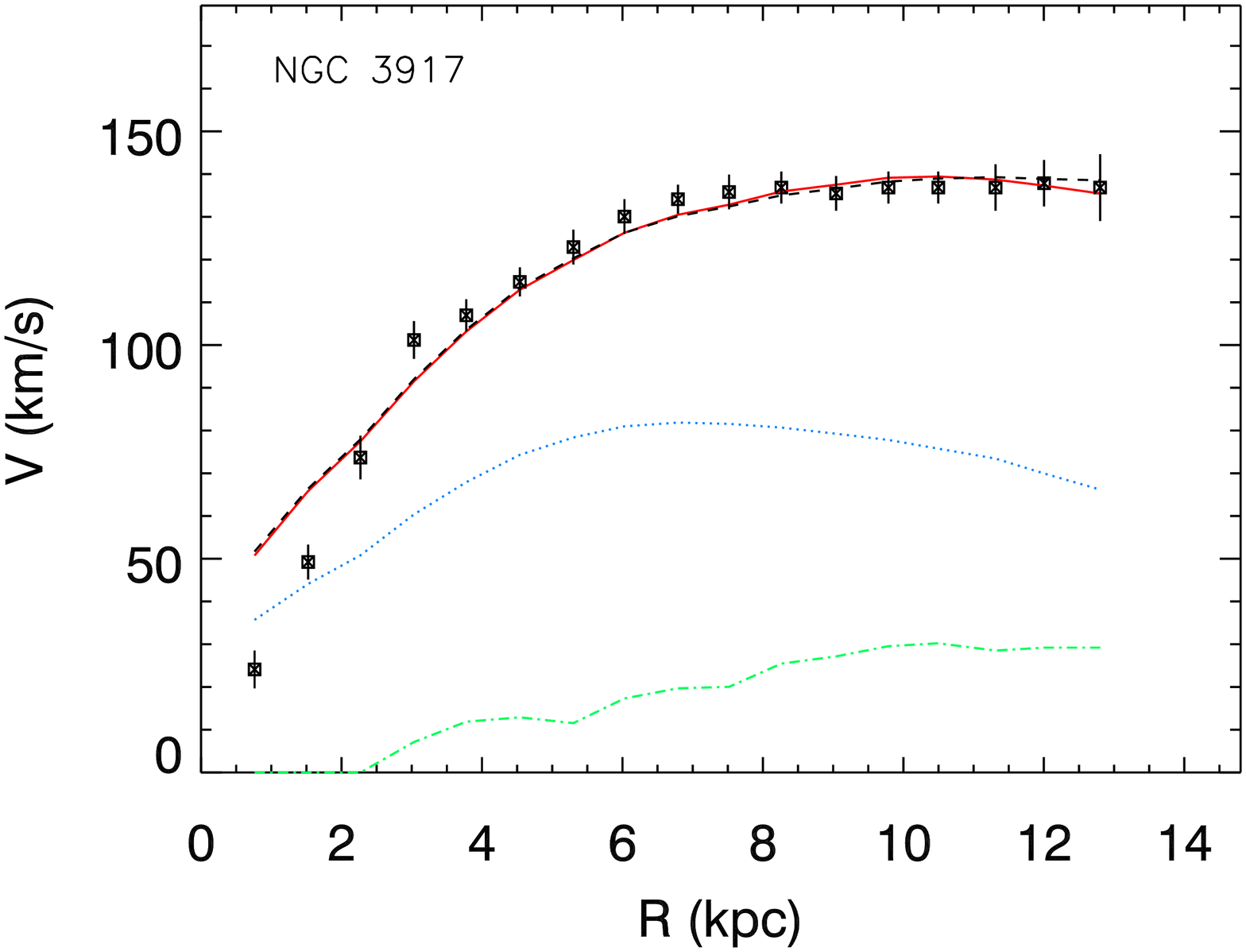}
  \includegraphics[angle=90,width=0.45\textwidth]{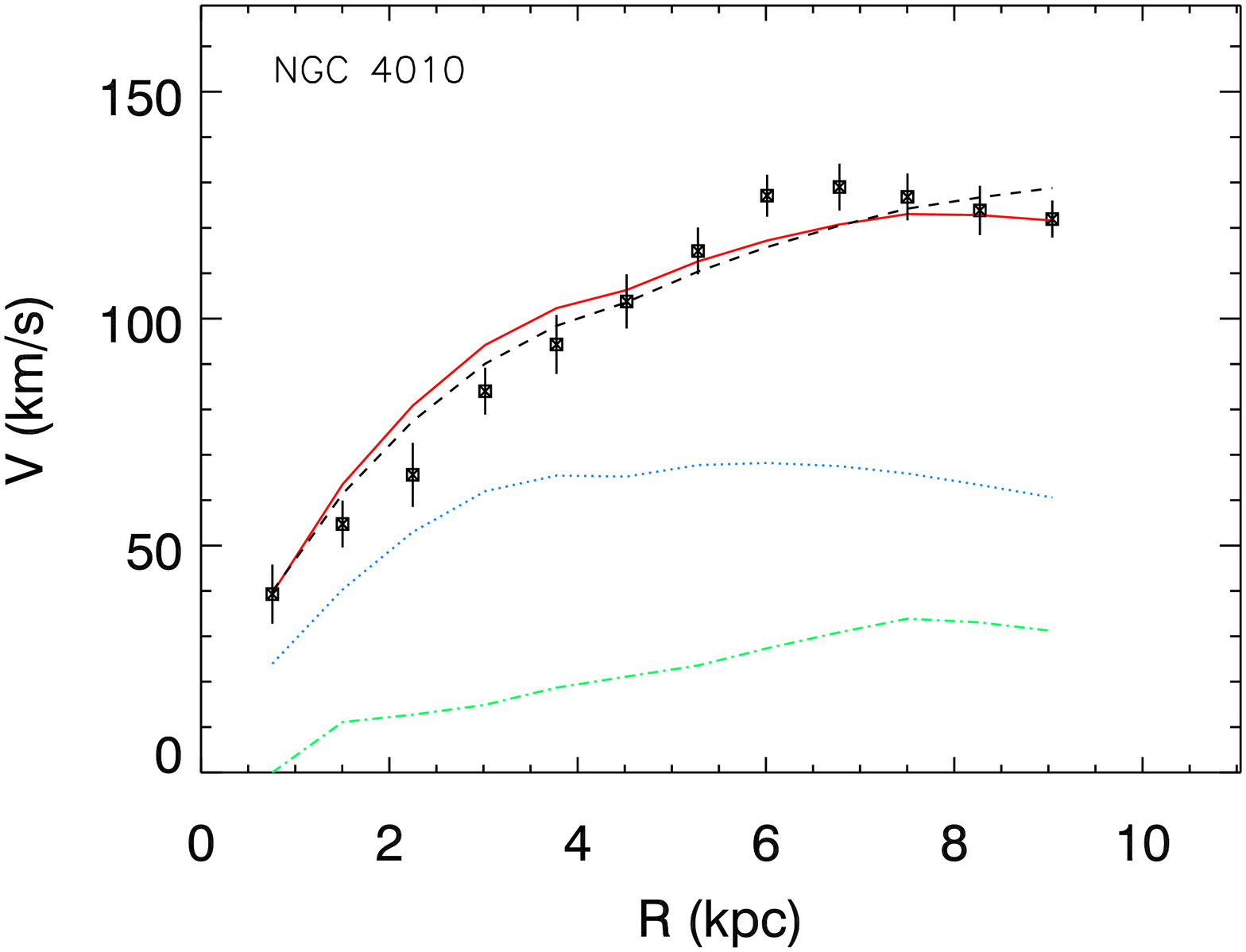}\vspace{3mm}\\
  \includegraphics[angle=90,width=0.45\textwidth]{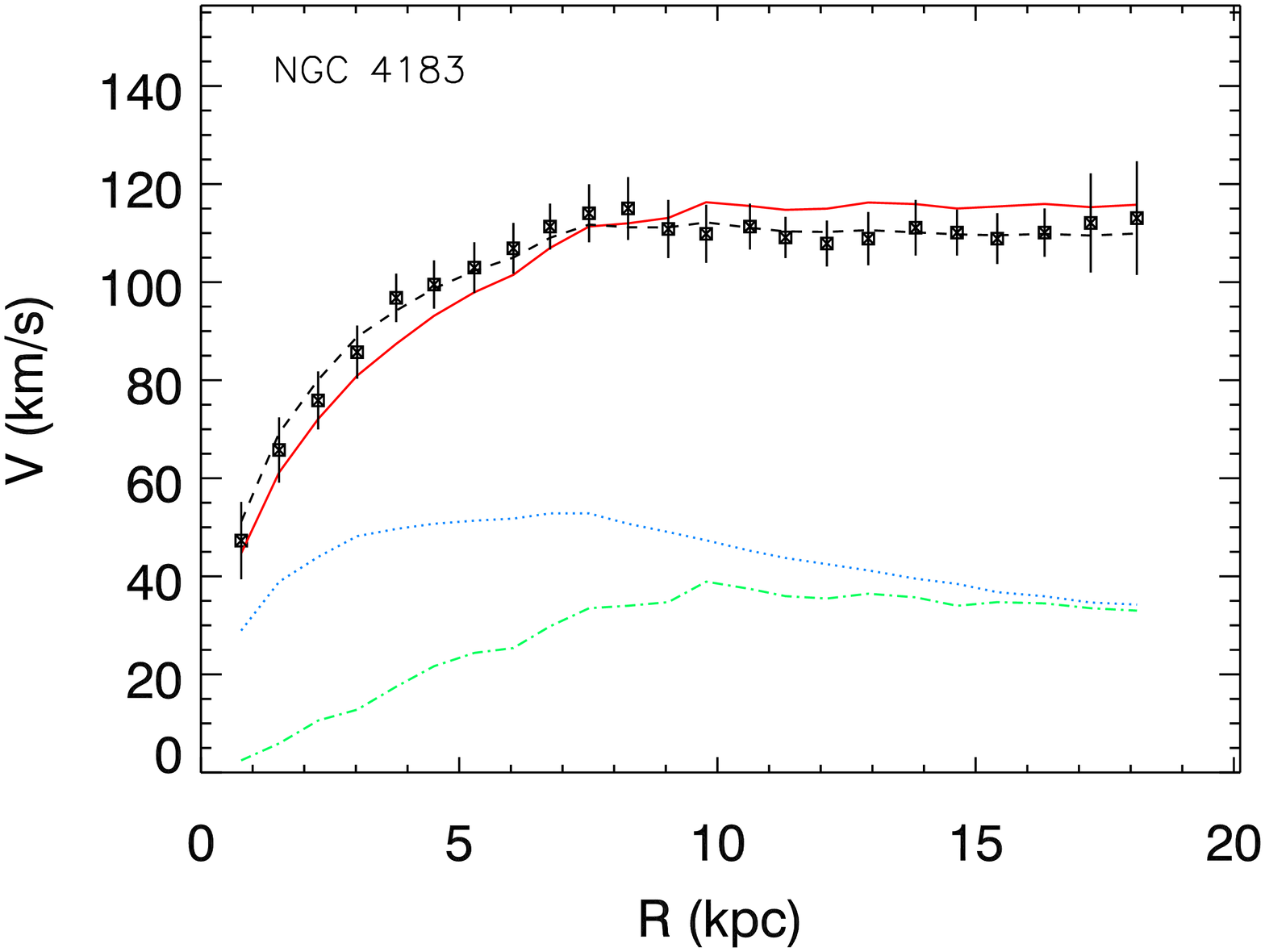}
  \includegraphics[angle=90,width=0.45\textwidth]{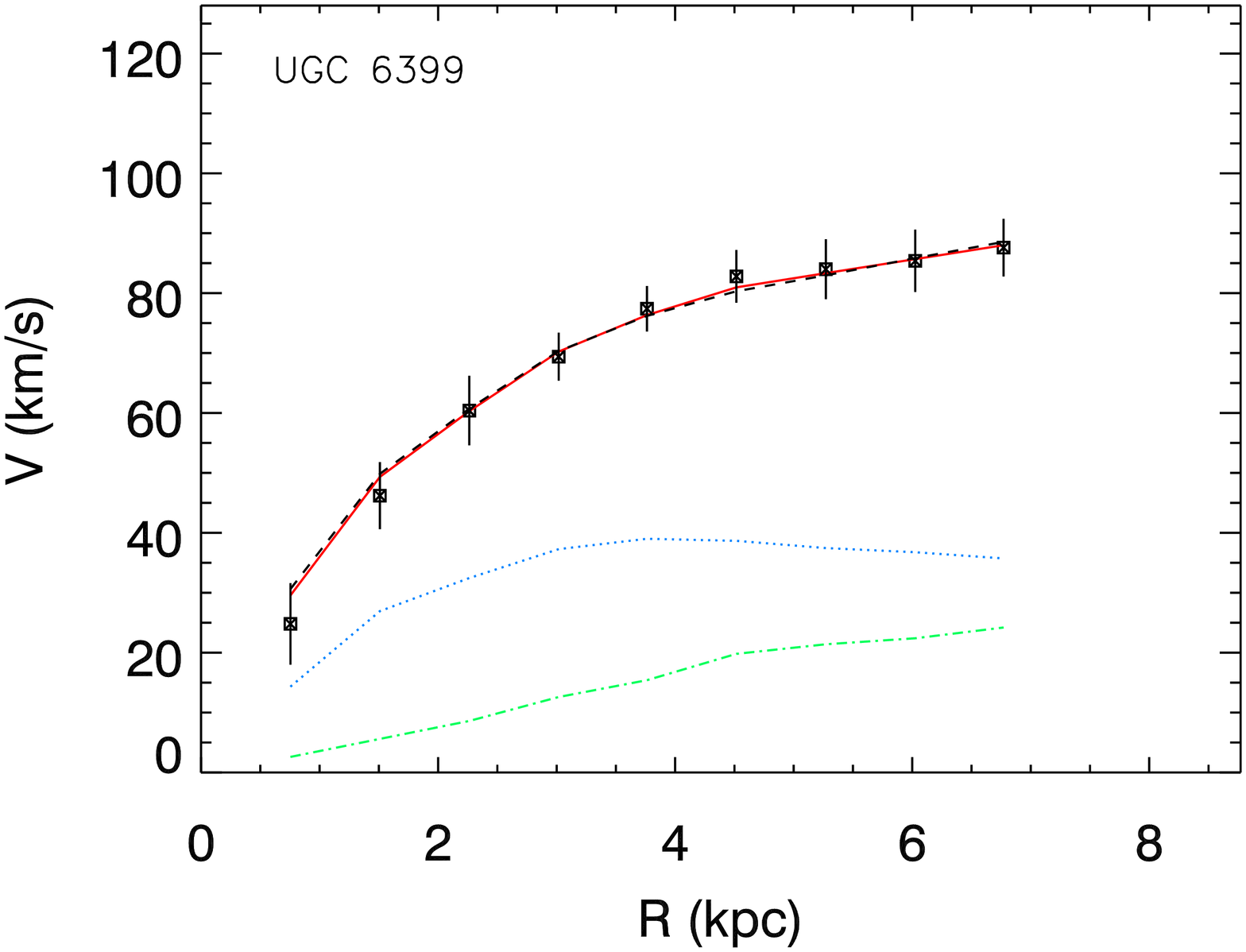}\vspace{3mm}\\
  \includegraphics[angle=90,width=0.45\textwidth]{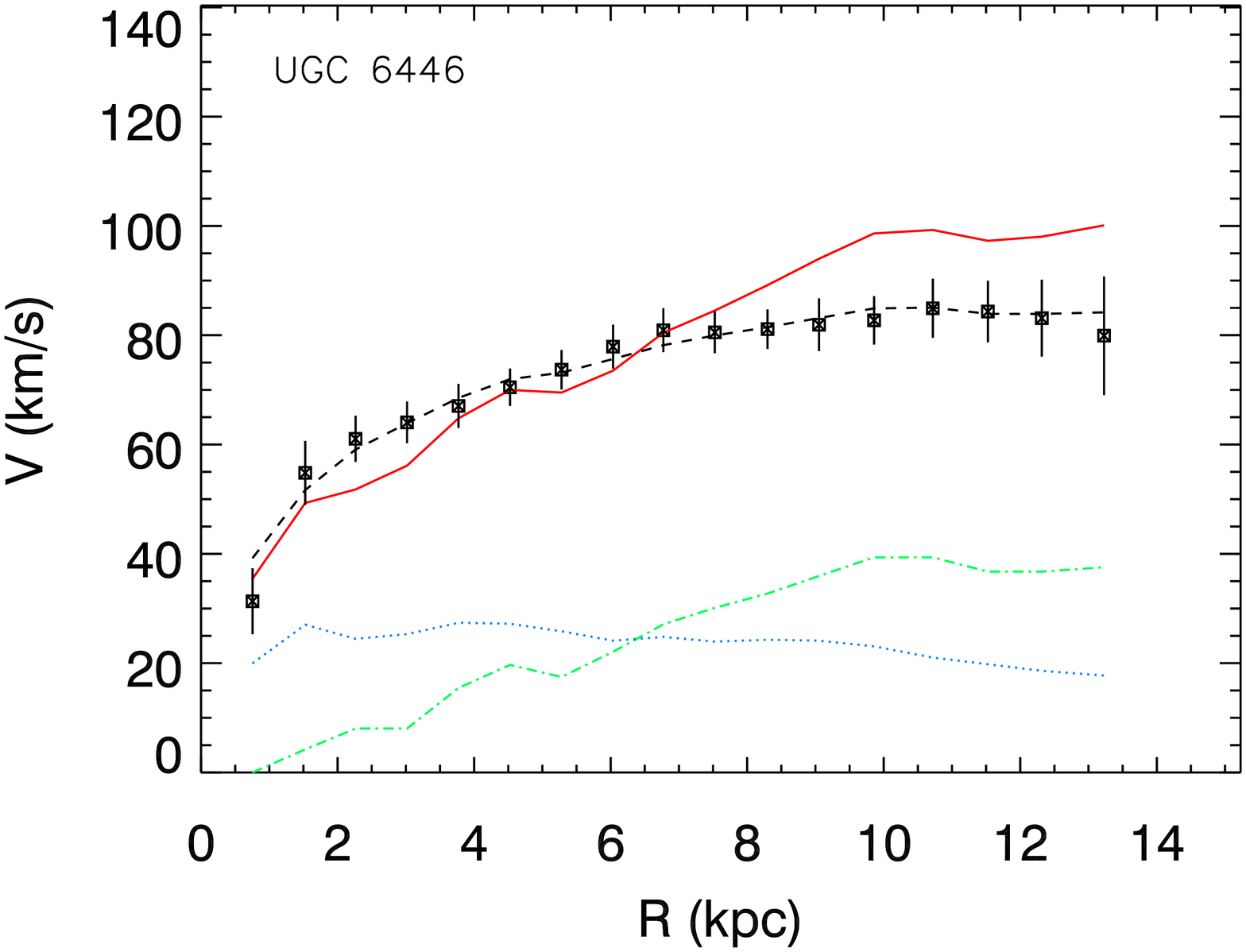}
  \includegraphics[angle=90,width=0.45\textwidth]{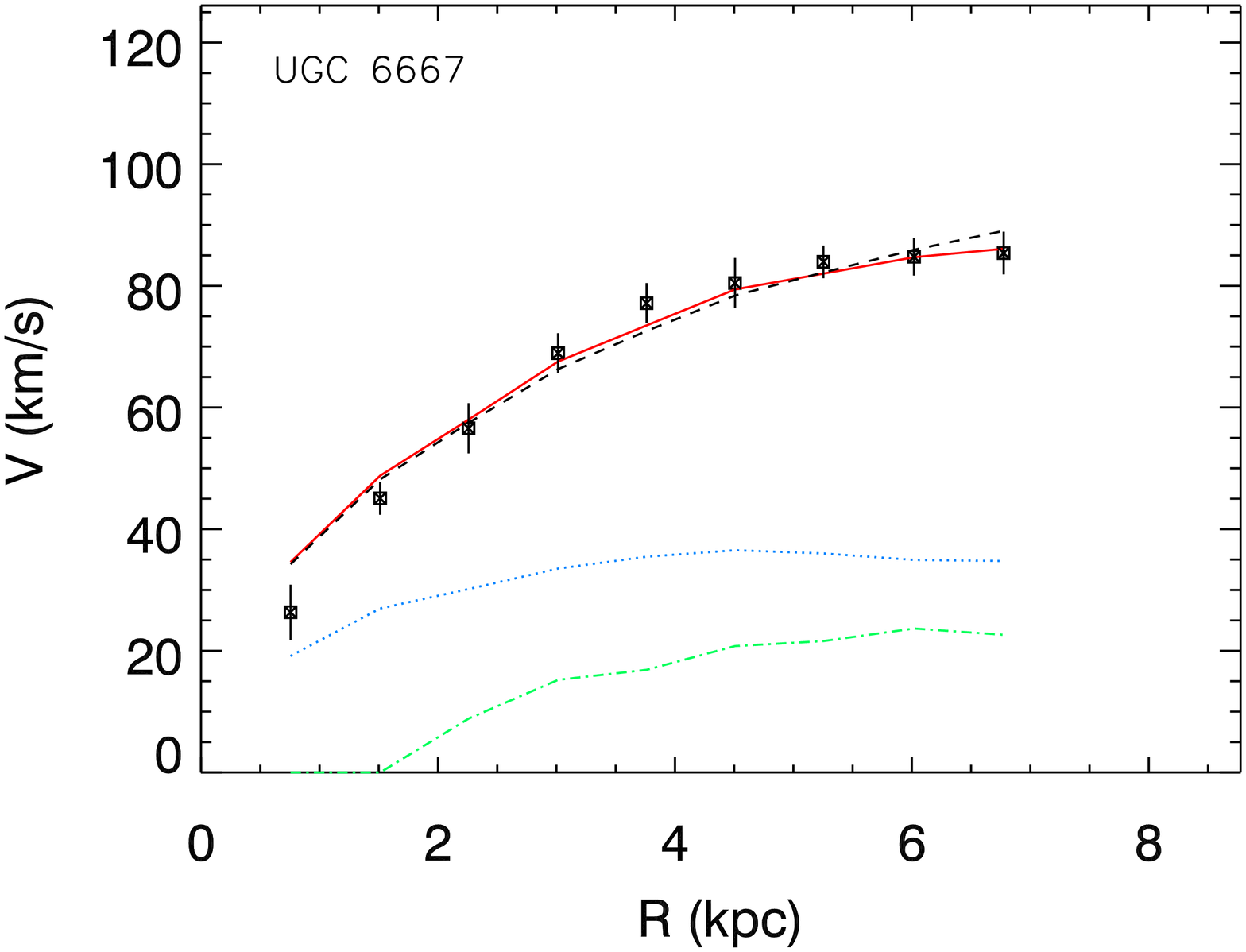}
  \caption{LSB galactic rotation curves. See Figure~\ref{fig:hsb} for description.}
  \label{fig:lsb}
\end{figure}

\begin{figure}[!ht]
  \includegraphics[angle=90,width=0.45\textwidth]{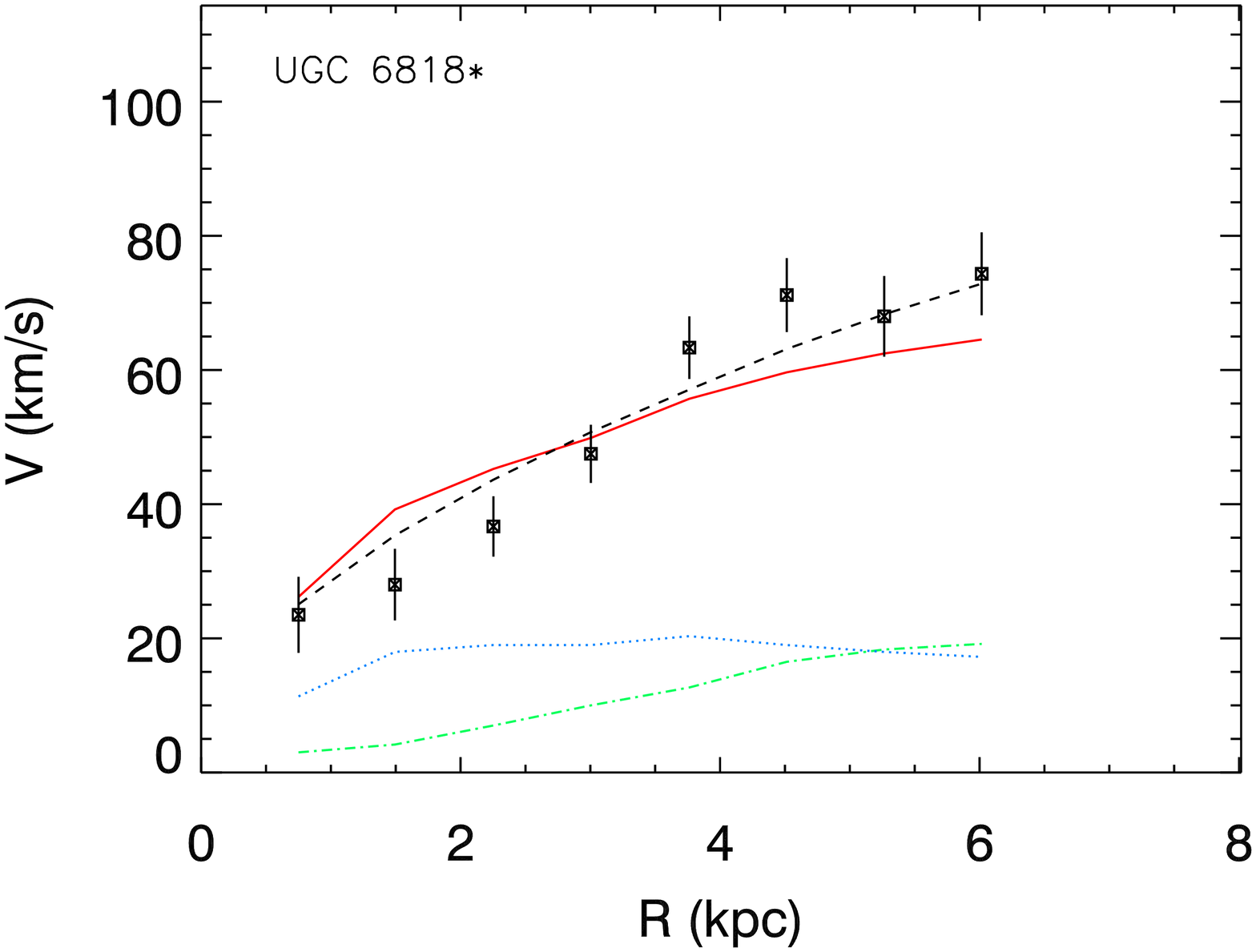}
  \includegraphics[angle=90,width=0.45\textwidth]{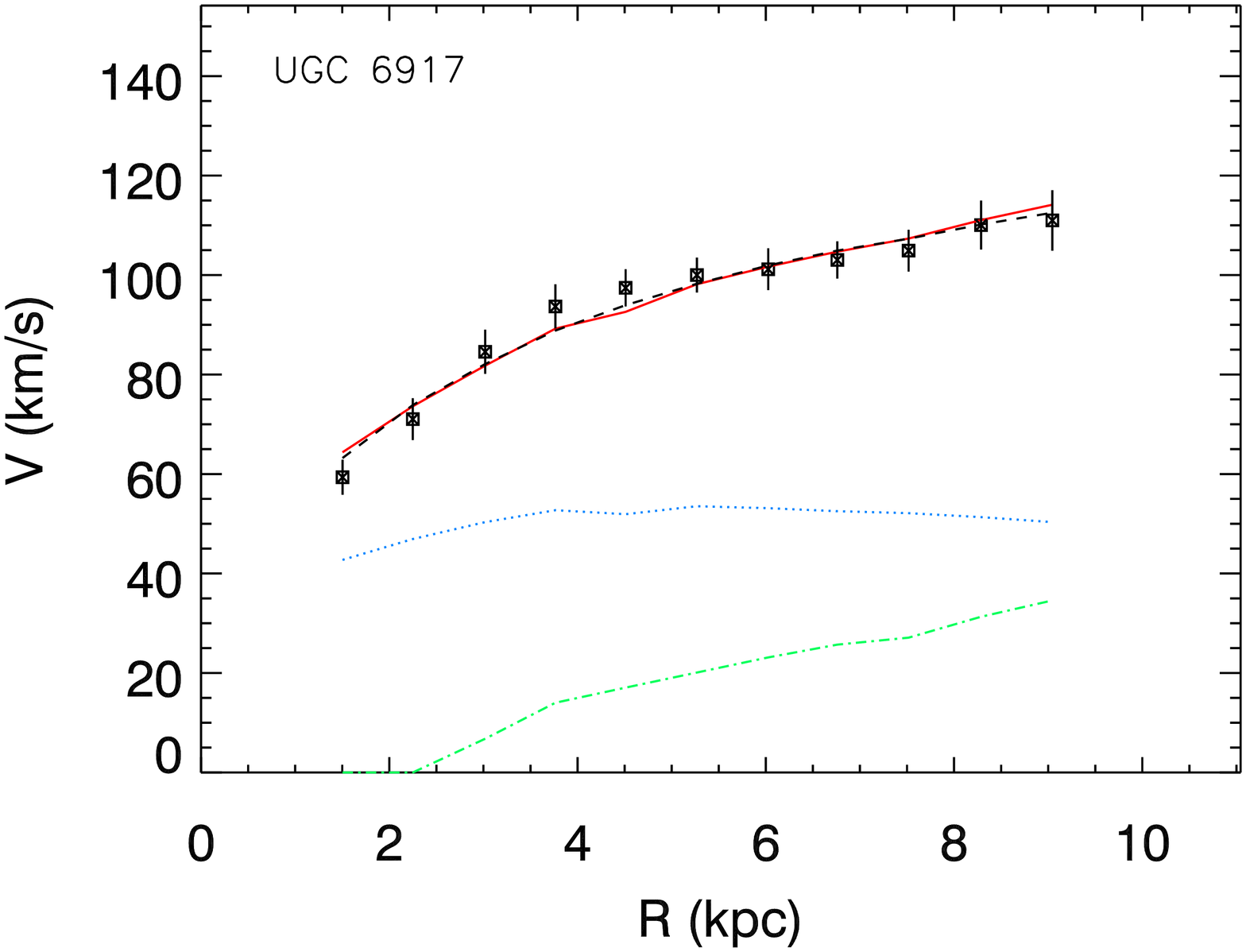}\vspace{3mm}\\
  \includegraphics[angle=90,width=0.45\textwidth]{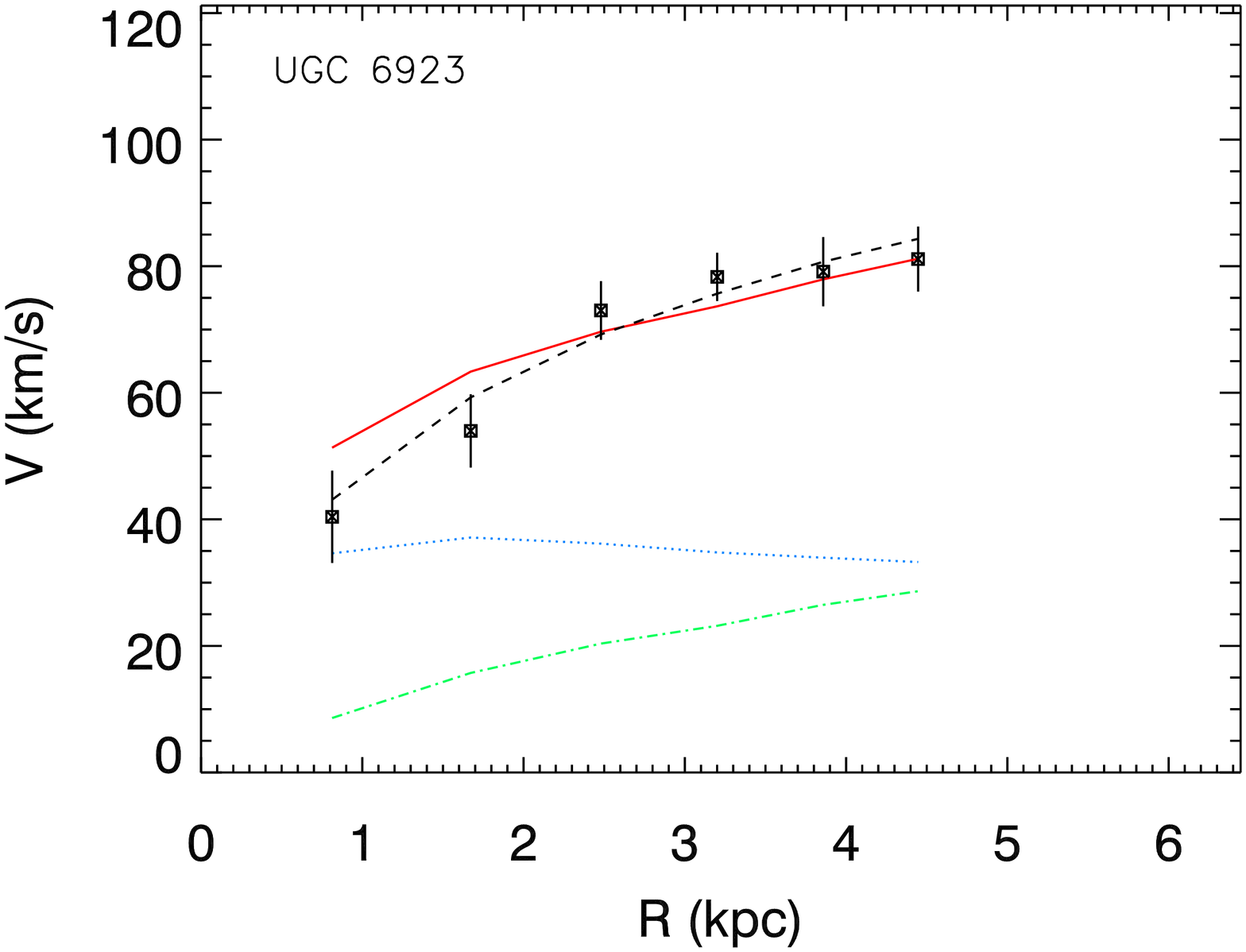}
  \includegraphics[angle=90,width=0.45\textwidth]{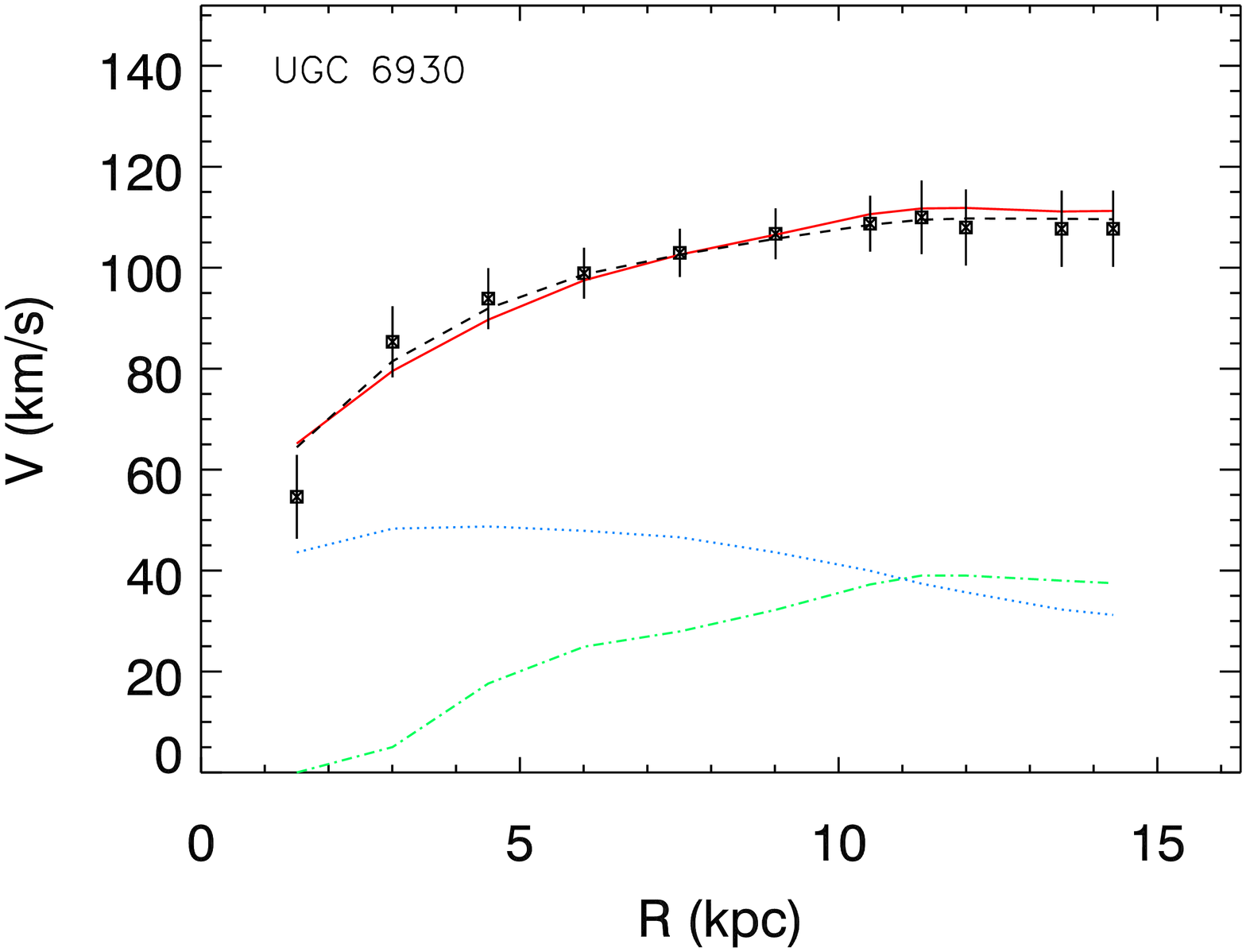}\vspace{3mm}\\
  \includegraphics[angle=90,width=0.45\textwidth]{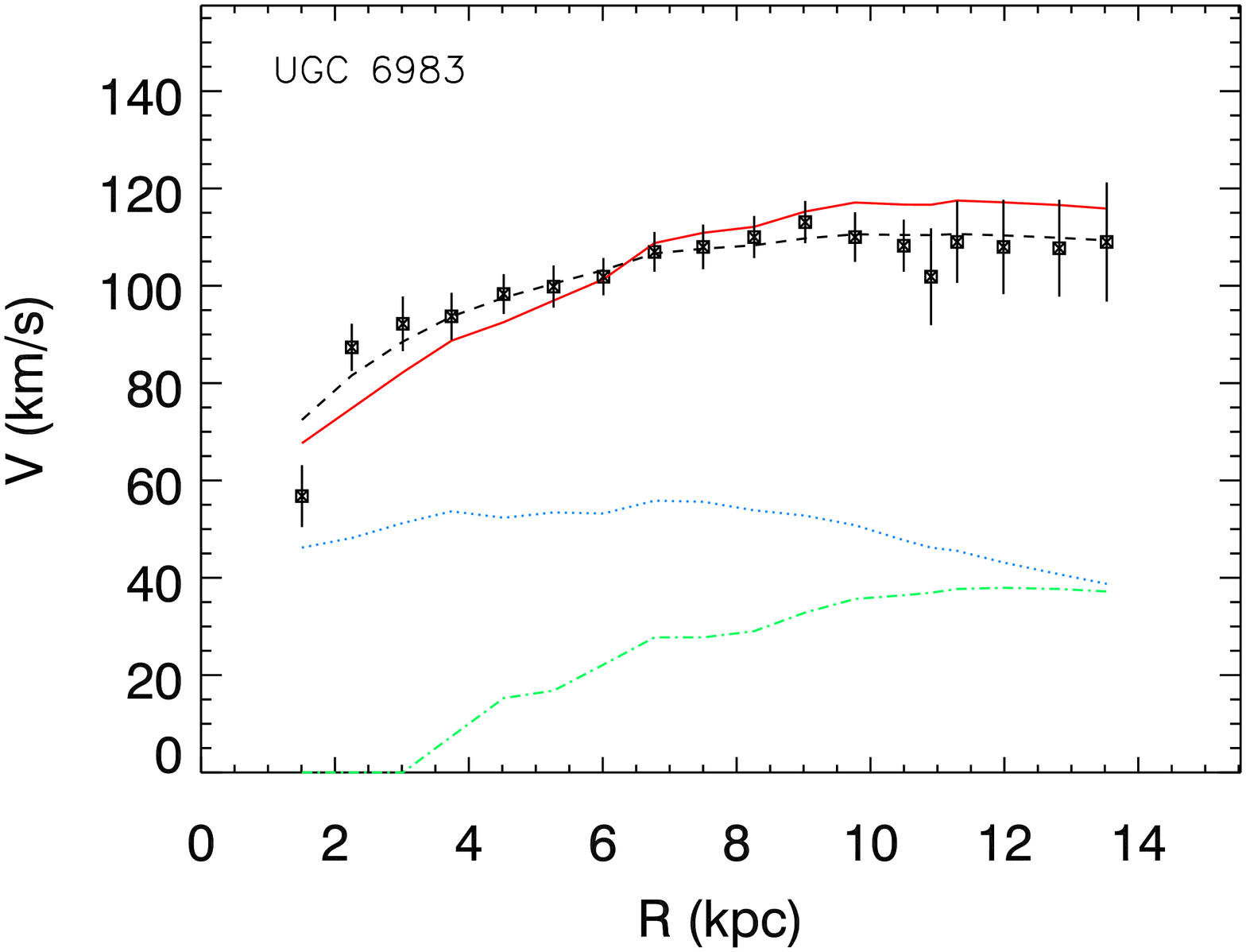}
  \includegraphics[angle=90,width=0.45\textwidth]{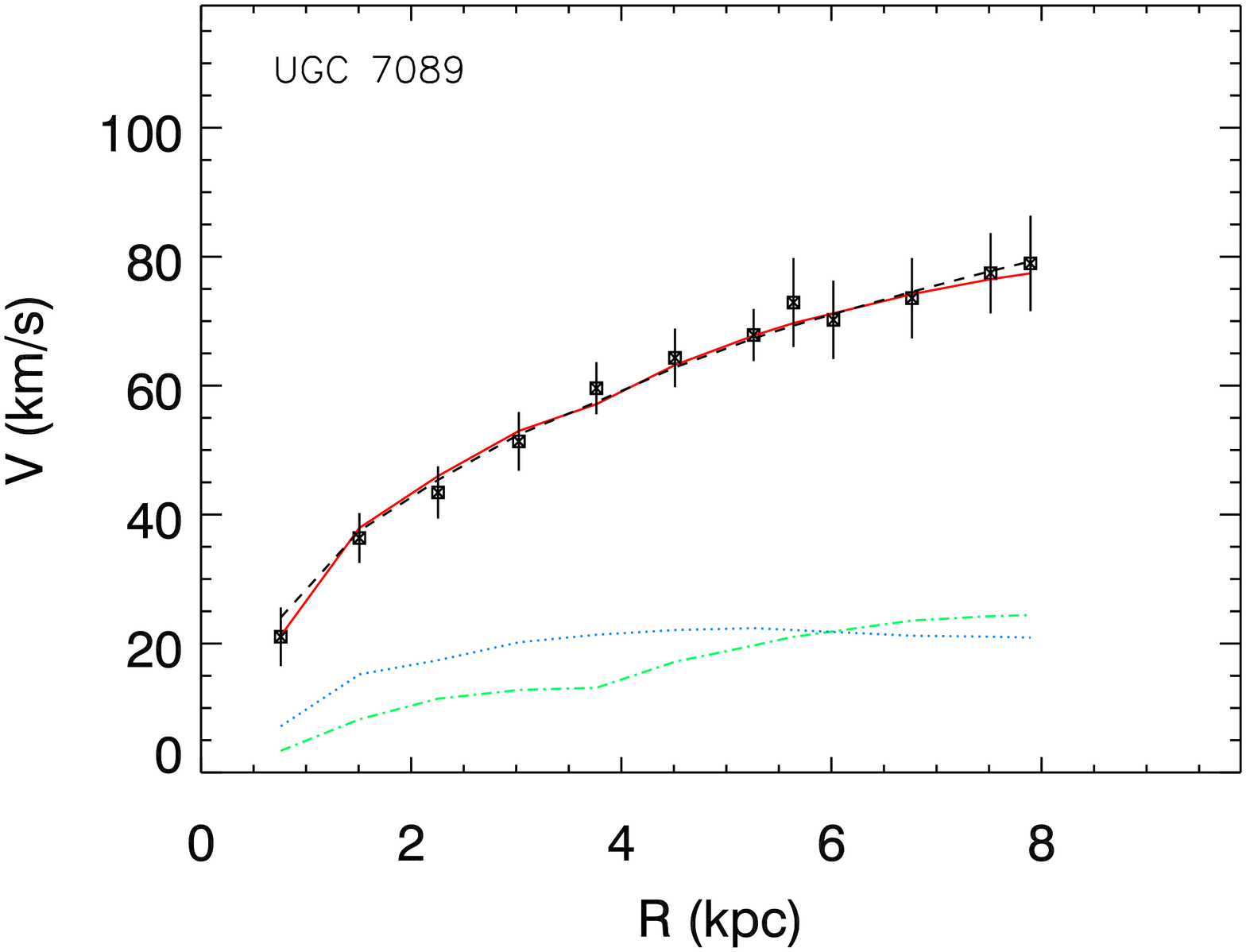}
  \caption{Figure~\ref{fig:lsb} continued.}
\label{fig:lsblast}
\end{figure}

\begin{figure}[!ht]
  \includegraphics[angle=90,width=0.45\textwidth]{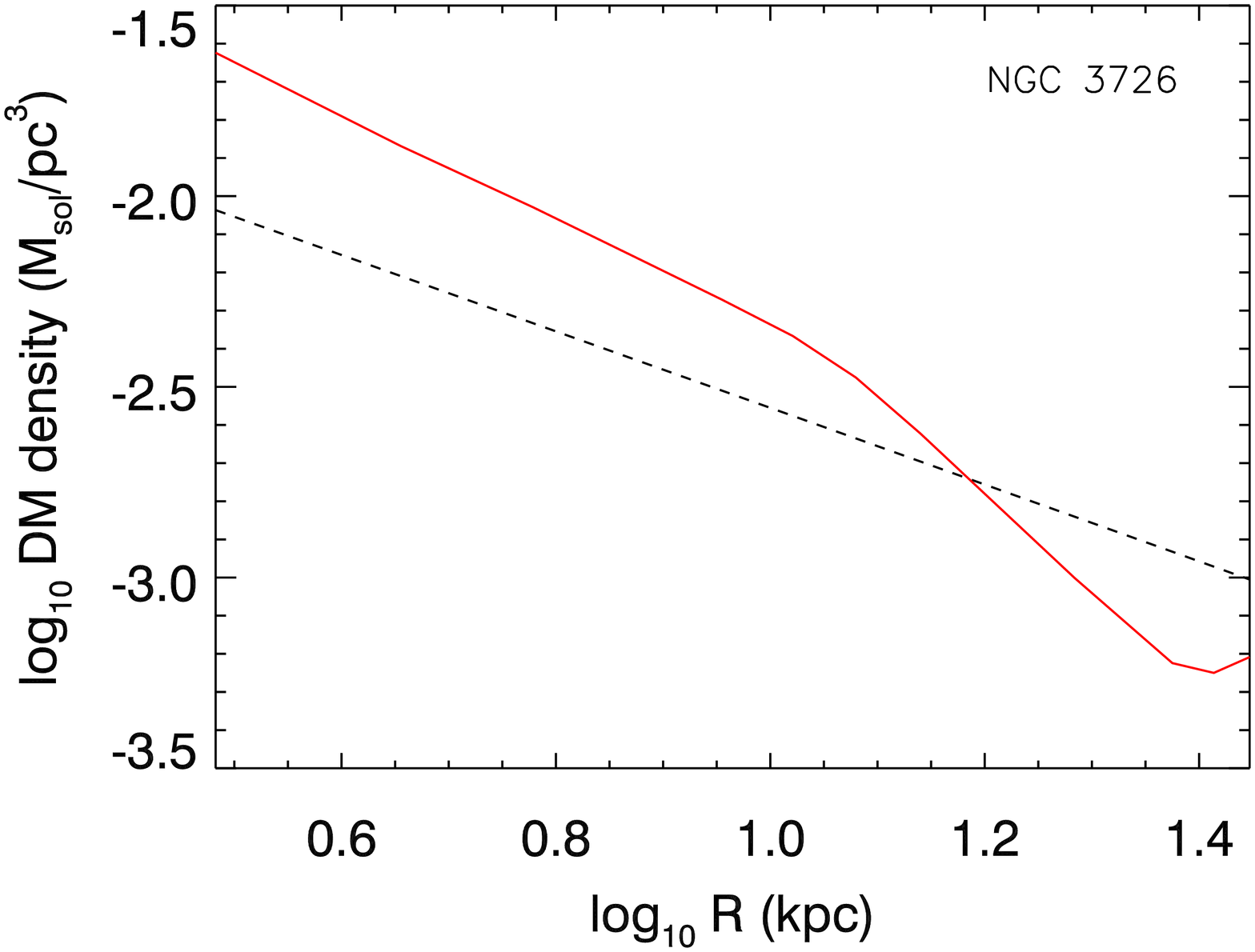}
  \includegraphics[angle=90,width=0.45\textwidth]{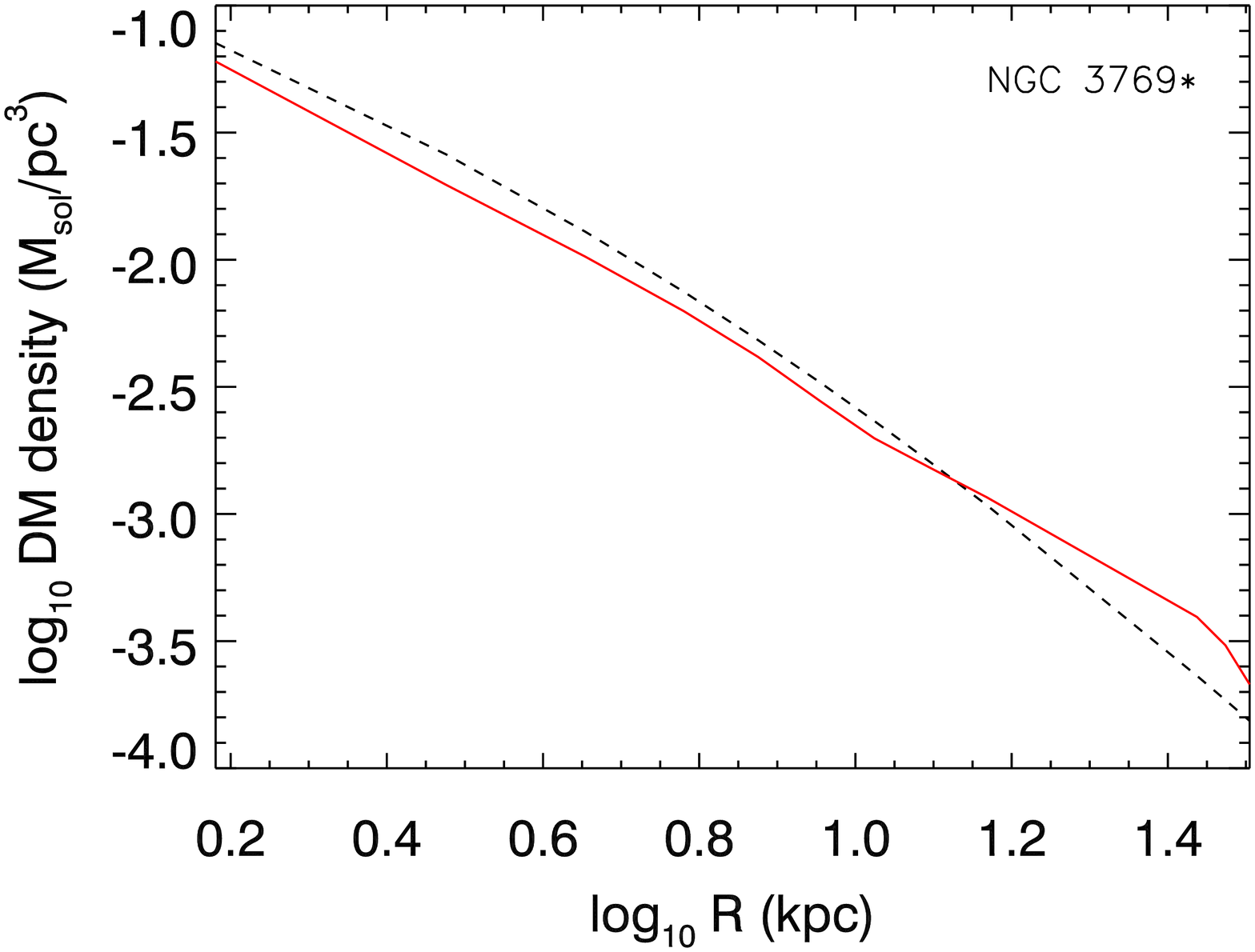}\vspace{3mm}\\
  \includegraphics[angle=90,width=0.45\textwidth]{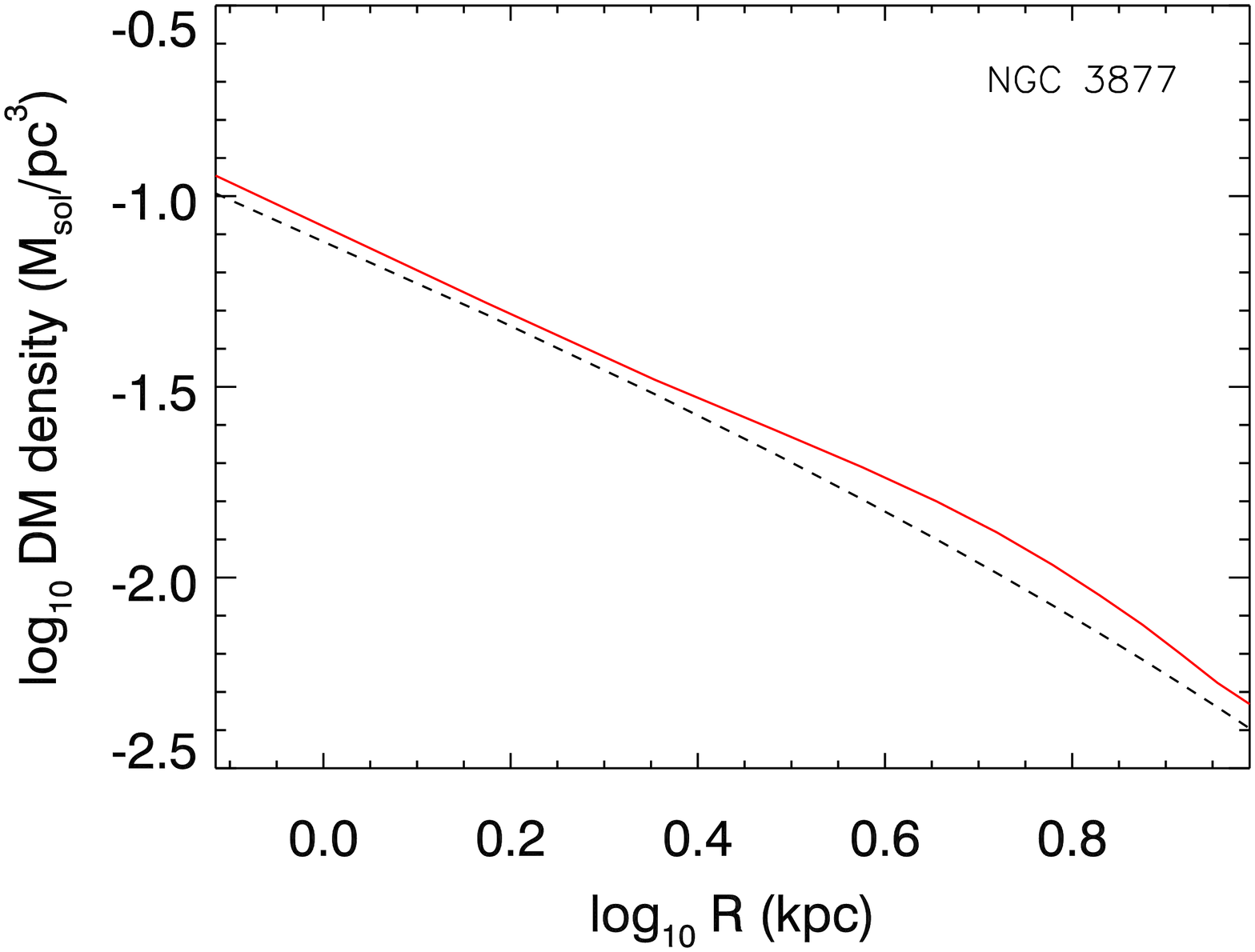}
  \includegraphics[angle=90,width=0.45\textwidth]{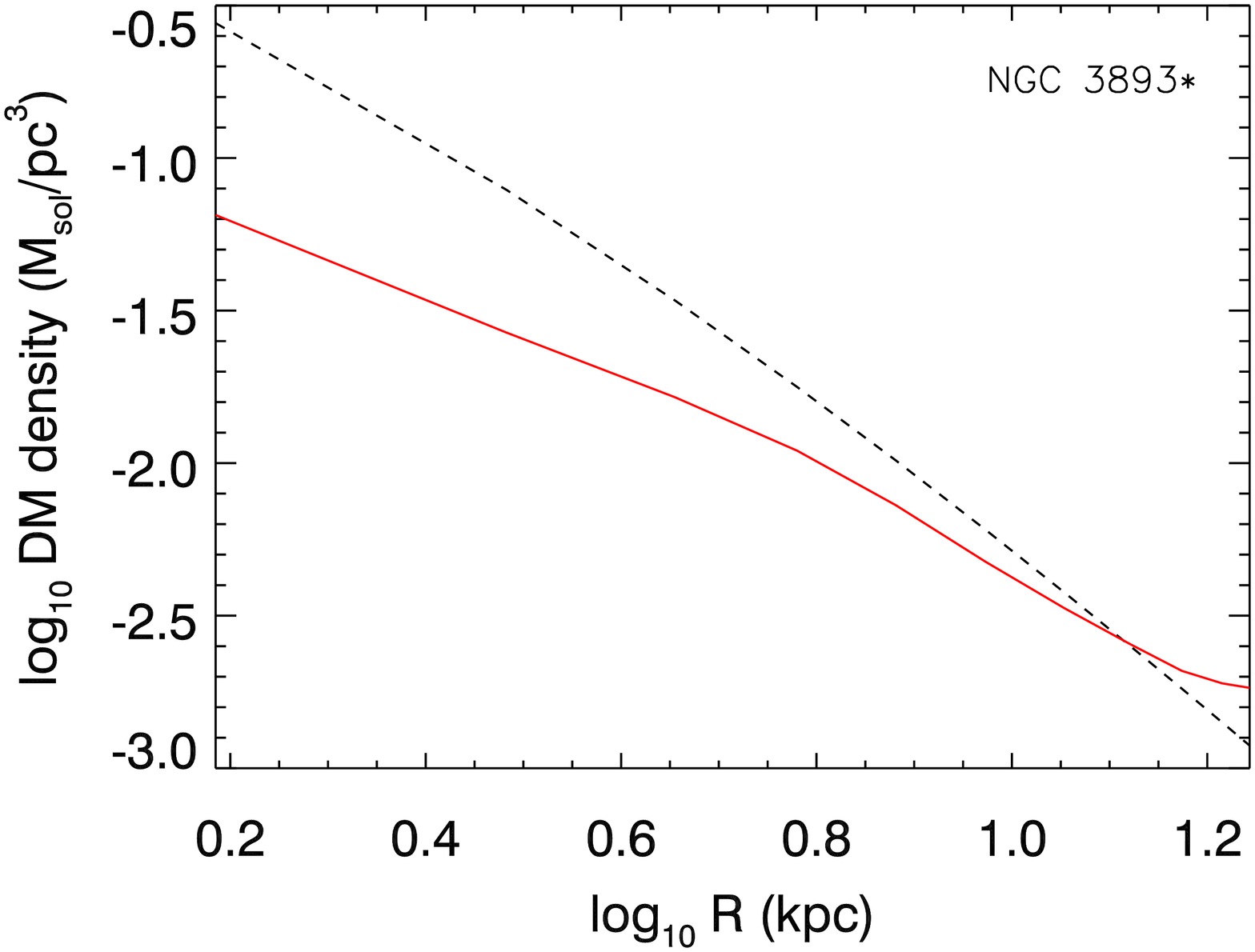}\vspace{3mm}\\
  \includegraphics[angle=90,width=0.45\textwidth]{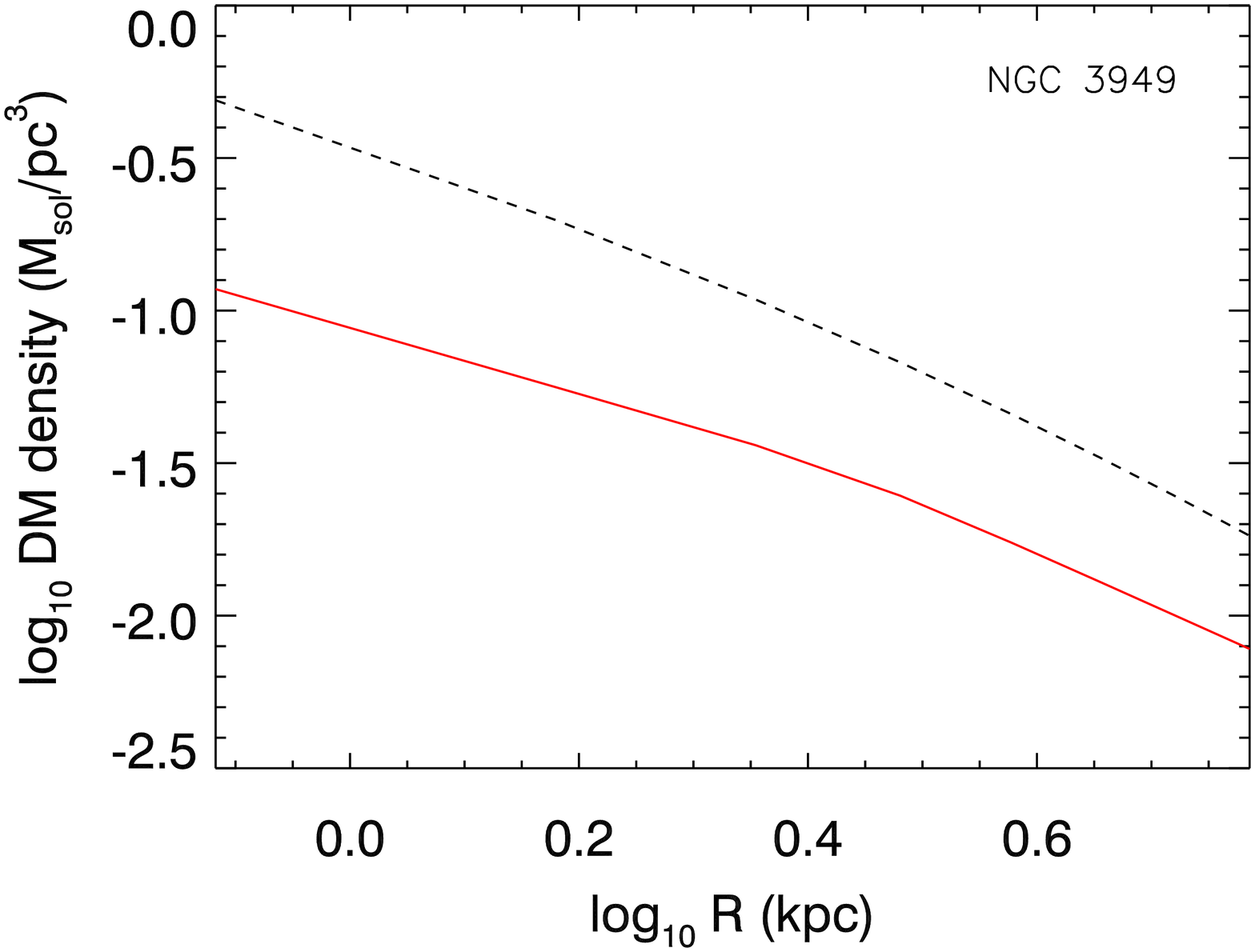}
  \includegraphics[angle=90,width=0.45\textwidth]{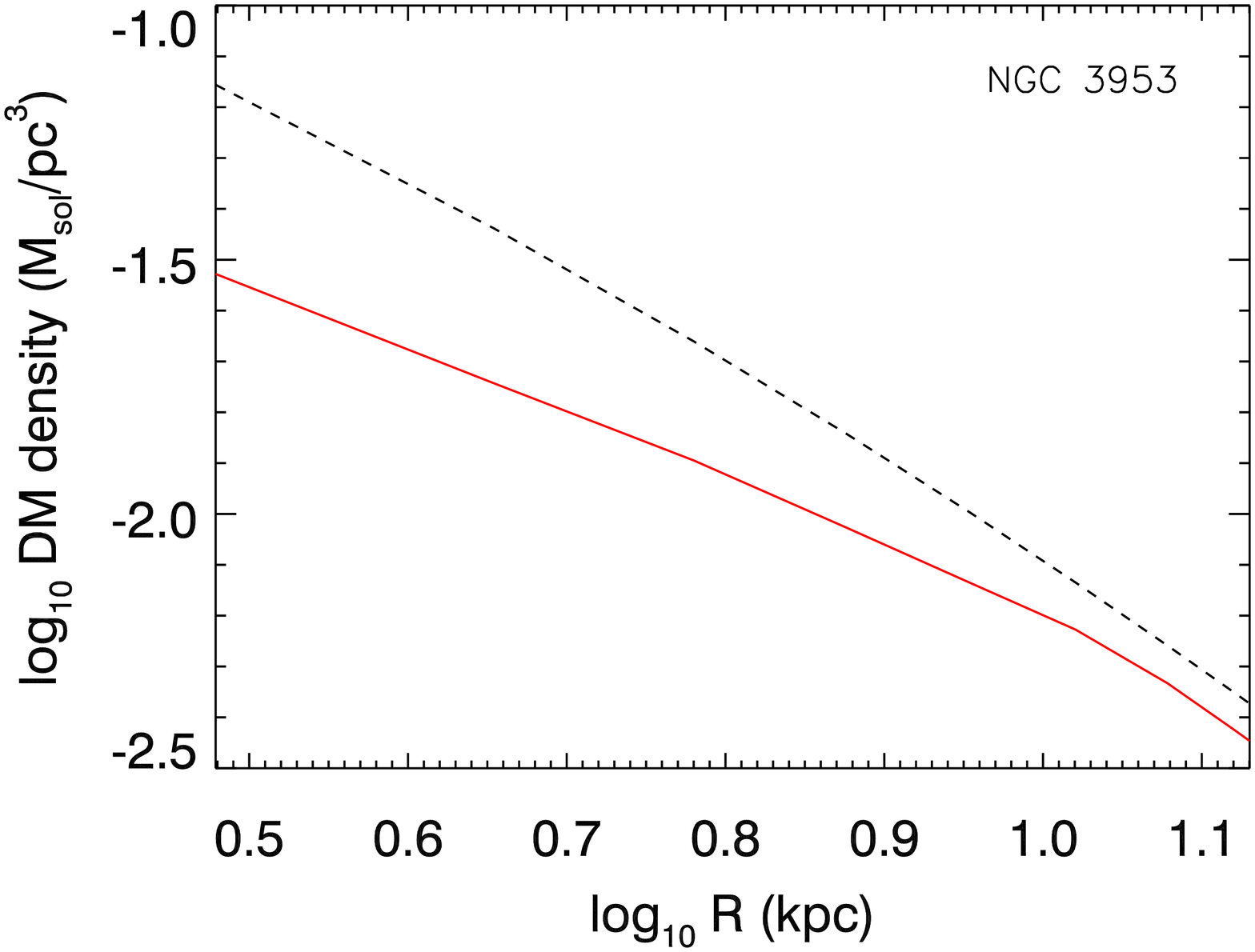}
  \caption{HSB dark matter density profiles. Density profiles for MDM and CDM are depicted as solid and dashed lines, respectively.}
  \label{fig:hsb_dens}
\end{figure}

Shown in Figures~\ref{fig:hsb_dens}---\ref{fig:hsb_denslast} and \ref{fig:lsb_dens}---\ref{fig:lsb_denslast}, are dark matter density profiles predicted by MDM (solid lines) and CDM (dashed lines) for HSB and LSB galaxies, respectively. The MDM density is given by
%
%
\begin{equation}
\rho'(r) \;=\; \left( \frac{a_c}{r} \right)^2 \frac{d}{dr} \left( \frac{M}{a^2} \right).
\end{equation}

For the CDM fits, we use the \citet{nfw} density profile:

\begin{equation}
\rho'(r) \;=\; \dfrac{\rho_0}{\dfrac{r}{r_s} \left( 1+\dfrac{r}{r_s} \right)^2},
\label{eqn:rhonfw}
\end{equation}

in which 

\begin{equation}
r_s = \dfrac{r_{200}}{c}
\end{equation}

Here, $r_{200}$ designates the `edge' of the halo, within which objects are assumed to be virialized, usually taken to be the boundary at which the halo density exceeds 200 times that of the background \citep{nfw}. The parameter $c$ is a dimensionless number that indicates how centrally concentrated the halo is. The velocity curves are then determined by

\begin{equation}
v(r) \;=\; v_{200}\;\sqrt{\frac{\ln(1+cx)-cx/(1+cx)}{x \left[ \ln(1+c)-c/(1+c) \right]}},
\label{eqn:vnfw}
\end{equation}

where $v_{200}$ is the Newtonian velocity at $r_{200}$ \citep{nfw}. Equation \eqref{eqn:vnfw} is fit to the data with $c$, $v_{200}$, and $\alpha = M/L~(\ge\,0)$ as free parameters. Values of the fitting parameters for each galaxy are given in Table~1. Rotation curves and density profiles for CDM are shown as dashed lines in the figures. For several galaxies, CDM fits predict very small $M/L$ ratios (formally approaching zero). In nearly a third of the galaxies in our sample, the estimated virial mass of the dark matter halo is too large. For NGC~4389, $v_{200} \approx 1000$~km~s$^{-1}$, leading to a virial mass of $\approx 5 \times 10^{14} M_\odot$. Since rotational velocities in our sample are less than 300~km~s$^{-1}$, we expect virial masses $\sim 10^{12} M_\odot$ \citep{nfw}. Furthermore, we note that many of the fits require $c,v_{200}$ pairs that do not agree with the $c-v_{200}$ relation shown in \cite{nfw} (see also \citealt{deBlok02}). The relation between $c$ and $v_{200}$ depends on the cosmology used, and a study by \citet{McGaugh07} found that dark matter concentrations predicted by the NFW profile for a range of measured cosmological parameters disagree with observations over a large range of rotation velocities (see Figure~10 of that paper).

\begin{figure}[!ht]
  \includegraphics[angle=90,width=0.45\textwidth]{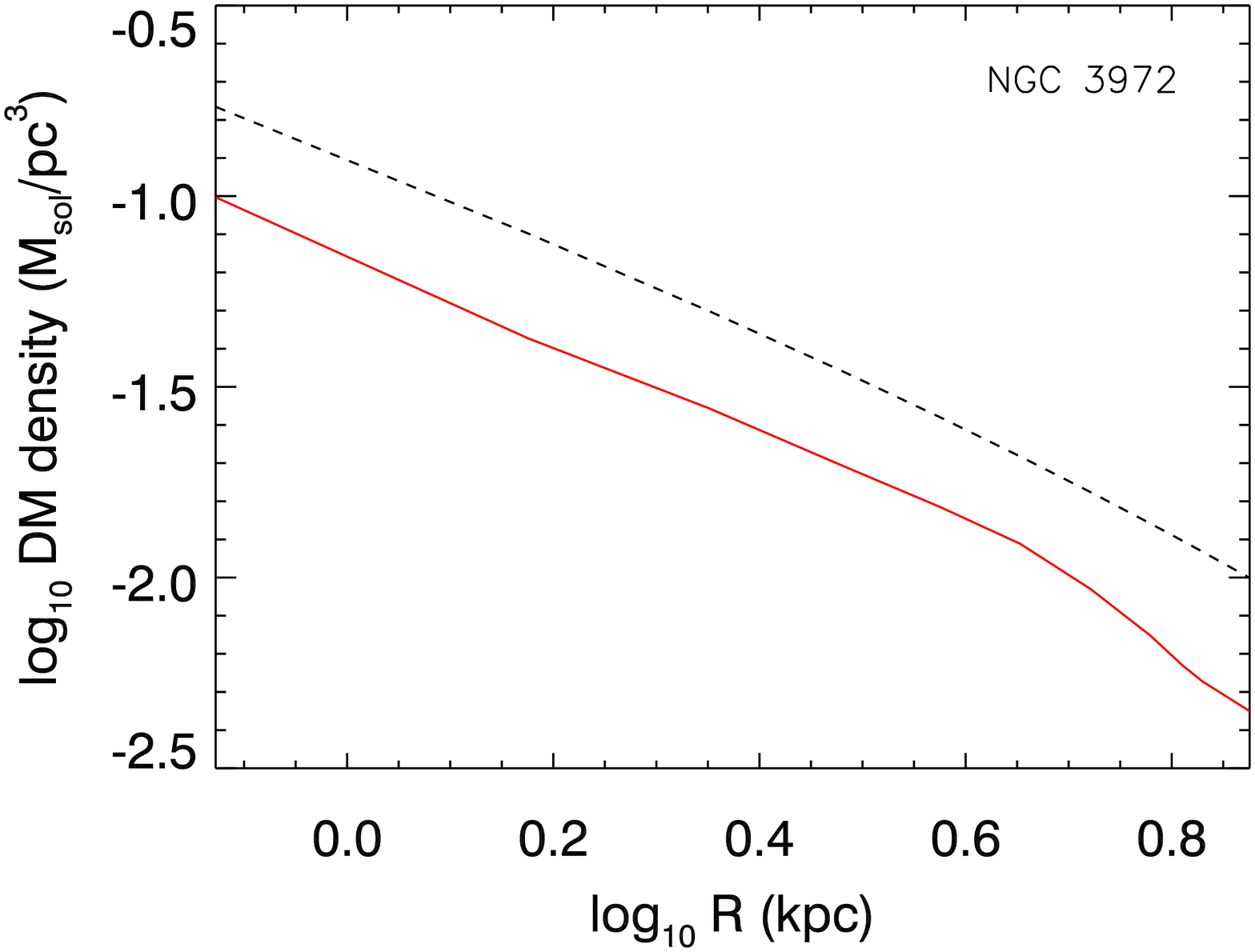}
  \includegraphics[angle=90,width=0.45\textwidth]{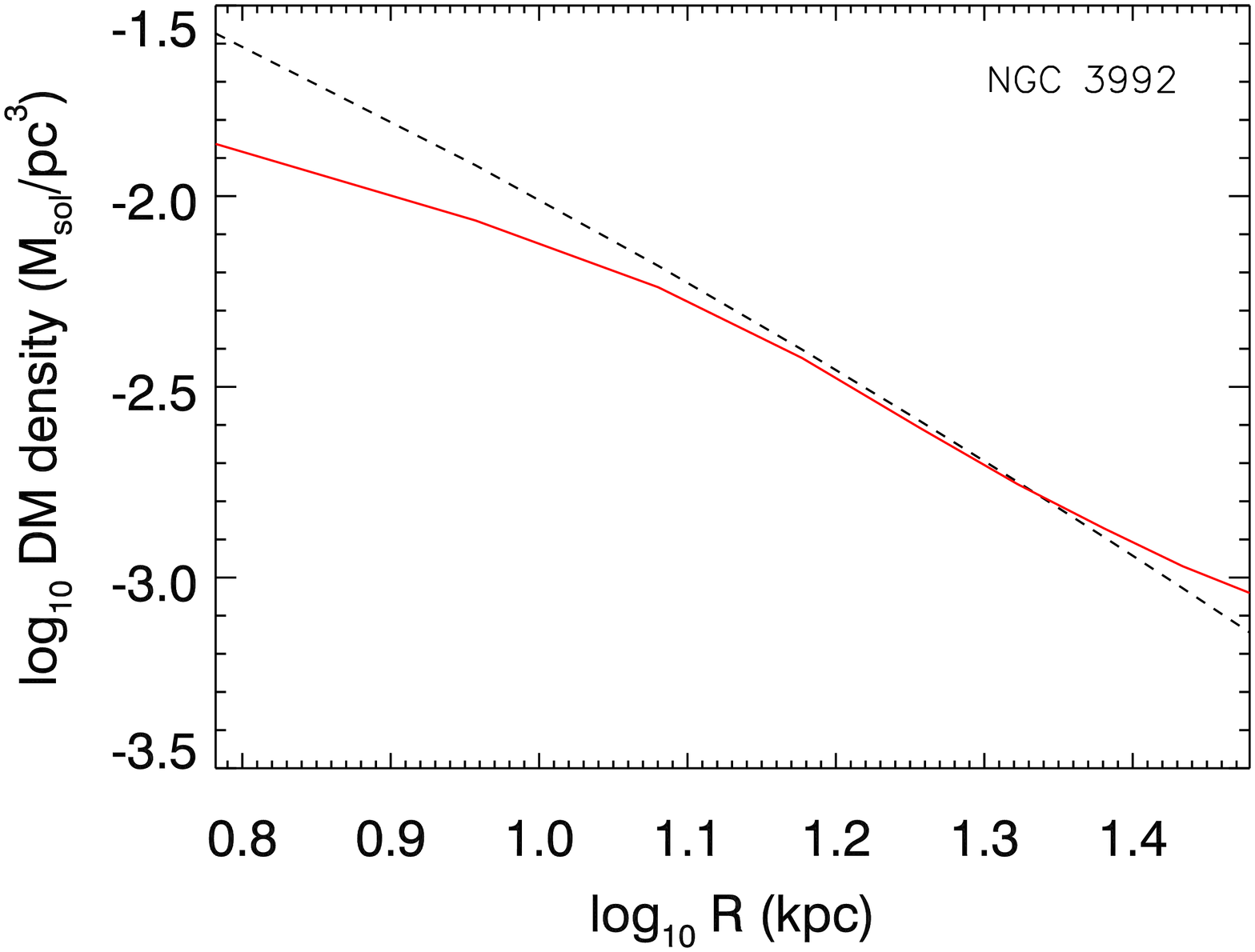}\vspace{3mm}\\
  \includegraphics[angle=90,width=0.45\textwidth]{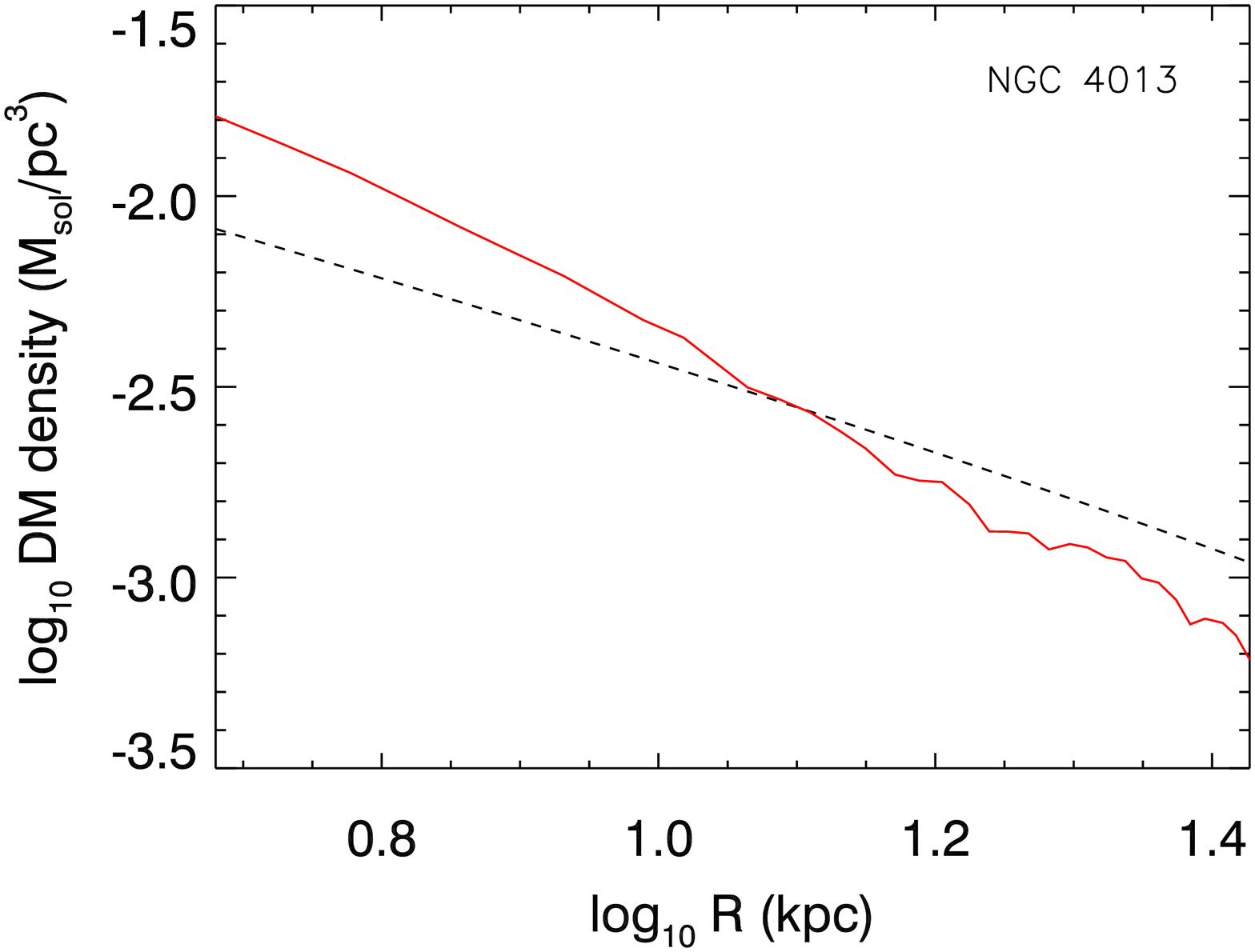}
  \includegraphics[angle=90,width=0.45\textwidth]{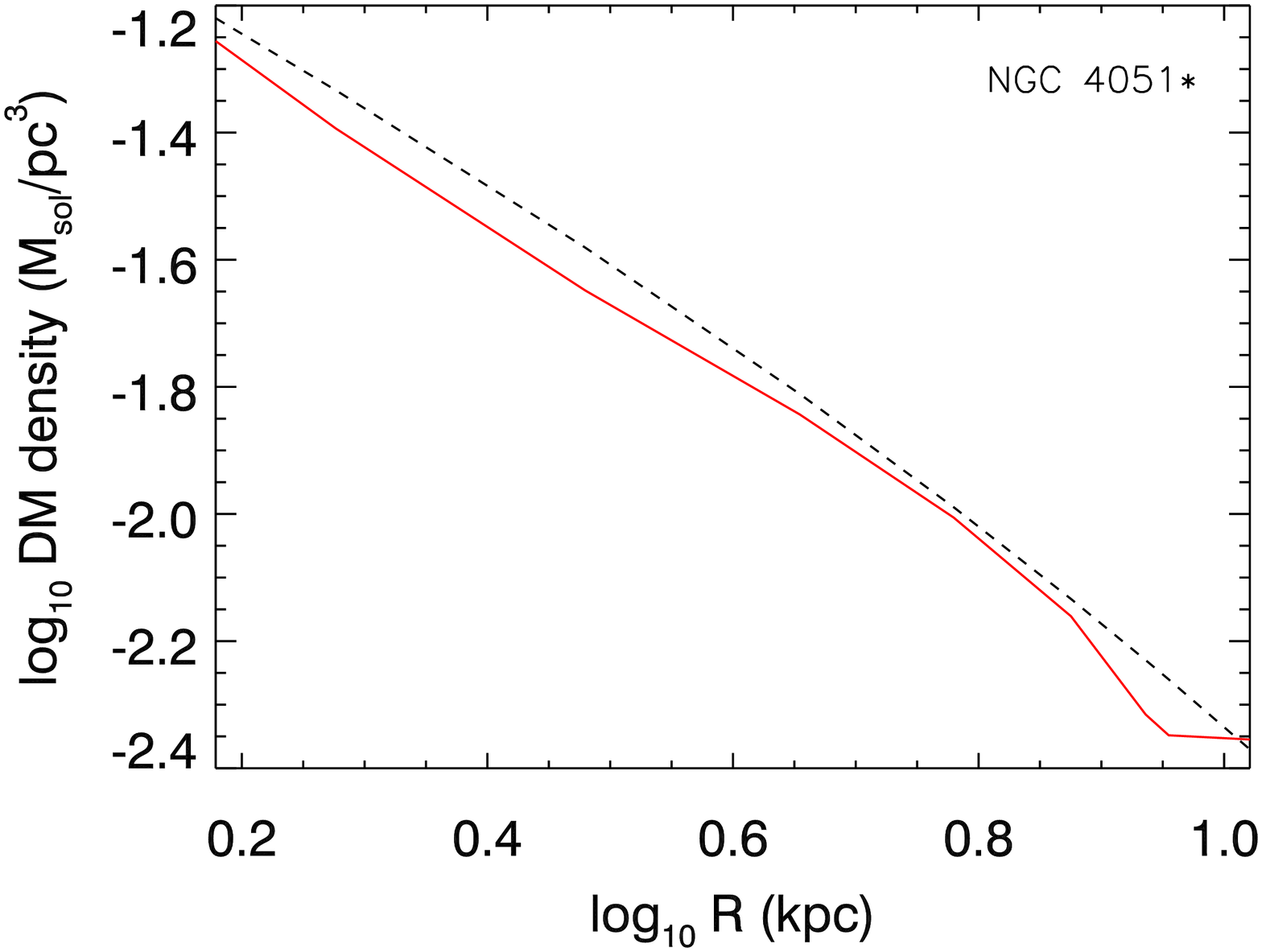}\vspace{3mm}\\
  \includegraphics[angle=90,width=0.45\textwidth]{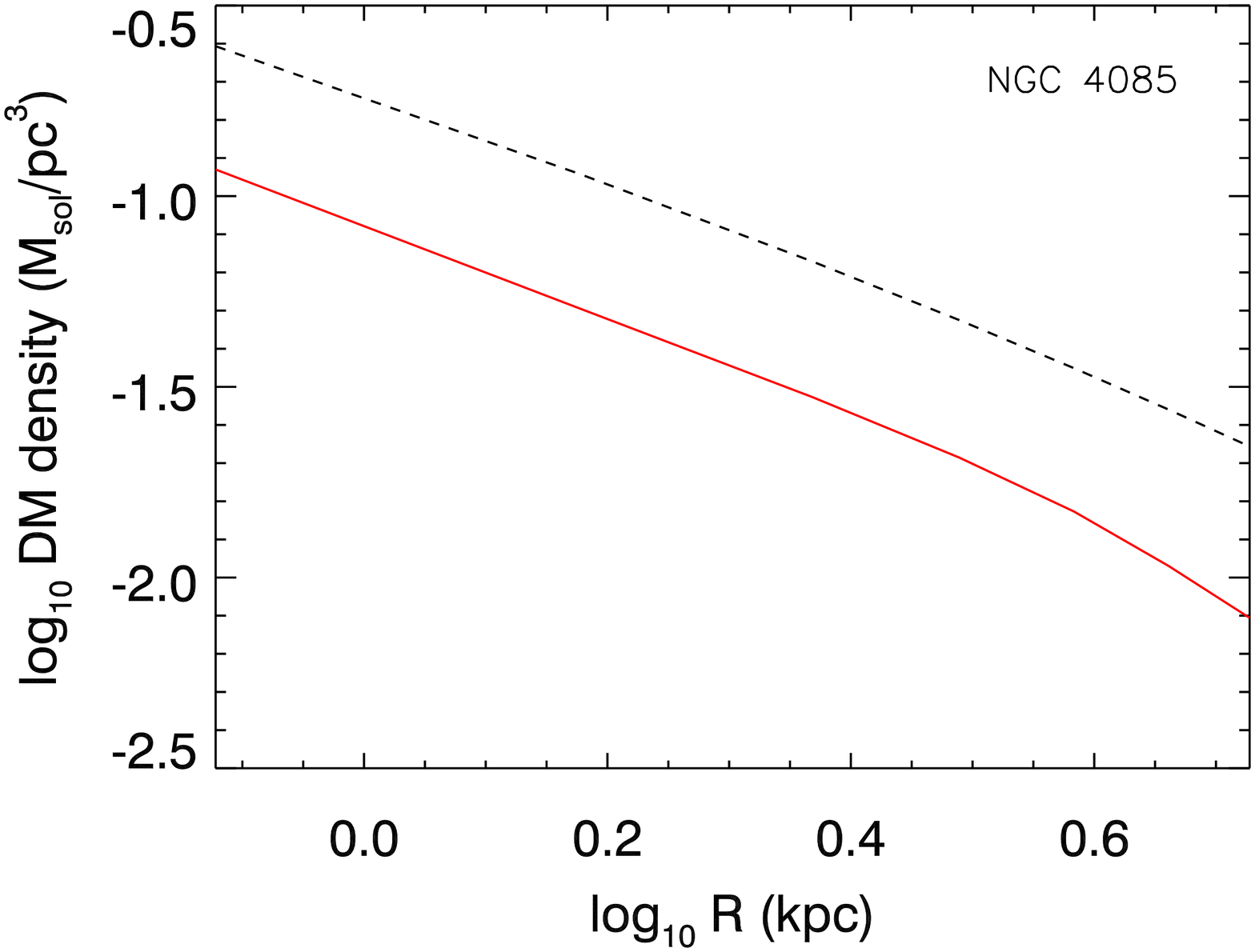}
  \includegraphics[angle=90,width=0.45\textwidth]{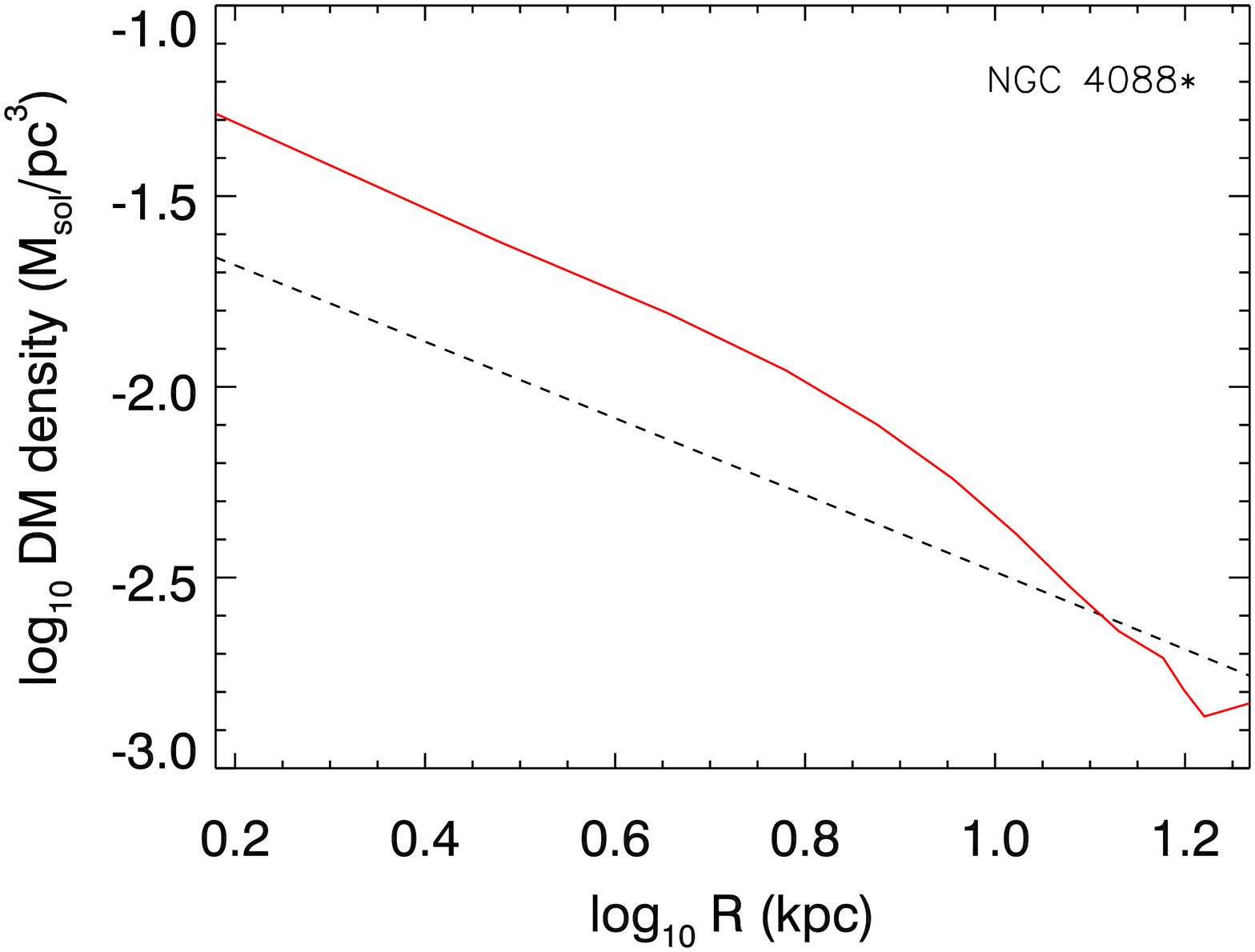}
  \caption{Figure~\ref{fig:hsb_dens} continued.}
\end{figure}

\begin{figure}[!ht]
  \includegraphics[angle=90,width=0.45\textwidth]{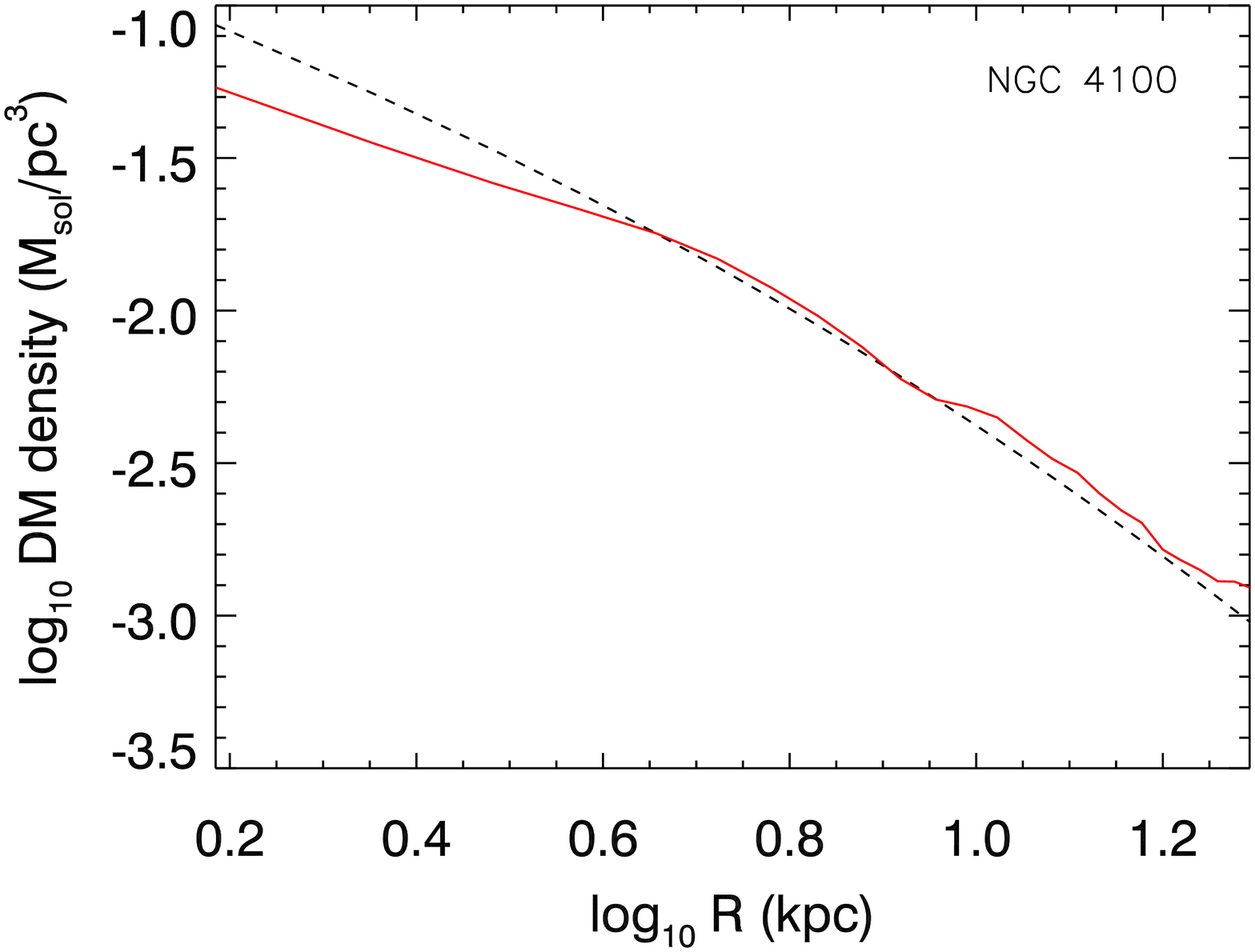}
  \includegraphics[angle=90,width=0.45\textwidth]{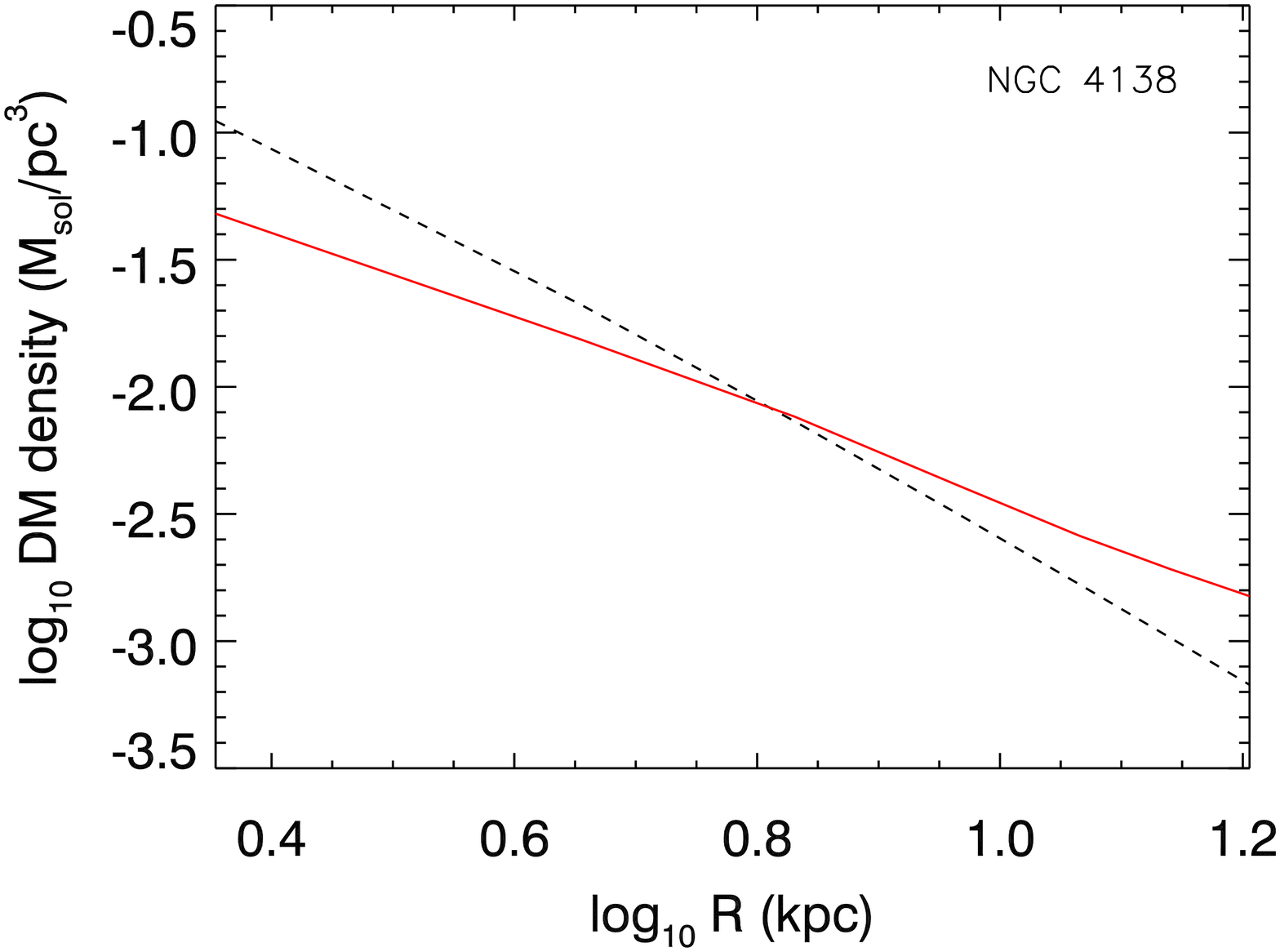}\vspace{3mm}\\
  \includegraphics[angle=90,width=0.45\textwidth]{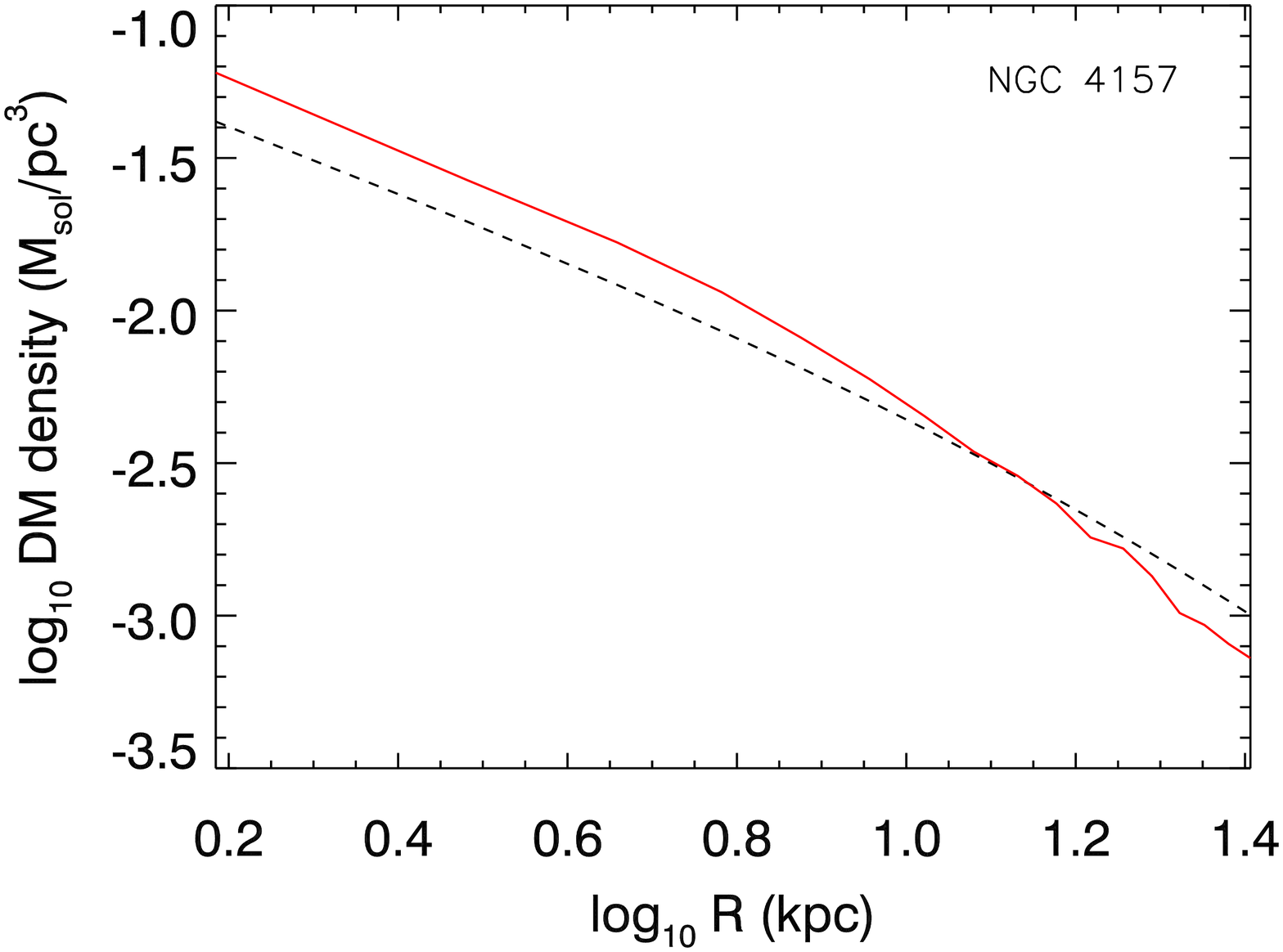}
  \includegraphics[angle=90,width=0.45\textwidth]{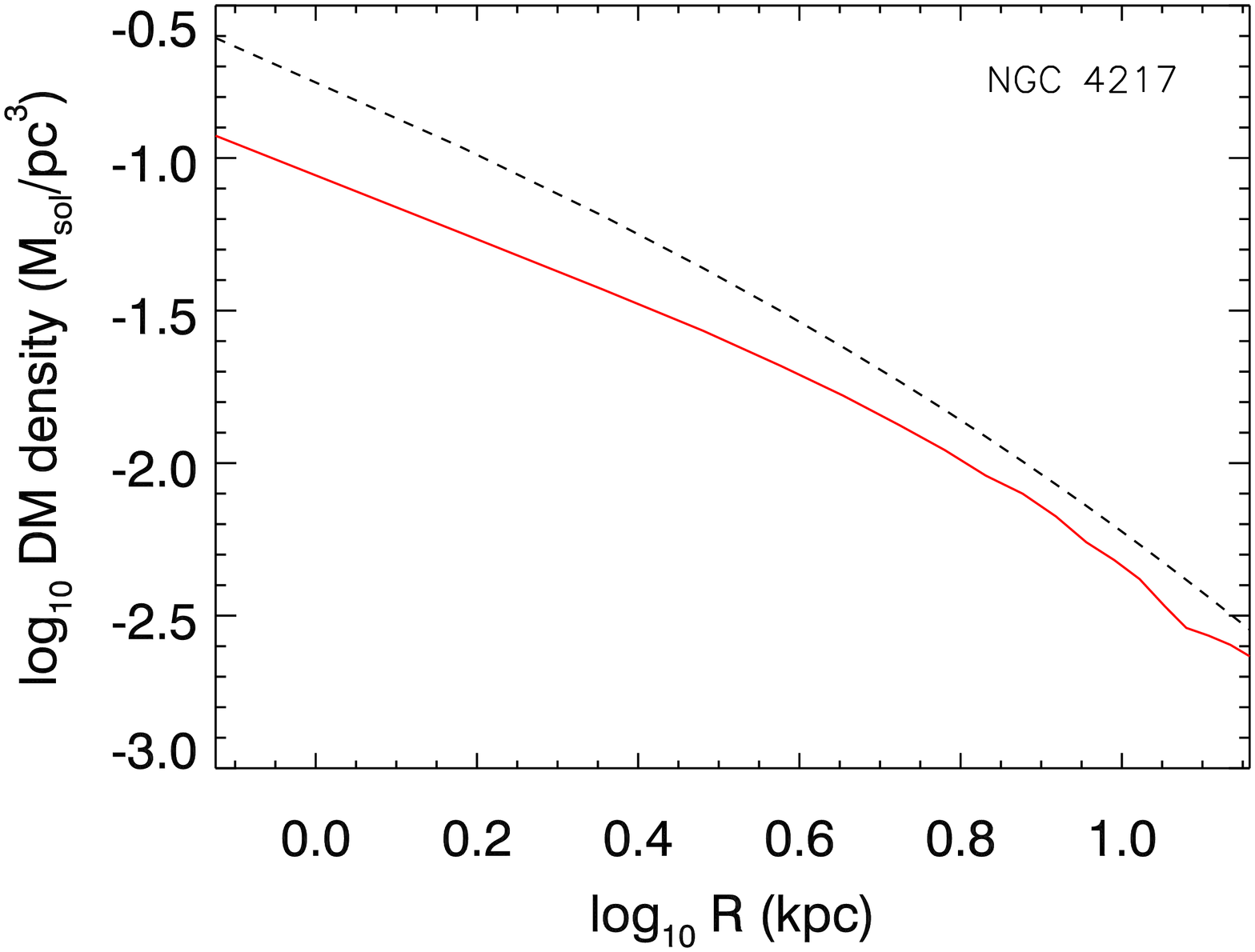}\vspace{3mm}\\
  \includegraphics[angle=90,width=0.45\textwidth]{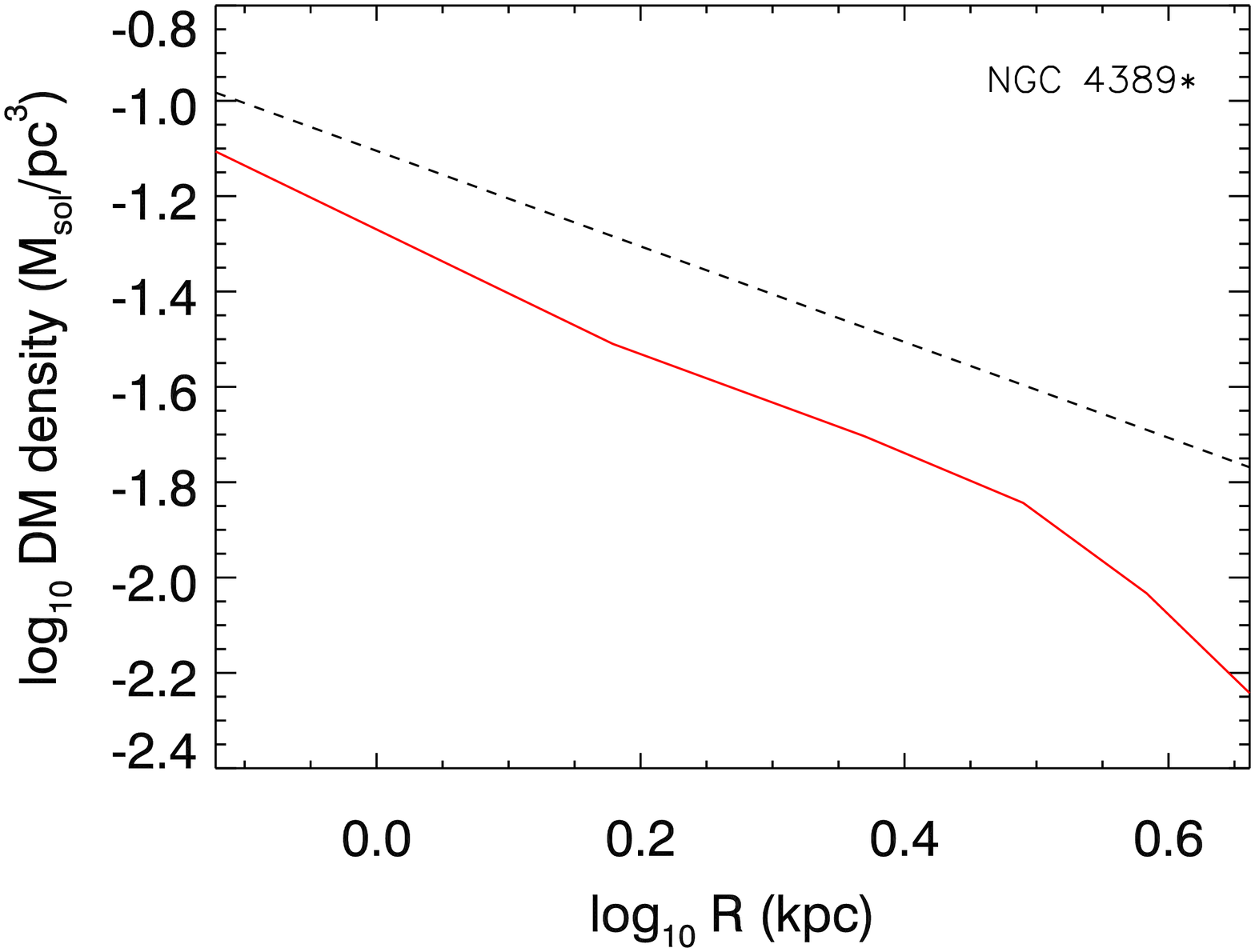}
  \includegraphics[angle=90,width=0.45\textwidth]{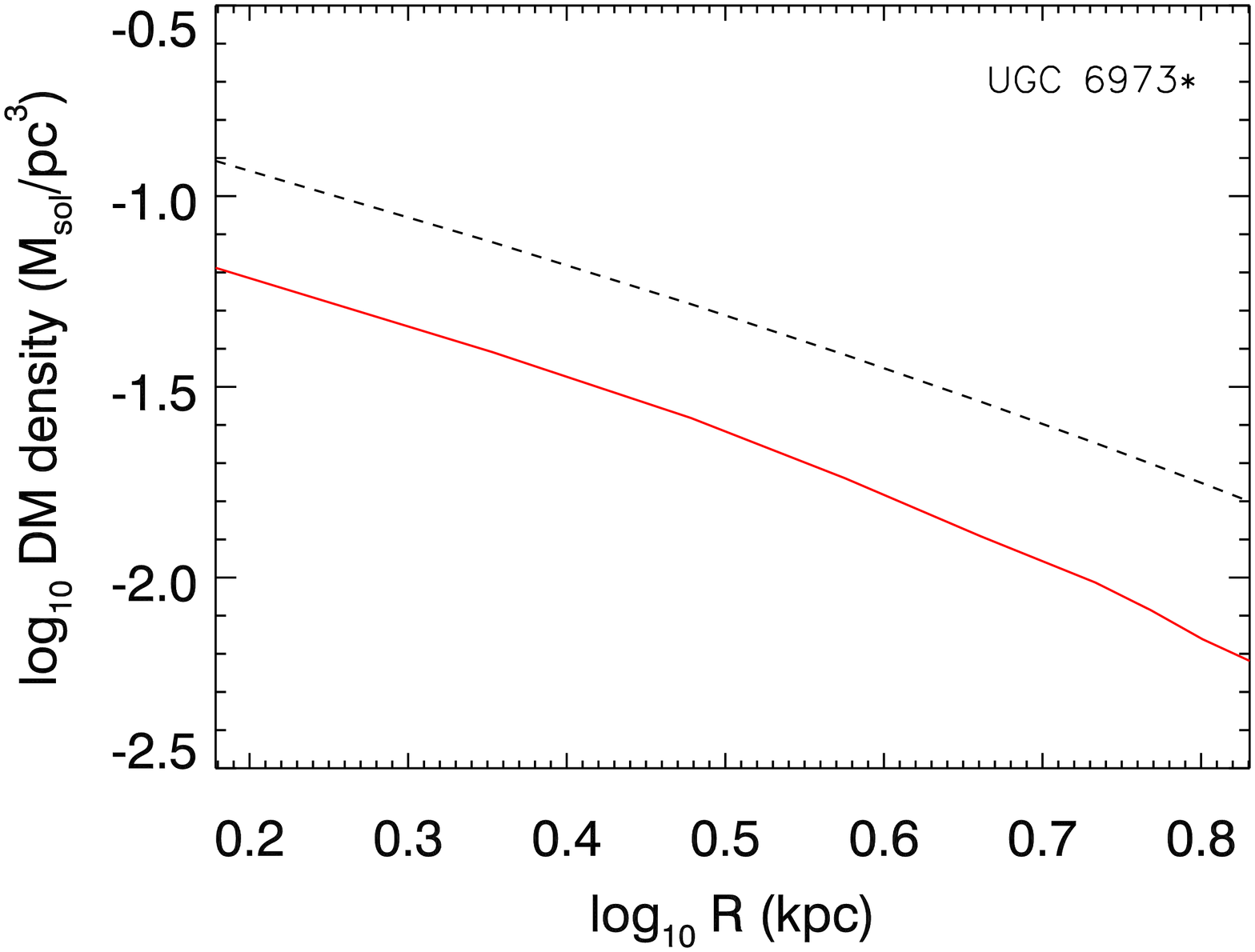}
  \caption{Figure~\ref{fig:hsb_dens} continued.}
\label{fig:hsb_denslast}
\end{figure}

\begin{figure}[!ht]
  \includegraphics[angle=90,width=0.45\textwidth]{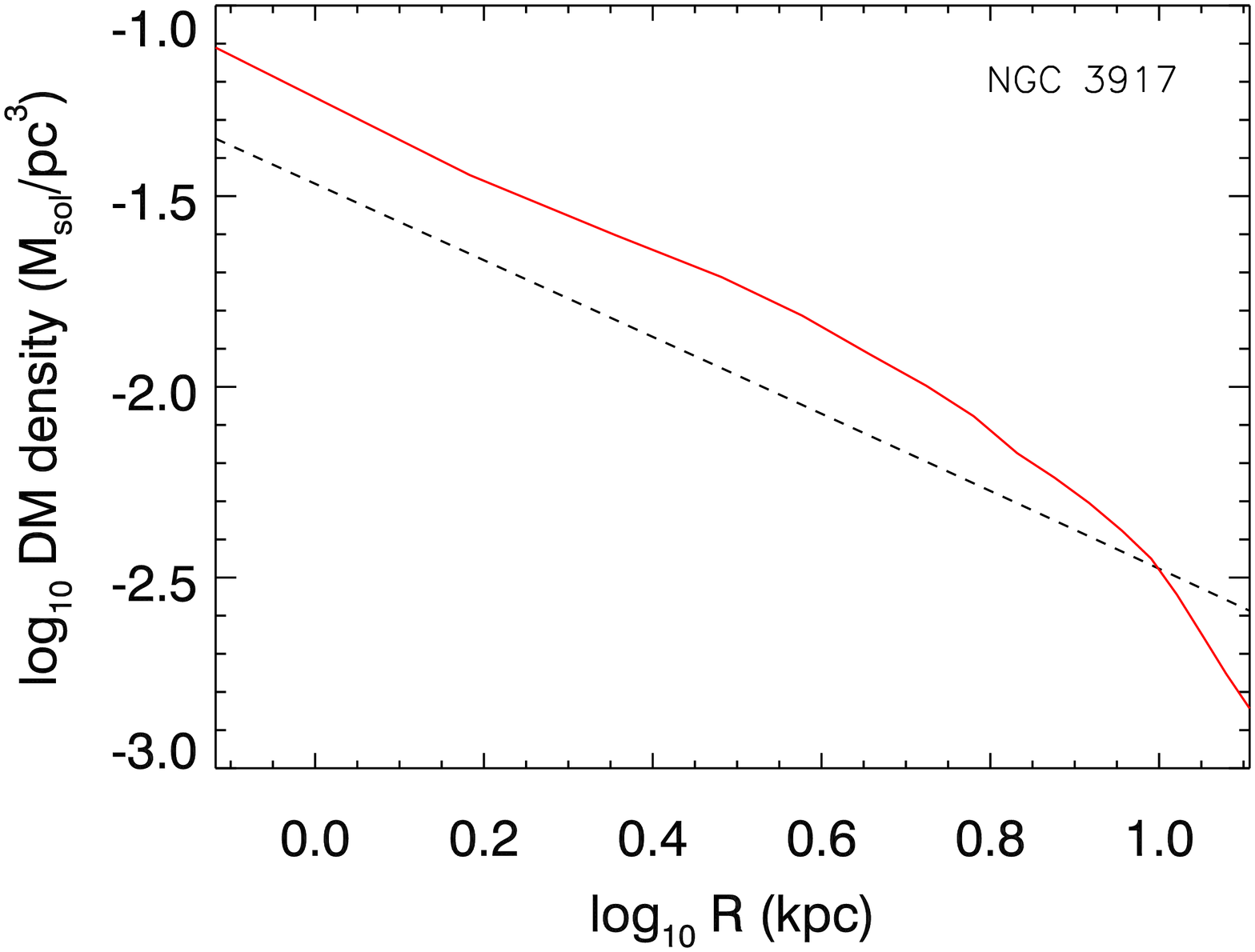}
  \includegraphics[angle=90,width=0.45\textwidth]{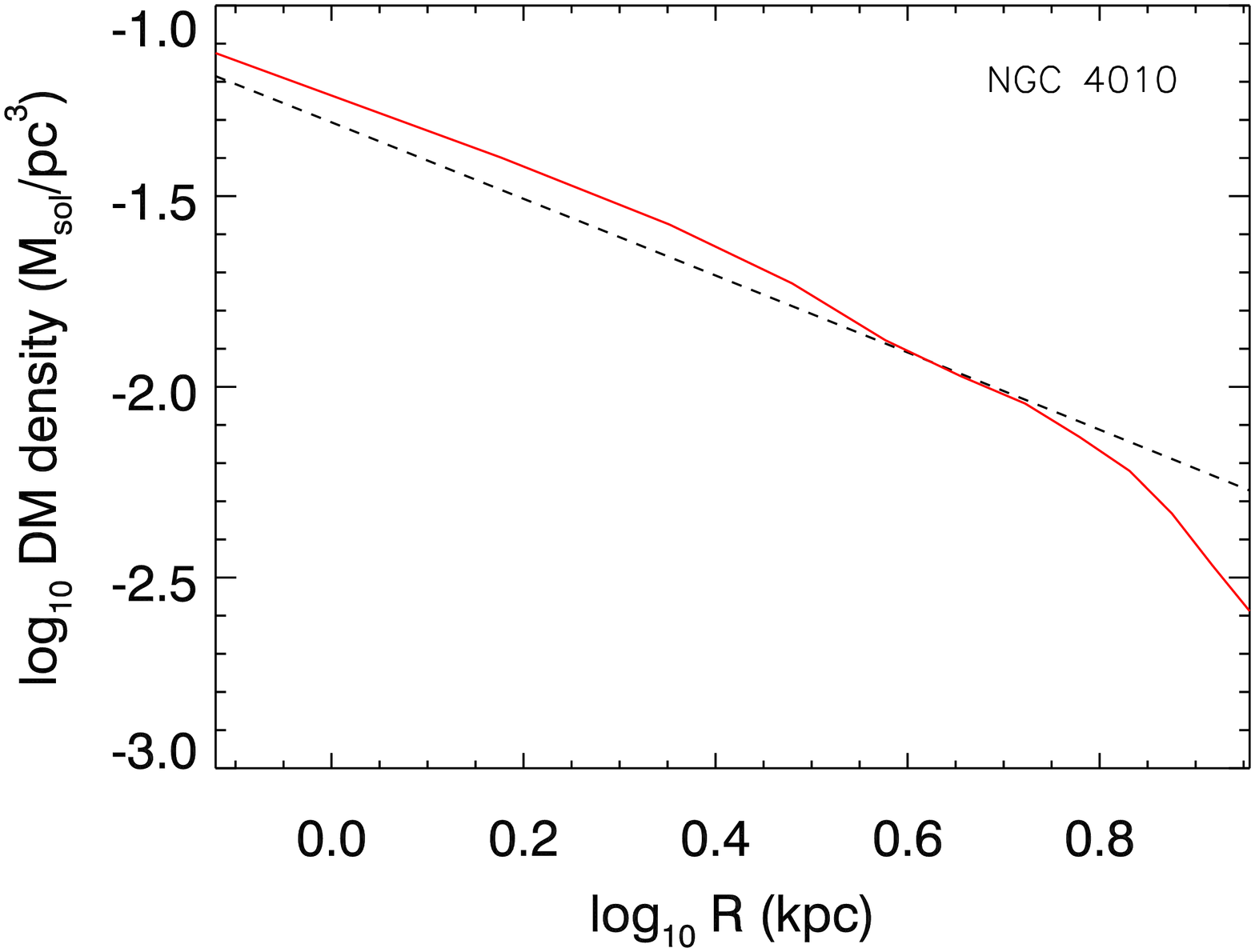}\vspace{3mm}\\
  \includegraphics[angle=90,width=0.45\textwidth]{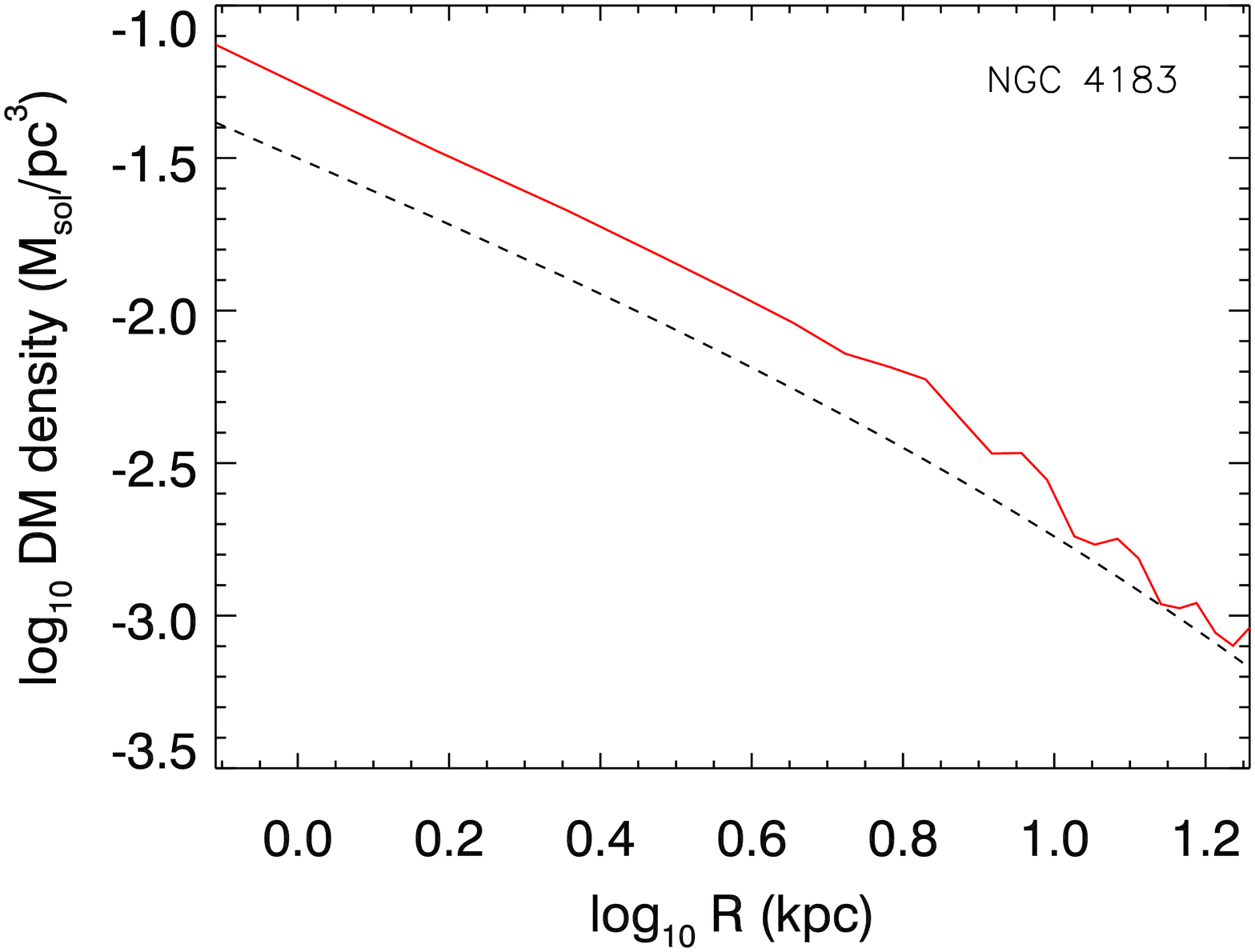}
  \includegraphics[angle=90,width=0.45\textwidth]{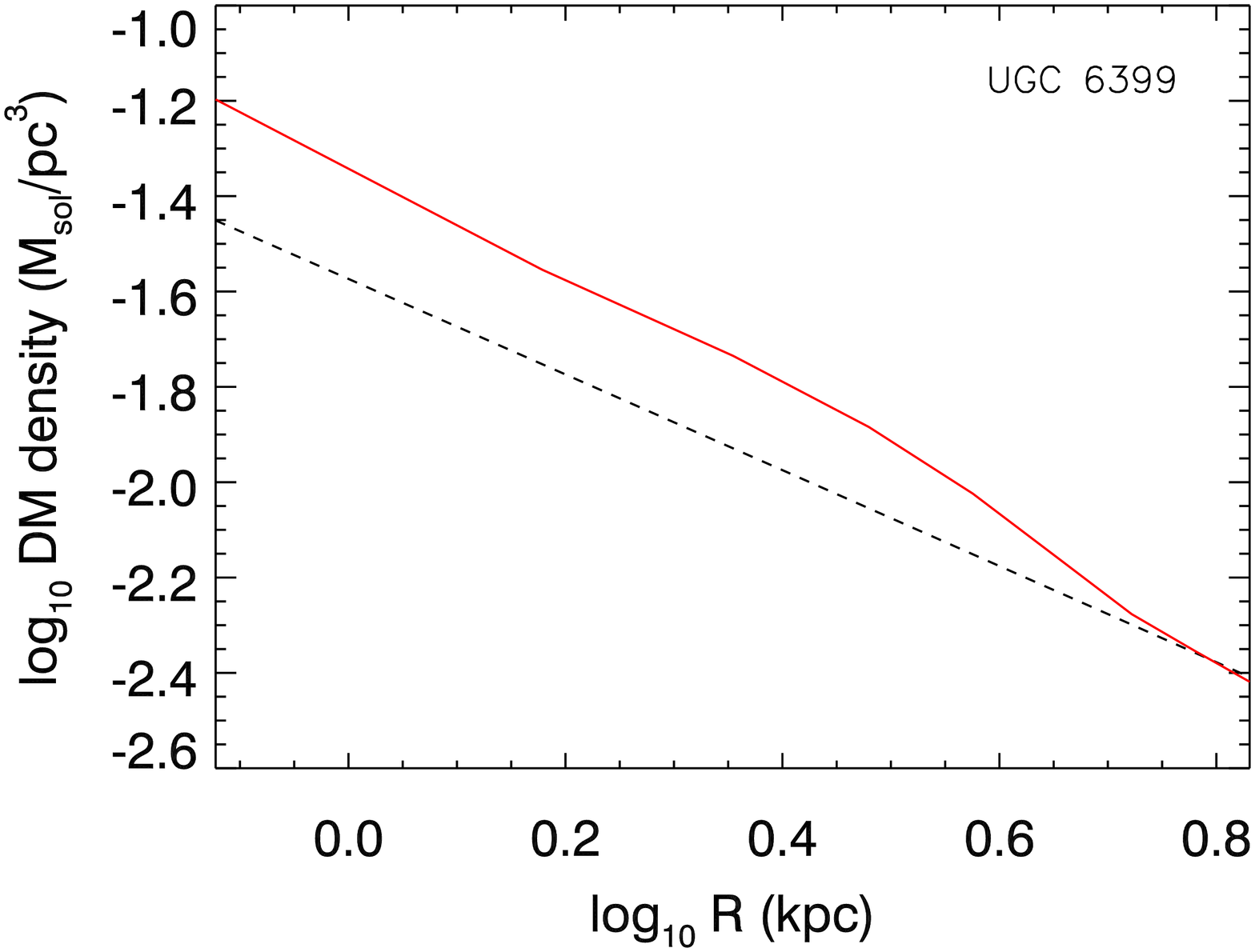}\vspace{3mm}\\
  \includegraphics[angle=90,width=0.45\textwidth]{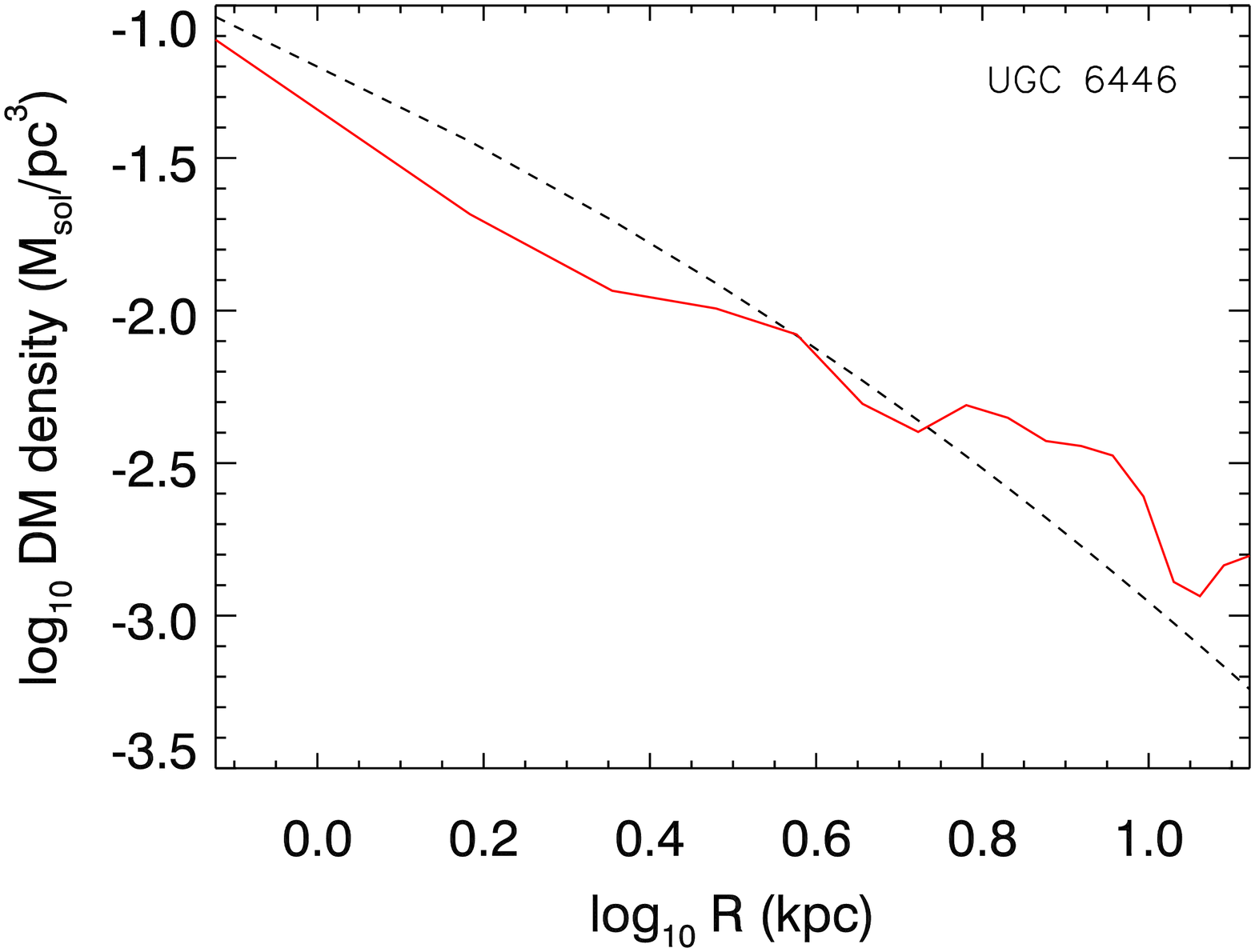}
  \includegraphics[angle=90,width=0.45\textwidth]{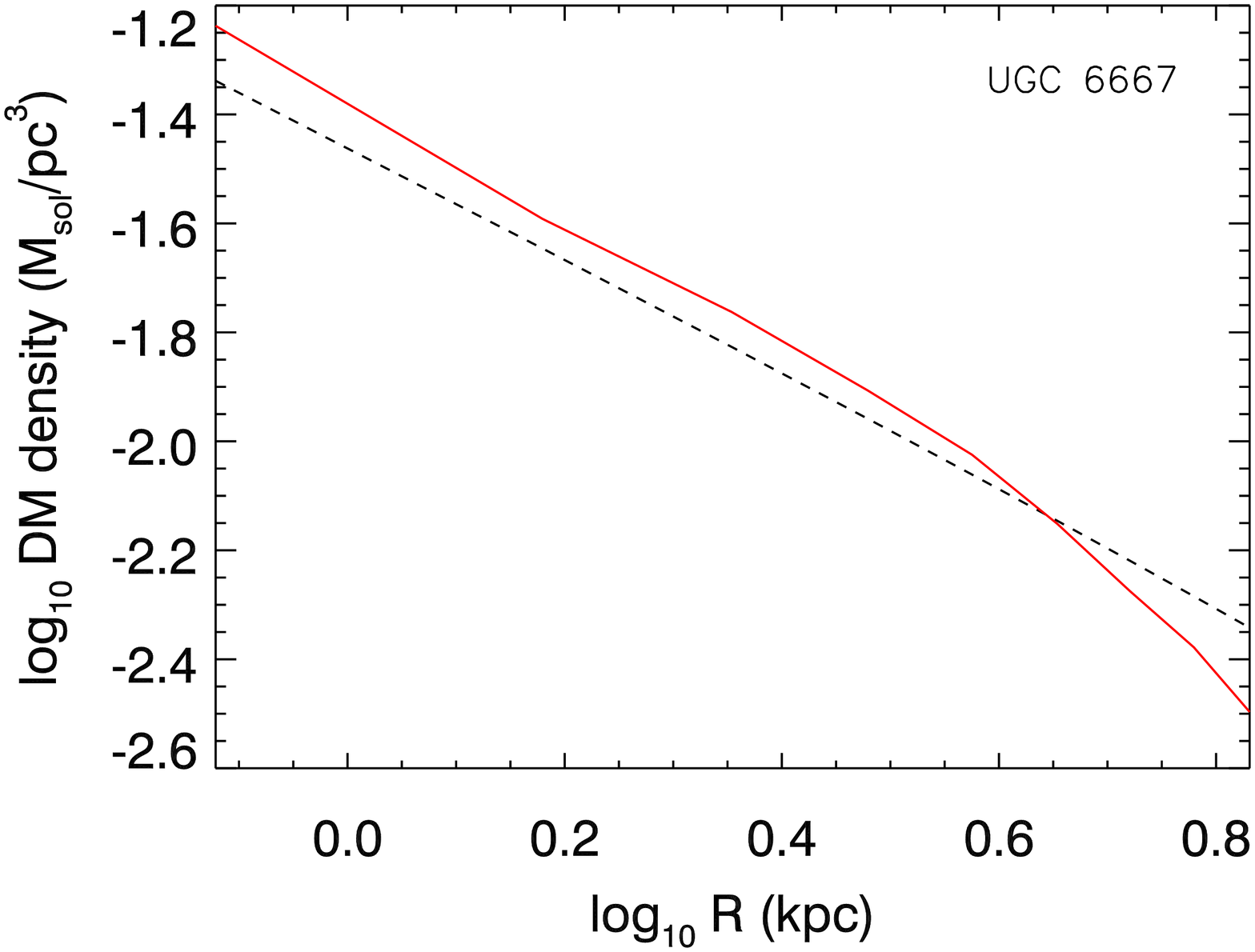}
  \caption{LSB dark matter density profiles. See Figure~\ref{fig:hsb_dens} for description.}
  \label{fig:lsb_dens}
\end{figure}

\begin{figure}[!ht]
  \includegraphics[angle=90,width=0.45\textwidth]{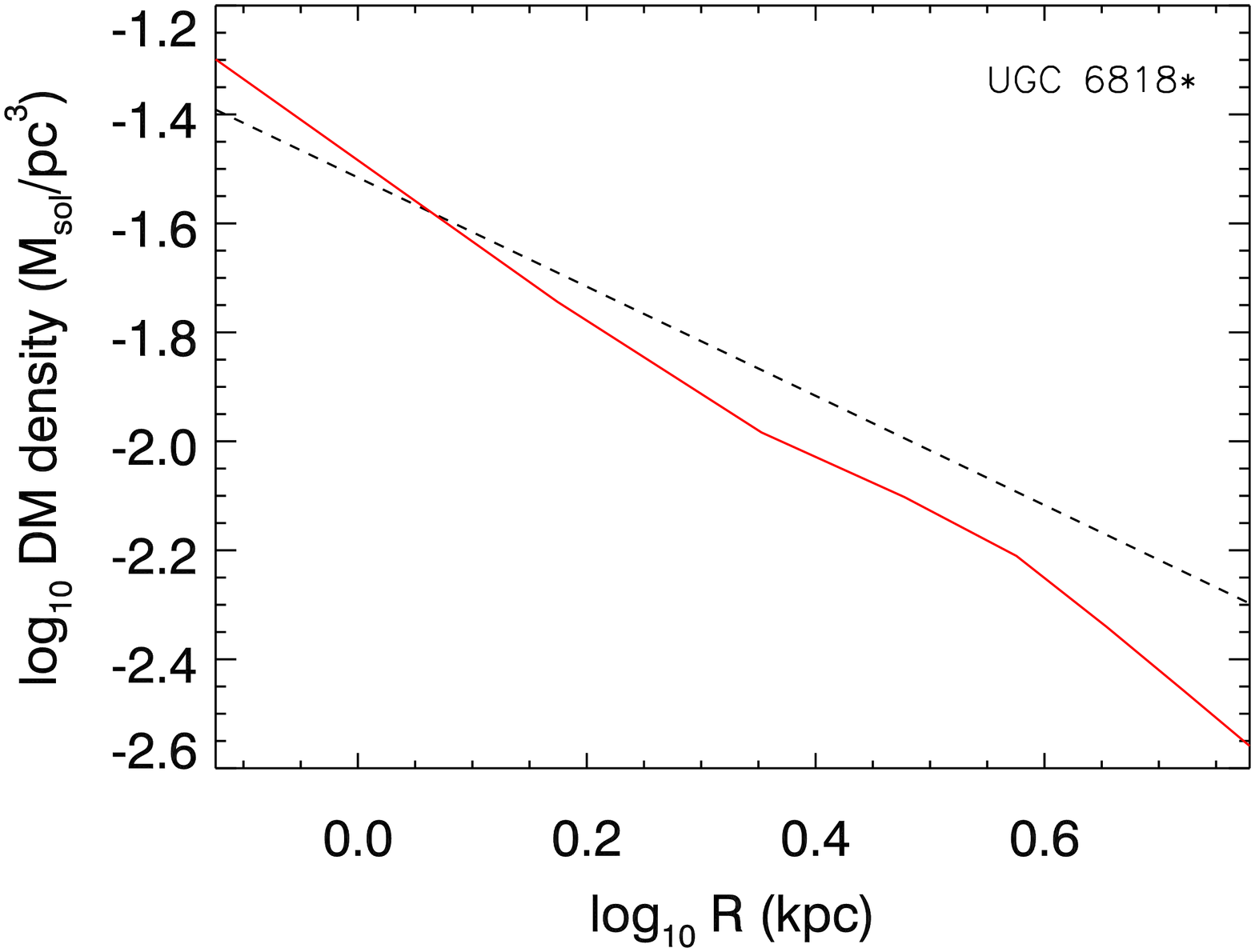}
  \includegraphics[angle=90,width=0.45\textwidth]{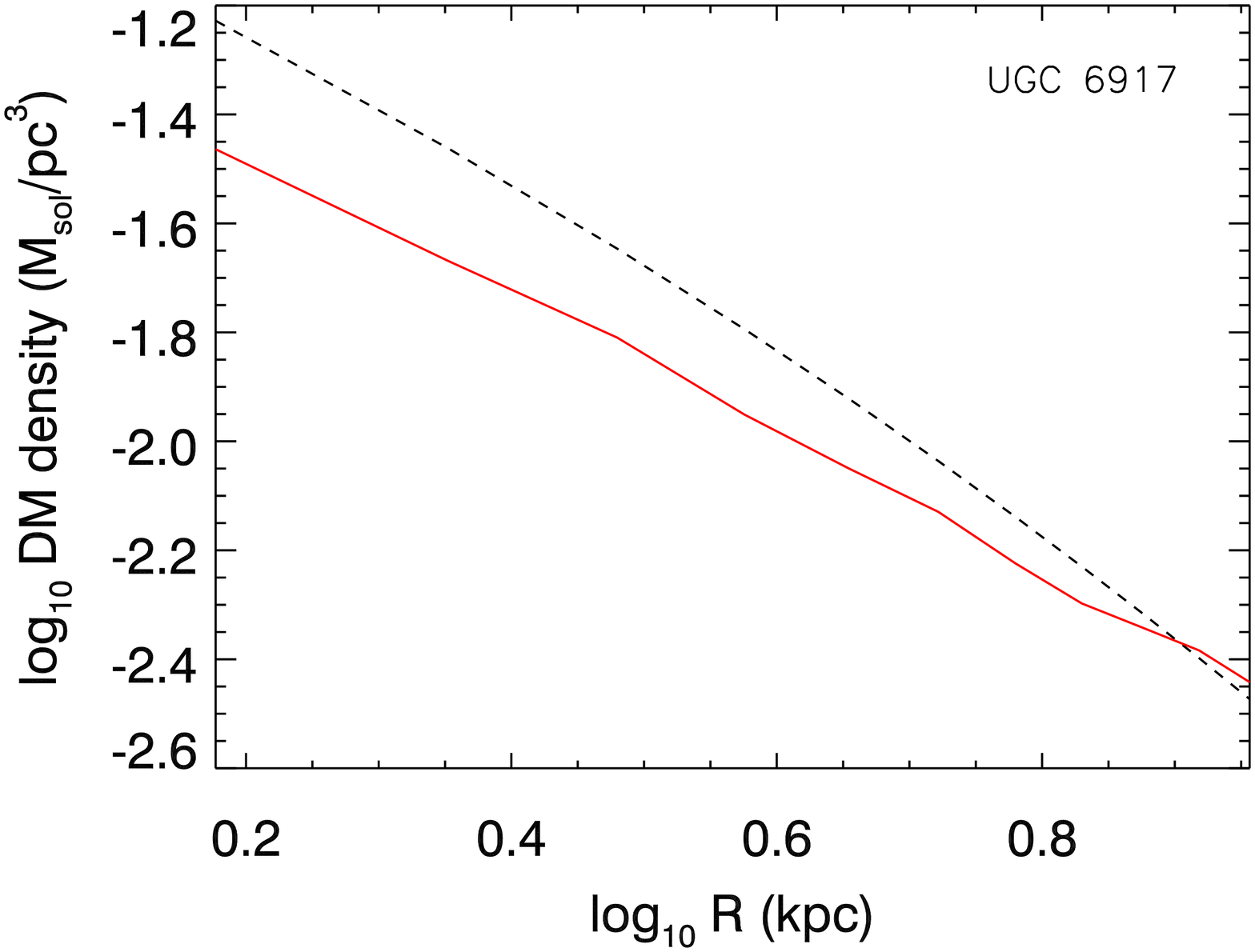}\vspace{3mm}\\
  \includegraphics[angle=90,width=0.45\textwidth]{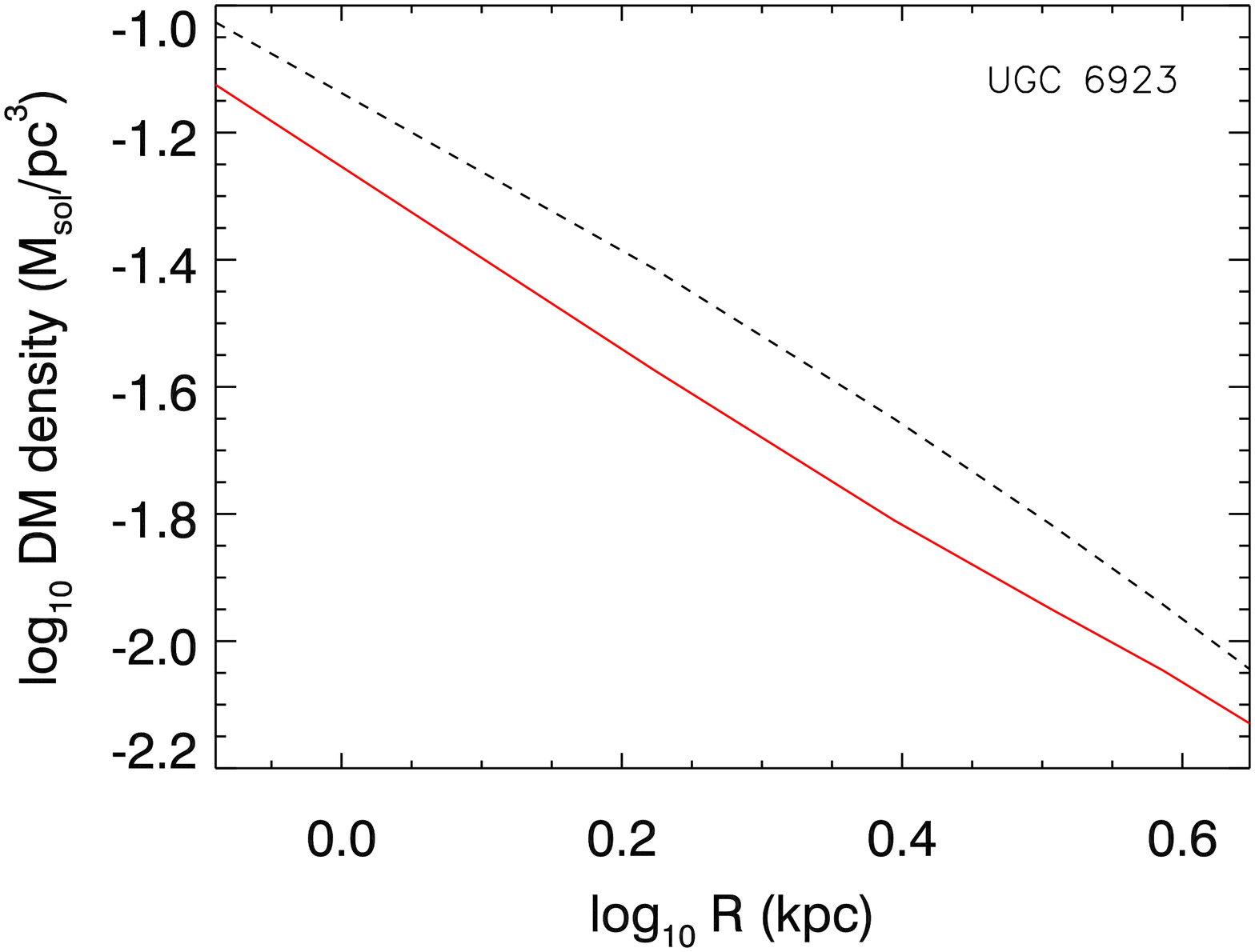}
  \includegraphics[angle=90,width=0.45\textwidth]{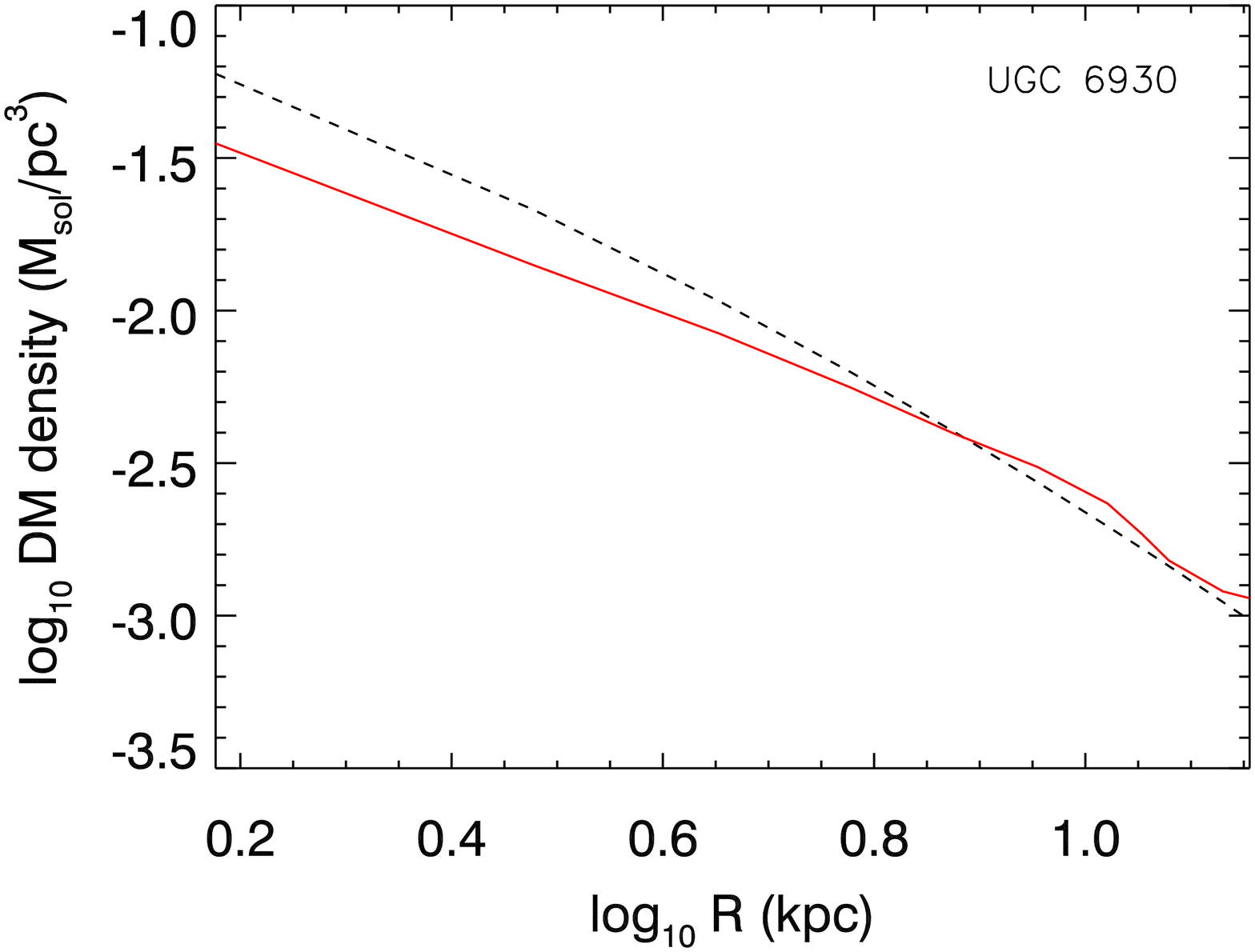}\vspace{3mm}\\
  \includegraphics[angle=90,width=0.45\textwidth]{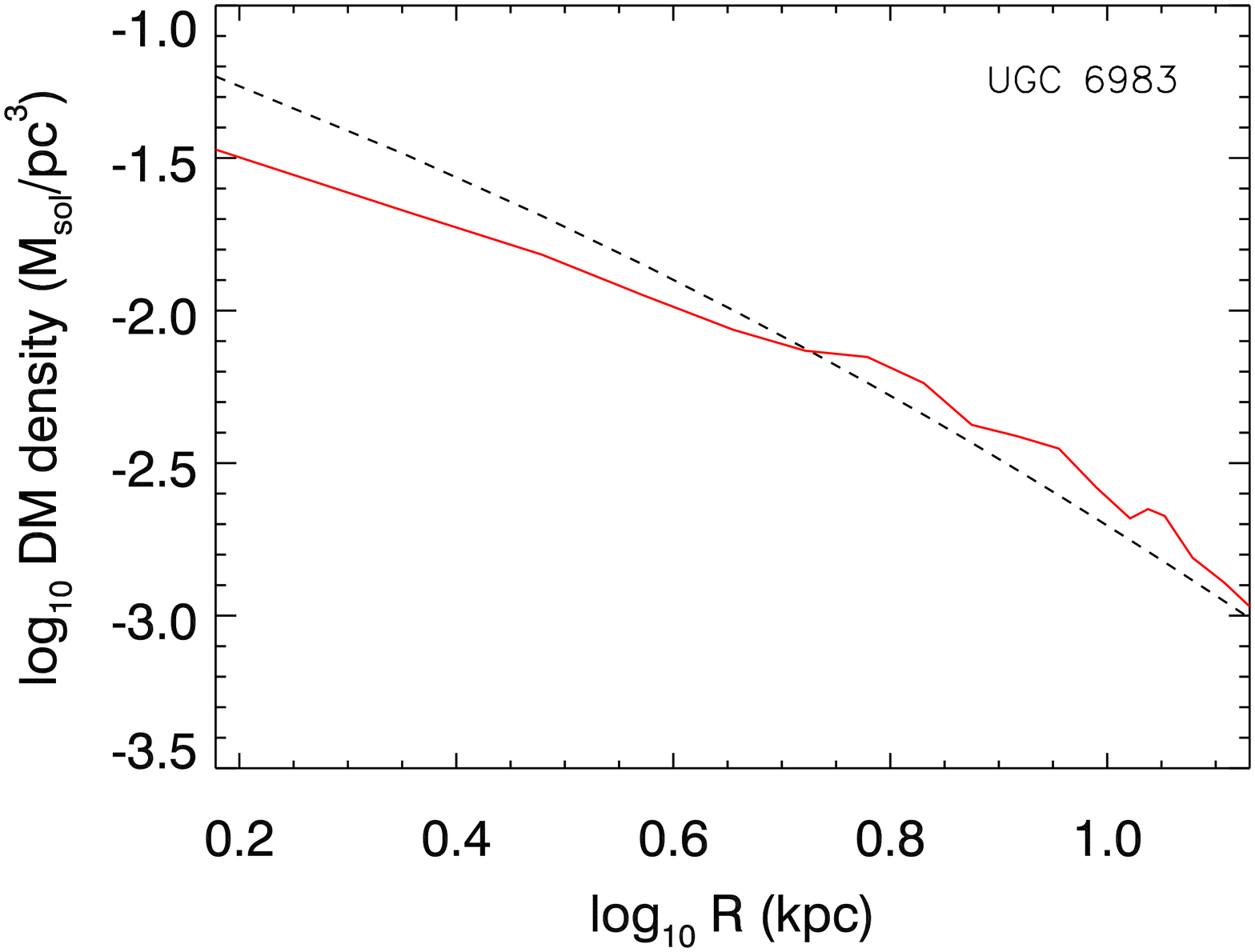}
  \includegraphics[angle=90,width=0.45\textwidth]{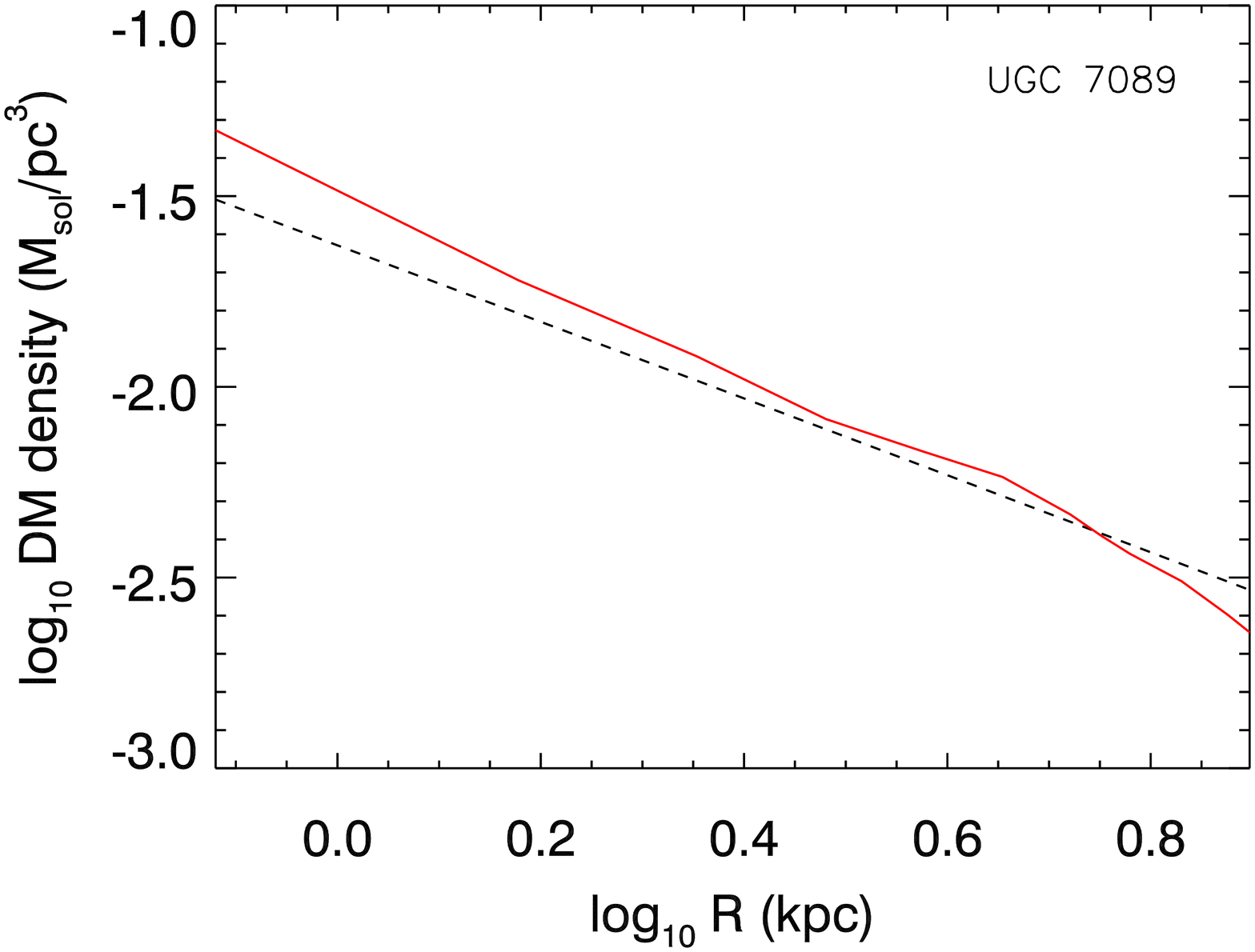}
  \caption{Figure~\ref{fig:lsb_dens} continued.}
\label{fig:lsb_denslast}
\end{figure}


\section{Summary and Discussion}

In this paper, we tested the inertial properties of  MONDian dark matter \citep{HMN, HMN2, HMN3} by fitting  rotation curves in a sample of 30 local galaxies. We compared rotation curves derived from MDM, MOND, and CDM. All three models fit the data well. While the fits for MDM are indistinguishable from the fits for MOND over the range of observed radii, these models are distinguishable from rotation curves at very small radii, as shown in Figure~\ref{MONDvsMDM}. Both MDM and MOND require a single fitting parameter, the mass-to-luminosity ratio $M/L$. We find that $M/L$ is very similar for both models. We also compared the dark matter profiles for MDM and CDM for each of the galaxies. The profiles differ at small $r$ in the range of observed radii but are quite similar at large $r$. Our results show that MDM is a viable model at the galactic scale. We will study constraints from galaxy clusters and the CMB in forthcoming papers.

The interpretation of the above data is unique from the point of view of MDM. On one hand we have one parameter fits to the galactic rotation curves that cannot be distinguished from the MOND fits.
However, the MDM rotation curves arise from the distribution of dark matter, unlike those in MOND, which does not include dark matter, but modifies inertia instead. The force given in Equation~\eqref{modG} arising from our MDM mass profile (Equation~\ref{Mdark}) can also be interpreted as a modification of inertia \citep{HMN}. However, consistency with the holographic formulation of gravity requires the dark matter interpretation \citep{HMN}. Furthermore, the success of CDM at large scales (e.g. galaxy clusters and the CMB) suggests that the dark matter interpretation is a better choice for explaining astronomical data at all scales.


On the other hand, our dark matter candidate is not the standard
CDM candidate.
An essential difference between the dark matter in MDM and the various proposals for the CDM particle
is that, while the latter presumably obeys the ordinary statistics (i.e.,
either Bose or Fermi statistics), the former has been proposed
(Ho, Minic \& Ng 2012) to obey the infinite statistics \citep{Doplicher71,Doplicher74,Govorkov83,Greenberg90}, also known as
the quantum Boltzmann statistics. This is the only known consistent
statistics in greater than two space dimensions without the Gibbs
factor, as described by the Cuntz algebra. If MDM quanta obey infinite statistics (in some sense a middle quantum statistics between the usual Bose-Einstein and Fermi-Dirac statistics), they should not be described by a local quantum field theory of particles.

Besides obeying different statistics, MDM has some unusual inertial properties (see Equation~\eqref{modG}). This provides a natural explanation for the measured underdensity of dark matter in the solar system (e.g., \citealt{moni}; but see \citealt{Bovy} who find a density consistent with stanard expectations). The so-called ``External Field Effect'' \citep[e.g.,][]{mond,external} is built into the inertial properties of MDM, so we do not expect to see dark matter effects in the Solar system.

\acknowledgments

\section*{Acknowledgments}

We would like to thank the referee for many useful comments that have greatly improved the presentation of the material in this paper. We would also like to thank Philip Mannheim and Louie Strigari for illuminating discussions.
CMH and YJN were supported in part by the
US Department of Energy under contract DE-FG05-85ER40226 and DE-FG02-06ER41418, respectively.
DM and TT were supported by the U.S. Department of Energy under contract DE-FG05-92ER40677, Task A. DM was also supported by the U.S. Department of Energy under
contract DE-FG02-13ER41917. TT was also supported by the World Premier International Research Center Initiative (WPI Initiative), MEXT, Japan.

\appendix

\section{Entropic Gravity and MONDian Dark Matter}

Our original proposal \citep{HMN} is based on the intriguing relationship between gravity and thermodynamics \citep{verlinde,Jacob95,Padmanabhan,Smolin}. Inspired by this relationship, Verlinde \citep{verlinde} recently showed that the canonical Newton's laws could be derived from the point of view of holography \citep{hawking,holography,Susskind,adscft}. Verlinde invoked the first law of thermodynamics and proposed the concept of entropic force
\bea
F_\mathrm{entropic} \;=\; T\; \dfrac{\Delta S}{\Delta x} \;,
\eea
where $\Delta x$ is an infinitesimal spatial displacement of a particle with mass $m$ from the heat bath with temperature $T$. Incorporating Bekenstein's original arguments for the entropy $S$ of black holes \citep{bekenstein}, Verlinde demands that $\Delta S = 2\,\pi\, k_B (m \, c/\hbar)\, \Delta x$.
Together with the Unruh temperature formula,
$
k_B \,T = \hbar \,a/( 2\,\pi\, c),
$
associated with a uniformly accelerating (Rindler) observer
\citep{unruh,Davies}, this
leads to Newton's second law
$F_\mathrm{entropic} = T\, \nabla_x S= m \,a$.

To determine the expression for $a$, Verlinde considered an
imaginary quasi-local (spherical) holographic screen of area $A = 4\,\pi\,r^2$
with temperature $T$. He then assumed the equipartition of energy
$E=\frac{1}{2}\, N \,k_B \,T$ with $N$ being
the total number of degrees of freedom (bits) on the screen given by
$N = A\,c^3/(G\, \hbar)$. Using
the energy relation $E=M\,c^2$,
Verlinde obtained $2 \,\pi\, k_B\, T = G\, M /r^2$,
and with the help of the Unruh
temperature formula, recovered exactly the non-relativistic Newton's law
of gravity, namely $a = G \,M /r^2$.

Since we live in an accelerating universe (in
accordance with the $\Lambda$CDM model), it is well-motivated to generalize
Verlinde's proposal to de Sitter (dS) space.
This was precisely the starting point of our derivation in \citet{HMN}. For convenience,
we set $\hbar = c= 1$ henceforth. First of all, the Unruh temperature, as measured by an inertial observer in
de Sitter space with a positive cosmological constant $\Lambda$,
is given by $T_{dS} = a_0/(2\,\pi \,k_B)$ where $a_0=\sqrt{\Lambda/3}$ \,\citep{hawking}. Note that
$\Lambda$ is related to the Hubble scale $H$ through $\Lambda \sim 3 \,H^2$.\, Numerically,
it turns out that $a_0$ is related to Milgrom's critical acceleration as
\bea
\label{azeroac}
a_0 \;\approx\; 2 \,\pi\, a_c \;,
\eea
and so we set $a_c = a_0/(2\pi )$ for simplicity. The corresponding Unruh temperature as measured by a
non-inertial observer with acceleration $a$ is
\citep{deser,Jacob98}
\begin{equation}
T_{dS+a} \;=\; \frac{1}{2\,\pi\, k_B}\,\sqrt{\,a^2+a_0^2\,}\;.
\end{equation}
Consequently, we can define the \emph{net}
temperature as measured by a non-inertial observer
(due to some matter sources that cause the acceleration $a$\,) to be
\bea
\label{net}
\tilde{T}
\;\equiv\; T_{dS+a} - T_{dS}
\;=\; \frac{1}{2\,\pi\, k_B} \left(\,\sqrt{a^2+a_0^2} - a_0 \,\right)
\;.
\eea
Milgrom has suggested that the difference between the
Unruh temperatures as measured by non-inertial and inertial observers in de Sitter space,
namely $2\pi k_B \Delta T =\sqrt{a^2+a_0^2} - a_0$,\,
might give the correct behaviors of the interpolating function for the Newtonian
acceleration and his modified acceleration at very small accelerations \citep{interpol}.
As we will see, adopting
Verlinde's entropic force point of view allows us to
justify his intuition naturally.

Following Verlinde's entropic approach, the force acting on the test
mass $m$ with acceleration $a$ in de Sitter space is given by
\begin{equation}
F_\mathrm{entropic}
\;=\; \tilde{T}\, \nabla_x S
\;=\; m\, \left(\,\sqrt{a^2+a_0^2}-a_0\,\right)\,.
\end{equation}
In order to derive an explicit form for $a$, we apply Verlinde's holographic approach
by invoking an imaginary holographic
screen of radius $r$. Then, using Equation~\eqref{net}, we can write
\begin{equation}
\sqrt{a^2+a_0^2}-a_0
\;=\; 2 \,\pi\, k_B\, \tilde{T}
\;=\; 2\,\pi \,k_B \,\left(\,\frac{2 \tilde{E}} {N\, k_B}\,\right)
\;=\; 4\,\pi \,\left(\,\frac{\tilde{M}} {A / G}\,\right)
\;=\; \frac{G\,\tilde{M}}{r^2}
\;,
\label{GMtilde}
\end{equation}
where $\tilde{M}$ represents the \emph{total} mass enclosed within the volume
$V = 4 \pi r^3 / 3$.

A necessary step is to determine $\tilde{M}$. In \citet{HMN}, we proposed that $\tilde{M} = M+M'$, where $M'$ is some unknown mass in addition to the ordinary mass $M$ enclosed within the volume $V=4\,\pi\, r^3/3$. We interpret $M'$ to be the total dark matter mass within the volume.
We thus have
\begin{equation}
\label{holographic}
\sqrt{a^2+a_0^2}-a_0 \;=\; \frac{G\,(M+M')}{r^2}\;.
\end{equation}
We also proposed a profile for the dark matter mass:\footnote{This mass profile is not unique and can be generalized. For example,
$M'/M~=~\left[~\lambda~\frac{a_0}{a}+\frac{1}{\pi}~\left(~\frac{a_0}{a}~\right)^2~\right]$,
where $\lambda$ is a constant,
is a more general expression that gives the correct asymptotic behavior.}
%
%
\begin{equation}
\label{Mdark}
M' \;=\; \frac{1}{\pi}\,\left(\,\frac{a_0}{a}\,\right)^2\, M \,.
\end{equation}
As a result, using Equations~\eqref{holographic} and \eqref{Mdark},
the entropic force is given by
\begin{equation}
\label{modG}
F_{\mathrm{entropic}}
\;=\; m \,\left(\,\sqrt{a^2+a_0^2}-a_0\,\right)
\;=\; m\,a_N
\,\left[\,1+\frac{1}{\pi}\,\left(\,\frac{a_0}{a}\,\right)^2\,\right]\,.
\end{equation}
By solving the above cubic equation, we can obtain a general
solution for $a$ as a function of $r$.
One could also find a general solution of
$(a_0/a)^2$ and substitute it into
Equation~\eqref{Mdark}
to obtain $M'$ as a function of $r$.
We have used Equation~\eqref{modG} to determine galactic rotation curves in section~\ref{sec:data}.
Before doing so, it is informative to state some general theoretical expectations of our approach.

For $a \gg a_0$, we have $ F_\mathrm{entropic} \approx m \,a \approx m\,a_N
$,
and hence $a \approx a_N = G\,M/r^2$, which is the usual
Newtonian acceleration without dark matter. For $a \ll a_0$,
we have $ F_{entropic} \approx  m \,a^2/(2\,a_0) \approx m\,a_N\,(1/\pi) \,(a_0/a)^2$.
Solving for $a$, we get
\begin{equation}
\label{acceleration}
a \;=\; \left(\,2\, a_N \,a_0^3 / \pi \, \right)^{\frac{1}{4}}\;,
\qquad \mbox{for $a \ll a_0$}\,.
\end{equation}
In order to fit the galactic rotation curves well, Milgrom
requires the force in the regime $a \ll a_0$ to be
\begin{equation}
\label{F_Milgrom}
F_\mathrm{Milgrom}
\;=\; m\,\sqrt{a_N\,a_c}
\;=\; m\,\sqrt{\frac{a_N\,a_0}{2\,\pi}}\,.
\end{equation}
But we recall that for $a \ll a_0$, our scheme predicts
$F_\mathrm{entropic}\approx m \,\frac{a^2}{2\,a_0}$. Substituting
$a= \left(\,2\, a_N \,a_0^3 / \pi \, \right)^{\frac14}$ into
this expression leads to
\begin{equation}
\label{F_smallA}
F_\mathrm{entropic}
\;\approx\; m \,\frac{a^2}{2\,a_0}
\;=\; m\, \sqrt{\frac{a_N\,a_0}{2\,\pi}}
\;=\; F_\mathrm{Milgrom}
\;.
\end{equation}
In conclusion, if we take the total mass of dark matter enclosed
by the volume $V = 4 \pi r^3 / 3$\ to be
given by Equation~\eqref{Mdark}
we can actually derive MOND, even though we have a non-trivial
mass profile.

To study the rotation curves, we note that our scheme requires
$F_\mathrm{centripetal} = m \,a^2/(2\,a_0)$\,
for $a \ll a_0$.
The terminal velocity $v$ is then determined
from $m \,v^2/r = F_\mathrm{centripetal} = m \,a^2/(2\,a_0)$
with $a$ \,given by Equation~\eqref{acceleration}.
Note that in this region, the acceleration $a$ is not related to the velocity $v$ in the
usual fashion as $a = v^2/r$.
That is, $a$ is not the time derivative of $v$.
This leads to
\begin{equation}
v
\;=\; \left(\,\frac{G\,M\,a_0}{2\,\pi}\,\right)^{1/4}
\;= v_{\infty}\; M^{1/4}\,.
\end{equation}
Therefore, we predict flat rotation curves and the Tully-Fisher
relation.

In retrospect, we see that with Equation~\eqref{Mdark},
we can write the entropic force in
the following two apparently different forms:
\begin{equation}
\label{duality}
\frac{G\,(M+M')\,m}{r^2}
\;=\; \frac{G\,M\,m}{r^2}\,\left[\,1+\frac{1}{\pi}\,\left(\,\frac{a_0}{a}\,\right)^2\,\right]\;.
\end{equation}
Interestingly, the LHS of Equation~\eqref{duality} implies that there is no modification of gravity but there is dark matter,
while the RHS implies that there is no dark matter but that there is modification of gravity.
Therefore, according to our scheme,
dark matter and modification of gravity could just be two different manifestations of the same physics at the galactic scale.
We interpret this as a CDM-MOND duality at the galactic scale --- i.e. dark matter with the profile
given by Equation~\eqref{Mdark} could behave as if there existed a modification of gravity but no dark matter.
We thus call our proposal ``MONDian dark matter.''

\section{Comparison of interpolating functions for MOND and MDM} 
\begin{figure}[h] 
\begin{center} 
\includegraphics[width=8cm]{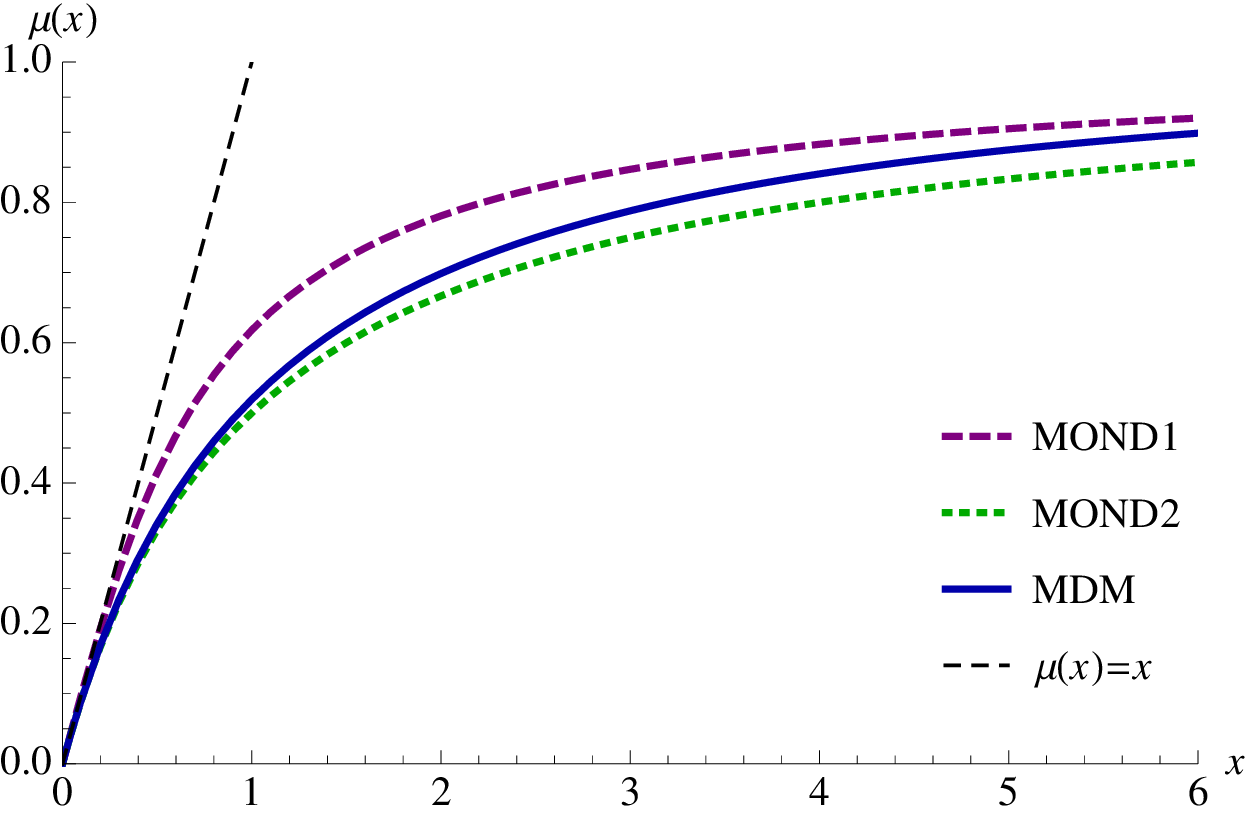} 
\caption{Comparison of the effective interpolating functions $\mu(x)$ for two MOND models 
and MDM.} 
\label{InterpolatingFunction} 
\end{center} 
\end{figure}

As we have argued in the main text, with $x=a_{obs}/a_c$, where $a_{obs}$ is the observed acceleration, the predictions of MOND with interpolating function $\mu(x)$ that enters into $F=ma_{obs}\mu(x)$ 
can be reproduced exactly using the dark matter profile $M'(r)$ calculated from Equation~(7). 
Conversely, we can take any dark matter profile and invert Equation~(7) to 
calculate the corresponding interpolating function in MOND.
For MDM, the dark matter profile is determined from Equation~(18), where $f(r) = M'(r)/M$ as defined in Equation~(17). The corresponding interpolating function is
\begin{equation}
\mu_\mathrm{MDM}(a/a_c) = \frac{1}{1+f(r)}.
\end{equation}
Using the dark matter profile given in Equation~(A8), we obtain
\begin{equation}
\mu_\mathrm{MDM}(a/a_c) = \left[1+4 \pi \left( \frac{a_c}{a} \right)^2 \right]^{-1},
\label{mdminterp}
\end{equation}
where we have used $a_0 = 2 \pi a_c$. As mentioned in Appendix~A, the acceleration parameter $a$ in the MDM mass profile is not the observed acceleration. In order to properly compare the interpolating functions, we write Equation~(\ref{mdminterp}) in terms of the observed acceleration.
From Equations~(20) and (A7), we have $a_{obs} = v^2/r = \sqrt{a^2 + a_0^2} - a_0$, which implies that $a^2 = a_{obs}^2 + 2 a_0 a_{obs}$. Substituting this expression for $a$ into Equation~(\ref{mdminterp}) yields
\begin{equation}
\mu_\mathrm{MDM}(x) = \left(1+\frac{4 \pi}{x^2+4 \pi x} \right)^{-1}.
\label{obsmdminterp}
\end{equation}
This function is plotted in Figure~(\ref{InterpolatingFunction}) along with the interpolating functions for the two MOND cases discussed in the main text. Just as in MOND, $\mu(x) \approx 1$ in the range $x \gg 1$, and $\mu(x) \approx x$ in the range $x \ll 1$ for MDM, as expected.

\clearpage
\begin{deluxetable}{lcccccccc}
 \tablewidth{1.0\textwidth}
 \tablecolumns{9}
 \tabletypesize{\tiny}
 \tablecaption{Galactic Properties}
 \tablehead{
 \colhead{Name} &
 \colhead{$M_B$} &
 \colhead{$M_{MDM}$} &
 \colhead{$M_{CDM}$} &
 \colhead{$\alpha_{MDM}$} &
 \colhead{$\alpha_{MOND}$} &
 \colhead{$\alpha_{CDM}$} &
 \colhead{$c$} &
 \colhead{$v_{200}$} \\
 \colhead{} &
 \colhead{($10^{10} M_\odot$)} &
 \colhead{($10^{10} M_\odot$)} &
 \colhead{($10^{10} M_\odot$)} &
 \colhead{($M_\odot /L_\odot$)} &
 \colhead{($M_\odot /L_\odot$)} &
 \colhead{($M_\odot /L_\odot$)} &
 \colhead{} &
 \colhead{(km~s$^{-1}$)}  \\
  }
 \startdata
 NGC 3726     & 2.83  & 12.4 & 13.5 & 0.52 $\pm$ 0.01 & 0.50 $\pm$ 0.01 & 0.89 $\pm$ 0.03 & 0.4 $\pm$ 0.01  &  991 $\pm$ 93 \\
 NGC 3769*     & 1.23 & 9.77 & 8.35 & 0.45 $\pm$ 0.02 & 0.43 $\pm$ 0.02 & 0.22 $\pm$ 0.12 & 22.5 $\pm$ 4.9 &  76 $\pm$ 4 \\
 NGC 3877     & 3.24 & 3.63 & 2.97 & 0.54 $\pm$ 0.01 & 0.52 $\pm$ 0.01 & 0.61 $\pm$ 0.08 & 11.0 $\pm$ 2.3 & 117  $\pm$ 13 \\
 NGC 3893*     & 3.91 & 8.21 & 10.9 & 0.72 $\pm$ 0.02 & 0.69 $\pm$ 0.01 & 0.00 $\pm$ 0.01 & 60.0 $\pm$ 2.9 & 90  $\pm$ 2 \\
 NGC 3917 (L) & 1.56 & 3.90 & 3.43 & 0.76 $\pm$ 0.01 & 0.74 $\pm$ 0.01 & 1.17 $\pm$ 0.06 & 1.4  $\pm$ 0.2 &  557  $\pm$ 54 \\
 NGC 3949     & 1.50 & 1.50 & 3.98 & 0.47 $\pm$ 0.01 & 0.45 $\pm$ 0.01 & 0.00 $\pm$ 0.00 & 36.4 $\pm$ 3.3 &  104  $\pm$ 7 \\
 NGC 3953     & 8.27 & 7.57 & 10.3 & 0.71 $\pm$ 0.01 & 0.68 $\pm$ 0.01 & 0.34 $\pm$ 0.12 & 29.0 $\pm$ 5.1 &  122  $\pm$ 4 \\
 NGC 3972     & 1.02 & 1.73 & 3.18 & 0.69 $\pm$ 0.02 & 0.66 $\pm$ 0.02 & 0.00 $\pm$ 0.00 & 13.4 $\pm$ 1.3 &  143  $\pm$ 14 \\
 NGC 3992     & 13.2 & 25.6 & 25.4 & 1.64 $\pm$ 0.02 & 1.57 $\pm$ 0.02 & 0.92 $\pm$ 0.27 & 33.0 $\pm$ 7.8 &  129  $\pm$ 108 \\
 NGC 4010 (L) & 0.98 & 2.13 & 2.49 & 0.52 $\pm$ 0.02 & 0.50 $\pm$ 0.02 & 0.52 $\pm$ 0.09 & 1.8 $\pm$ 1.5 &  655 $\pm$ 224 \\
 NGC 4013     & 4.04 & 13.7 & 14.2 & 0.69 $\pm$ 0.01 & 0.66 $\pm$ 0.01 & 0.89 $\pm$ 0.02 & 3.7 $\pm$ 1.0 &  250  $\pm$ 111 \\
 NGC 4051*     & 2.71 & 3.76 & 3.94 & 0.57 $\pm$ 0.02 & 0.55 $\pm$ 0.02 & 0.52 $\pm$ 0.10 & 12.5 $\pm$ 3.8 &  122  $\pm$ 92 \\
 NGC 4085     & 0.90 & 1.06 & 2.46 & 0.55 $\pm$ 0.02 & 0.52 $\pm$ 0.02 & 0.00 $\pm$ 0.01 & 17.2  $\pm$ 2.1 &  146  $\pm$ 22 \\
 NGC 4088*     & 3.58 & 8.49 & 7.00 & 0.41 $\pm$ 0.01 & 0.39 $\pm$ 0.01 & 0.63  $\pm$ 0.10 & 0.9 $\pm$ 7.2 & 750  $\pm$ 334 \\
 NGC 4100     & 3.98 & 9.50 & 8.75 & 0.91 $\pm$ 0.01 & 0.88 $\pm$ 0.01 & 0.87 $\pm$ 0.05 & 20.6 $\pm$ 1.5 & 97   $\pm$ 2 \\
 NGC 4138     & 2.53 & 6.22 & 5.25 & 0.80 $\pm$ 0.03 & 0.76 $\pm$ 0.03 & 0.26 $\pm$ 0.12 & 101.5 $\pm$ 5.7 &  69  $\pm$ 3 \\
 NGC 4157     & 4.68 & 13.8 & 14.2 & 0.61 $\pm$ 0.01 & 0.57 $\pm$ 0.01 & 0.72 $\pm$ 0.07 & 7.8 $\pm$ 3.9 &  157  $\pm$ 182 \\
 NGC 4183 (L) & 0.95 & 4.69 & 3.49 & 0.60 $\pm$ 0.02 &  0.58 $\pm$ 0.02 & 1.36 $\pm$ 0.21 & 6.9 $\pm$ 3.2 & 93  $\pm$ 130 \\ 
 NGC 4217     & 3.94 & 6.46 & 8.83 & 0.60 $\pm$ 0.01 & 0.57 $\pm$ 0.01  & 0.38  $\pm$ 0.03 & 20.4 $\pm$ 0.5 &  116 $\pm$ 2 \\
 NGC 4389*     & 0.25 & 0.54 & 1.01 & 0.14 $\pm$ 0.01 & 0.13 $\pm$ 0.01 & 0.00 $\pm$ 0.00 & 1.6 $\pm$ 0.2 &  1110  $\pm$ 128 \\
 UGC 6399 (L) & 0.29 & 0.93 & 0.75 & 0.74 $\pm$ 0.04 & 0.72 $\pm$ 0.04 & 1.39  $\pm$ 0.26 & 1.0 $\pm$ 0.9 &  554 $\pm$ 114 \\
 UGC 6446 (L) & 0.53 & 2.55 & 1.72 & 0.64 $\pm$ 0.03 & 0.62 $\pm$ 0.03 & 0.00  $\pm$ 0.27 & 19.5 $\pm$ 2.1 &  53 $\pm$ 3 \\
 UGC 6667 (L) & 0.27 & 0.90 & 0.92 & 0.66 $\pm$ 0.03 & 0.64 $\pm$ 0.03 & 0.83 $\pm$ 0.22 & 4.0 $\pm$ 1.0 &  194  $\pm$ 58 \\
 UGC 6818* (L) & 0.09 & 0.49 & 0.68 & 0.26 $\pm$ 0.03 & 0.25 $\pm$ 0.03 & 0.00  $\pm$ 0.00 & 0.8 $\pm$ 0.1 &  732 $\pm$ 78 \\
 UGC 6917 (L) & 0.78 & 1.95 & 2.27 & 0.94 $\pm$ 0.03 & 0.91 $\pm$ 0.03 & 0.00 $\pm$ 0.12  & 17.9 $\pm$ 1.4 &  81 $\pm$ 4 \\
 UGC 6923 (L) & 0.20 & 0.48 & 0.62 & 0.55 $\pm$ 0.04 & 0.53 $\pm$ 0.04 & 0.00 $\pm$ 0.02 & 16.3 $\pm$ 3.5 &  71  $\pm$ 17 \\
 UGC 6930 (L) & 0.79 & 3.32 & 3.38 & 0.75 $\pm$ 0.05 & 0.73 $\pm$ 0.05 & 0.00 $\pm$ 0.16 & 20.9 $\pm$ 2.2 &  71  $\pm$ 5 \\
 UGC 6973*     & 1.62 & 1.77 & 3.64 & 0.49 $\pm$ 0.01 & 0.46 $\pm$ 0.01 & 0.31  $\pm$ 0.03 & 18.8 $\pm$ 6.2 &  141 $\pm$ 121 \\
 UGC 6983 (L) & 0.91 & 3.31 & 2.94 & 1.23 $\pm$ 0.04 & 1.19 $\pm$ 0.04 & 0.63 $\pm$ 0.55 & 21.9 $\pm$ 3.9 &  67  $\pm$ 2 \\
 UGC 7089 (L) & 0.19 & 0.91 & 0.90 & 0.33 $\pm$ 0.03 & 0.32 $\pm$ 0.03 & 0.54  $\pm$ 0.34 & 1.0 $\pm$ 0.6 &  486 $\pm$ 113 \\

 \enddata
\label{tab:galaxies}
\tablecomments{Sample of galaxies from \citet{Sanders98}. Asterisks denote galaxies with disturbed velocity fields \citep{Sanders98}. LSBs are marked with an (L), and the rest are HSBs. For each galaxy, the baryonic, MDM, and CDM masses contained within the observed radii are given in columns 2---4, respectively. Mass-to-light ratios ($\alpha$) in the K' band are given for MDM, MOND, and CDM in columns 5---7, respectively.}
\end{deluxetable}
\clearpage



\begin{thebibliography}{99}

\bibitem[Aharony et al. (2000)]{adscft}
Aharony, O., Gubser, S. S., Maldacena, J., Ooguri, J. H., and Oz, Y. 2000,
Phys. Rept. {\bf 323}, 183.

\bibitem[Angus et al. (2007)]{Angus07} Angus, G.~W., Shan, 
H.~Y., Zhao, H.~S., and Famaey, B.\ 2007, ApJL, {\bf 654}, L13.


\bibitem[Begeman (1987)]{Begeman87}
Begeman, K. G. 1987, Ph.D. dissertation, Kapteyn Institute.

\bibitem[Begeman et al. (1991)]{Begeman91}
Begeman, K. G., Broeils, A. H., and Sanders, R. H. 1991,
MNRAS {\bf 249}, 523.

\bibitem[Bekenstein (1973)]{bekenstein}
Bekenstein, J. D. 1973, Phys. Rev. {\bf D7}, 2333.

\bibitem[Bekenstein (2004)]{teves}
Bekenstein, J. D. 2004, Phys. Rev. D {\bf 70}, 083509.

\bibitem[Bertone, Hooper \& Silk (2005)]{dark}
Bertone, G., Hooper, D. and Silk, J. 2005,
Phys. Rept. {\bf 405}, 279 and references therein.

\bibitem[Bidin et al. (2012)]{moni}
Bidin, C. M., Carraro, G., Mendez, R. A. and Smith, R. 2012,
arXiv:1204.3919 [astro-ph.GA].

\bibitem[Blanchet \& Le Tiec (2009)]{Blanchet}
Blanchet, L. and Le Tiec A. 2009, Phys. Rev. D {\bf 80} 023524.

\bibitem[Blanchet \& Novak (2011)]{external}
Blanchet, L. and Novak, J. 2011, MNRAS {\bf 412}, 2530.

\bibitem[Bottema et al. (2002)]{Bottema}
Bottema, R., Pesta{\~n}a, J.~L.~G., Rothberg, B., \& Sanders, R.~H. 2002, A\&A {\bf 393}, 453. 

\bibitem[Bovy \& Tremaine (2012)]{Bovy}
  Bovy, J., Tremaine, S. 2012,
  ApJ {\bf 756}, 89.

\bibitem[Catinella, Giovanelli \& Haynes (2006)]{Catinella} Catinella, B., 
Giovanelli, R., \& Haynes, M.~P.\ 2006, ApJ {\bf 640}, 751. 

\bibitem[Cen (2001)]{cen}
Cen, R. 2001, ApJ  {\bf 546}, L77.

\bibitem[Clowe, Gonzalez \& Markevitch (2004)]{bullet}
Clowe, D., Gonzalez A. \& Markevitch, M. 2004,
Ap J {\bf 604}, 596.

\bibitem[Cyburt (2004)]{Cyburt04} Cyburt, R.~H.\ 2004, Phys. Rev. D {\bf 70},
023505.

\bibitem[Davies (1975)]{Davies}
Davies, P. C. W. 1975, J. Phys. {\bf A8}, 609.

\bibitem[de Blok \& Bosma (2002)]{deBlok02}
de Blok, W.~J.~G., \& Bosma, A.\ 2002, \aap, 385, 816

\bibitem[Deser \& Levin (1997)]{deser}
Deser, S. and Levin, O. 1997, Class. Quant. Grav. {\bf 14}, L163.

\bibitem[Doplicher, Haag \& Roberts (1971)]{Doplicher71}
Doplicher, S., Haag, R. and Roberts, J. 1971, Commun. Math. Phys. {\bf 23}, 199.

\bibitem[Doplicher, Haag \& Roberts (1974)]{Doplicher74}
Doplicher, S., Haag, R. and Roberts, J. 1971, Commun. Math. Phys. {\bf 35}, 49.

\bibitem[Famaey \& Binney (2005)]{Famaey05}
Famaey, B. and Binney, J. 2005, MNRAS {\bf 363}, 603.

\bibitem[Famaey \& McGaugh (2012)]{fmrev}
Famaey, B. and McGaugh, S. 2012, Living Rev.\ Rel.\  {\bf 15}, 10.

\bibitem[Fermi-LAT (2012)]{fermi}
Fermi-LAT Collaboration 2012, arXiv:1205.2739.

\bibitem[Firmani et al.(2000)]{Firmani00} Firmani, C., D'Onghia,
E., Avila-Reese, V., Chincarini, G., \& Hern{\'a}ndez, X.\ 2000, MNRAS,{\bf 315}, L29.

\bibitem[Frenk \& White (2012)]{Frenk}
Frenk, C. and White, S. 2012, Annalen Phys. {\bf 524}, 507.

\bibitem[Gentile et al.(2007)]{Gentile}
Gentile, G., Famaey, B., Combes, F., Kroupa, P., Zhao, H.S. and Tiret, O. 2007, A\&A {\bf 472}, L25.

\bibitem[Greenberg (1990)]{Greenberg90}
Greenberg, O. W. 1990, Phys. Rev. Lett. {\bf 64}, 705.

\bibitem[Govorkov (1983)]{Govorkov83}
Govorkov, A. B. 1983, Theor. Math. Phys. {\bf 54}, 234.

\bibitem[Hawking (1975)]{hawking}
 Hawking, S. W. 1975, Comm. Math. Phys. {\bf 43}, 199.

\bibitem[Ho, Minic \& Ng (2010)]{HMN}
  Ho, C. M., Minic, D. and Ng, Y. J. 2010,
  Phys.\ Lett.\ B {\bf 693}, 567.

\bibitem[Ho, Minic \& Ng (2011)]{HMN2}
  Ho, C. M., Minic, D. and Ng, Y. J. 2011,
  Gen.\ Rel.\ Grav.\  {\bf 43}, 2567.

\bibitem[Ho, Minic \& Ng (2012)]{HMN3}
  Ho, C. M., Minic, D. and Ng, Y. J. 2012,
  Phys.\ Rev.\ D {\bf 85}, 104033.

\bibitem[Hojman, Kuchar \& Teitelboim (1976)]{Hojman}
Hojman, S. A., Kuchar, K. and Teitelboim, C. 1976, Annals Phys.\  {\bf 96}, 88.

\bibitem[Jacobson (1995)]{Jacob95}
  Jacobson, T. 1995, Phys. Rev. Lett. {\bf 75}, 1260.

\bibitem[Jacobson (1998)]{Jacob98}
Jacobson, T. 1998, Class. Quant. Grav. {\bf 15}, 251.

\bibitem[Kaplinghat \& Turner (2002)]{turner}
Kaplinghat, M. and Turner, M.S 2002, ApJ Lett. {\bf 569}, 19.

\bibitem[Klinkhamer \& Kopp (2011)]{Klinkhamer}
Klinkhamer, F. R. and Kopp, M., 2011, Mod. Phys. Lett. A {\bf 26}, 2783.

\bibitem[Kuchar (1974)]{Kuchar}
Kuchar, K. 1974, J.\ Math.\ Phys.\  {\bf 15}, 708.

\bibitem[Lineweaver et al.(1997)]{Lineweaver97}
Lineweaver, C.~H., Barbosa, D., Blanchard, A., \& Bartlett, J.~G.\ 1997, A\&A {\bf 322}, 365.

\bibitem[Markwardt (2009)]{mpfit}
Markwardt, C.~B.\ 2009, Astronomical Data Analysis Software and Systems XVIII, {\bf 411}, 251.

\bibitem[McGaugh et al.(2000)]{McGaugh}
McGaugh, S. S., et al. 2000, ApJ {\bf 533}, L99.

\bibitem[McGaugh et al.(2007)]{McGaugh07}
McGaugh, S.~S., de Blok, W.~J.~G., Schombert, J.~M., Kuzio de Naray, R., 
\& Kim, J.~H. 2007, ApJ {\bf 659}, 149.

\bibitem[McGaugh et al.(2010)]{McGaugh2}
McGaugh, S.S., Schombert, J.M., de Blok, W.J.G. and Zagursky, M.J. 2010, ApJ {\bf 708}, L14.

\bibitem[Milgrom (1983)]{mond}
Milgrom, M. 1983, ApJ {\bf 270}, 365, 371, 384.

\bibitem[Milgrom (1999)]{interpol}
Milgrom, M. 1999, Phys. Lett. {\bf A253}, 273.

\bibitem[Milgrom (2009)]{dsmond}
Milgrom, M. 2009, ApJ {\bf 698}, 1630.

\bibitem[Mor\'{e} (1978)]{minpack}
Mor\'{e}, J. J. 1978, Numerical Analysis {\bf 630} ed. G. A. Watson, Springer-Verlag: Berlin, 105.

\bibitem[Navarro, Frenk \& White (1996)]{nfw}
Navarro, J.~F., Frenk, C.~S., and White, S.~D.~M. 1996,
ApJ {\bf 462}, 563.

\bibitem[Padmanabhan (2010)]{Padmanabhan}
  Padmanabhan, T. 2010, Mod.\ Phys.\ Lett.\ A {\bf 25}, 1129.


\bibitem[Persic \& Salucci (1991)]{Persic1} Persic, M., \& Salucci, P.\ 1991, ApJ {\bf 368}, 60.

\bibitem[Persic, Salucci \& Stel (1996)]{Persic2}
Persic, M., Salucci, P., Stel, F. 1996, MNRAS {\bf 281} 27.

\bibitem[Planck (2013)]{planck}
  Planck Collaboration 2013, arXiv:1303.5075.

\bibitem[Profumo (2012)]{Profumo}
  Profumo, S. 2012, arXiv:1209.5702 [hep-ph].

\bibitem[Rubin, Ford \& Thonnard (1980)]{Rubin80}
Rubin, V.~C., Ford, W.~K.~J., \& .~Thonnard, N.\ 1980, ApJ {\bf 238}, 471.

\bibitem[Sanders \& Verheijen(1998)]{Sanders98}
Sanders, R.~H., and Verheijen, M.~A.~W. 1998, ApJ {\bf 503}, 97.

\bibitem[Shevchenko (2008)]{infstat}
Shevchenko, V. 2008, arXiv:0812.0185.

\bibitem[Smolin (2010)]{Smolin}
  Smolin, L. 2010, arXiv:1001.3668.

\bibitem[Sommer-Larsen (2006)]{Sommer06} Sommer-Larsen, J.\ 2006, 
ApJL {\bf 644}, 1.

\bibitem[Spergel \& Steinhardt(2000)]{Spergel00}
Spergel, D.~N., \& Steinhardt, P.~J.\ 2000, Phys. Rev. Lett. {\bf 84}, 3760.


\bibitem[Strigari (2012)]{Strigari}
Strigari, L.E. 2012, arXiv:1211.7090.

\bibitem[Susskind (1995)]{Susskind}
 Susskind, L. 1995, J. Math. Phys. {\bf 36}, 6377.

\bibitem[Swaters et al.(2003)]{Swaters03} Swaters, R.~A., Madore,
B.~F., van den Bosch, F.~C., \& Balcells, M.\ 2003, ApJ {\bf 583}, 732.

\bibitem[Swaters, Sanders \& McGaugh (2010)]{Swaters10}
Swaters, R.~A., Sanders, R.~H., \& McGaugh, S.~S.\ 2010, ApJ {\bf 718}, 380.

\bibitem['t Hooft (1993)]{holography}
 't Hooft, G. 1993, arXiv: gr-qc/9310026.

\bibitem[Tully et al.(1996)]{Tully96}
Tully, R.~B., Verheijen, M.~A.~W., Pierce, M.~J., Huang, J.-S.,
and Wainscoat, R.~J.\ 1996, AJ {\bf 112}, 2471.

\bibitem[Tully \& Fisher (1977)]{TF}
Tully, R. B. and Fisher, J. R. 1977, A\&A {\bf 54},
661.

\bibitem[Unruh (1976)]{unruh}
Unruh, W. G. 1976, Phys. Rev. {\bf D14}, 870.

\bibitem[van den Bosch \& Dalcanton (2000)]{vandenBosch}
  van den Bosch, F. C., Dalcanton, J. J.,
  arXiv:astro-ph/0007121.

\bibitem[Verheijen(1997)]{Verheijen97} Verheijen, M.~A.~W.\ 1997,
Ph.D.~dissertation, Kapteyn Institute.

\bibitem[Verlinde (2011)]{verlinde}
  Verlinde, E. 2011, J. High Energy Phys. {\bf 1104}, 029.

\end{thebibliography}
\end{document}